\documentclass[article,preprint,showkeys,secnumarabic,amsfonts,showpacs,amsmath,amssymb]{revtex4}
\usepackage{amsmath}
\usepackage{appendix}
\usepackage[dvips]{color}
\usepackage{epsfig}
\usepackage{graphicx}
\usepackage{amssymb,epsf}
\usepackage{epstopdf}
\usepackage{makeidx}
\usepackage{verbatim}
\usepackage{hyperref}
\usepackage{multirow}
\usepackage{natbib}
\usepackage{amsthm}

\usepackage{cancel}
\usepackage{enumitem}

\def\Box{\hbox{$\rlap{$\sqcup$}\sqcap$}}

\begin{document}
	\title{Non-singular Bouncing Cosmology in $f(R,G,T)$--Quintom model}
	
	\author{Farzad Milani}\email{fmilani@tvu.ac.ir}\affiliation{Department of Basic Sciences, Technical and Vocational University (TVU), Tehran, Iran.}
	\date{\today}
	
	\begin{abstract}
		We present a unified framework for non-singular bouncing cosmologies in modified gravity, combining $f(R,G,T)$ geometry with quintom scalar dynamics in a flat FLRW universe. While single-field models achieve phantom divide line (PDL) crossing and stable bounces, our $f(R,G,T)$-quintom coupling provides a novel implementation of a \textit{double} PDL crossing of $\omega_{\text{eff}}$ during the bounce. 
			
		We address stability concerns through Hamiltonian analysis, showing that FLRW symmetry constraints suppress Ostrogradsky instabilities by reducing higher-derivative terms to metric invariant. The scalar field equation of motion is explicitly derived, confirming cancellation of pathological modes.
			
		Numerical reconstruction of five $f(R,G,T)$ models confirms non-singular bounces with $\rho_{\text{eff}}>0$ and $c_s^2 \geq 0$, alongside parametric control over energy condition violations. Our work extends prior studies by: (1) unifying early-time bounce dynamics with late-time dark energy, (2) demonstrating a novel double-PDL crossing signature compatible with FLRW stability, and (3) establishing explicit ghost-free criteria for higher-derivative terms.
	\end{abstract}
	
	
	\maketitle
	
	\section{Introduction}\label{sec:Introduction}
	
	The recent observational data clearly indicate that the universe is undergoing accelerated expansion \cite{Riess1998Observational, perlmutter1999measurements, peiris2003first, spergel2007three}, presenting modern cosmology with the dual challenge of explaining late-time acceleration while avoiding initial singularities. While the $\Lambda$CDM model remains phenomenologically successful \cite{aghanim2020planck}, its theoretical shortcomings regarding the cosmological constant's magnitude have driven interest in alternatives. Among these, $f(R)$ gravity \cite{nojiri2011unified} and scalar-tensor theories \cite{faraoni1998conformal} have shown promise, particularly in contexts where conformal invariance---a property absent in General Relativity (GR)---plays a key role \cite{essen1990general}. Conformal Weyl gravity, for instance, leverages local conformal invariance of the metric \cite{takook2010linear} to address quantum-gravitational inconsistencies and galactic-scale anomalies \cite{mannheim1997galactic}.  
	
	The standard cosmological framework is based on the Einstein-Hilbert action:
	\begin{equation}
		\mathcal{S}_{EH} = \frac{1}{16\pi G_r} \int d^4 x \sqrt{-g} \, R,
	\end{equation}
	where $G_r$ is Newton's gravitational constant and $R$ is the Ricci scalar \cite{FeynmanGravitation1995}. Modifications to this action must reconcile two competing demands: theoretical consistency (e.g., avoiding Ostrogradsky instabilities) and observational viability (e.g., enabling phantom divide line (PDL) crossings). While single-field models can achieve PDL crossings \cite{hu2007models,deffayet2010imperfect} and stable bounces \cite{qiu2011bouncing,lin2011matter}, our work synthesizes three advances: (1) the geometric control of Gauss-Bonnet terms \cite{nojiri2005modified}, where $G=24H^2(\dot{H}+H^2)$; (2) the stability of $f(R)$-based PDL crossings \cite{bamba2008future}; and (3) the parametric flexibility of quintom models \cite{feng2005dark}. Crucially, the FLRW symmetry constraints $R=6(\dot{H}+2H^2)$ and $G=24H^2(\dot{H}+H^2)$ reduce higher-derivative terms to second order, suppressing ghosts as demonstrated in \cite{ye2019bounce,cai2012towards}  and consistent with the degeneracy conditions of DHOST theories \cite{langlois2017degenerate}. 
	
	Our model builds upon early two-field approaches by coupling the energy-momentum trace $T$ to curvature invariants. This coupling provides new degrees of freedom that enable novel double PDL crossings while helping to maintain $c_s^2 \geq 0$. This contrasts with single-field mechanisms which often require fine-tuned potentials \cite{vikman2005can} or non-canonical kinetic terms \cite{armendariz2000dynamical} to achieve a single crossing. The bouncing cosmology framework further avoids initial singularities through scale factor reversal, offering a cyclic universe alternative \cite{khoury2002big, elitzur2002big, sadeghi2009bouncing, bamba2014bounce}.
	
	\textbf{Stability as a cornerstone:} A key criticism of higher-derivative theories is the potential for ghost instabilities \cite{eliezer1989instability, Woodard2007avoiding, woodard2015theorem}. In Section \ref{sec:Stability}, we rigorously analyze the dynamical degrees of freedom in our $f(R,G,T)$-quintom system, showing that the FLRW background inherently suppresses pathological modes---a feature absent in generic higher-derivative proposals. This builds on healthy bounce constructions \cite{lin2011matter} while generalizing them to curvature-matter couplings.
	
	Observational support for these ideas comes from Type Ia supernovae \cite{smecker1991type, ruiz1995type, nomoto1997type}, Baryon Acoustic Oscillations (BAO) \cite{eisenstein2005detection}, and Cosmic Microwave Background (CMB) measurements \cite{peiris2003first, spergel2007three}. These align with theoretical models like quintessence \cite{ratra1988cosmological}, phantom energy \cite{caldwell2002phantom}, and hybrid approaches \cite{sen2002tachyon, gasperini2001quintessence, wei2005hessence}.

	In Section~\ref{sec:Model}, we derive the equations of motion and energy-momentum components for an $f(R, G, T)$-quintom system in a flat FLRW metric, ensuring conservation law compatibility. Section~\ref{sec:EFE-MFD} presents the modified Friedmann equations, while Section~\ref{sec:Weyl} shows how conformal Weyl gravity emerges as a limiting case. A comprehensive stability analysis, encompassing both ghost freedom and scalar perturbations, is conducted in Section~\ref{sec:Stability}. Following this, Section~\ref{sec:DBCM} numerically reconstructs five distinct $f(R, G, T)$ models, demonstrating their viable bounce dynamics and Phantom Divide Line (PDL) crossings. We conclude in Section~\ref{sec:SumCon} by highlighting how curvature-matter coupling provides a unified framework for early-time bounces and late-time acceleration, thereby setting the stage for future extended analyses.
	
	\section{The Model}\label{sec:Model}
	In modified gravity, we consider an extension of the Einstein-Hilbert action by introducing a general function of the Ricci scalar $R$, Gauss-Bonnet invariant $G$, and the trace of the matter stress-energy tensor $T \equiv g^{\mu\nu} T_{\mu\nu}^{(m)}$, coupled to a quintom scalar sector consisting of a phantom field $\phi(t)$ and a canonical field $\psi(t)$. The total action is given by:
	\begin{eqnarray}
		\mathcal{S} = \int d^4x \sqrt{-g} \left[ \frac{1}{2\kappa^2} f(R, G, T) + \Xi(\phi, \psi) + \mathcal{L}_{m} \right],
		\label{eq:action}
	\end{eqnarray}
	where $\kappa^2 \equiv 8\pi G_r/c^4$ is the gravitational coupling constant, $G_r$ is Newton's constant and $\mathcal{L}_{m}$ is the matter-radiation Lagrangian. 
	
	\subsection{Quintom Sector and FLRW Metric}
	The quintom Lagrangian combines the phantom, $\phi$, and canonical, $\psi$, fields together as:
	\begin{eqnarray}
		\Xi(\phi, \psi) = -\frac{1}{2} \partial_\mu \phi \partial^\mu \phi + \frac{1}{2} \partial_\mu \psi \partial^\mu \psi - V(\phi, \psi),
		\label{eq:quintom}
	\end{eqnarray}
	where $V(\phi, \psi)$ is the interaction potential. For a flat FLRW universe:
	\begin{eqnarray}
		ds^2 = -dt^2 + a^2(t) \sum_{i=1}^3 (dx^i)^2,
		\label{eq:flrw}
	\end{eqnarray}
	the general expressions for the Ricci scalar $R$ and the Gauss-Bonnet invariant $G$ in terms of the metric are given in Appendix~\ref{sec:Modified E-M-T}. For the FLRW metric, these reduce to:
	\begin{eqnarray}
		R &=& 6 \left( \dot{H} + 2H^2 \right), \label{eq:R} \\
		G &=& 24 H^2 \left( \dot{H} + H^2 \right). \label{eq:G}
	\end{eqnarray}
	In the FLRW background, $G$ can be expressed as a total derivative, $G = 12 \frac{d}{dt}(\dot{a}^2)$, which explains why a linear term in $G$ ($f_G = \text{const.}$) does not contribute to the bulk dynamics, as will be seen in the linear coupling model \ref{subsec:Linear Coupling Model}.
	
	\subsection{Scalar Field Dynamics}
	The equations of motion for $\phi$ (phantom) and $\psi$ (canonical) are:
	\begin{eqnarray}
		\Box \phi - V_{,\phi} + \frac{f_T}{2\kappa^2} \frac{\delta T}{\delta \phi} &=& 0, \label{eq:EOM_phi} \\
		\Box \psi + V_{,\psi} + \frac{f_T}{2\kappa^2} \frac{\delta T}{\delta \psi} &=& 0, \label{eq:EOM_psi}
	\end{eqnarray}
	where the coupling terms are:
	\begin{eqnarray}
		\frac{\delta T}{\delta \phi} &=& -2 \frac{\partial \mathcal{L}_m}{\partial \phi} + g^{\mu\nu} \frac{\partial T_{\mu\nu}^{(m)}}{\partial \phi}, \label{eq:dTdphi} \\
		\frac{\delta T}{\delta \psi} &=& -2 \frac{\partial \mathcal{L}_m}{\partial \psi} + g^{\mu\nu} \frac{\partial T_{\mu\nu}^{(m)}}{\partial \psi}. \label{eq:dTdpsi}
	\end{eqnarray}
	
	For minimal coupling ($\delta T/\delta \phi = \delta T/\delta \psi = 0$), these reduce to:
	\begin{eqnarray}
		\ddot{\phi} + 3H \dot{\phi} - V_{,\phi} &=& 0, \label{eq:minimal_phi} \\
		\ddot{\psi} + 3H \dot{\psi} + V_{,\psi} &=& 0. \label{eq:minimal_psi}
	\end{eqnarray}
	
	\subsection{Energy-Momentum Tensors}
	The total energy-momentum tensor (as shown in Appendix \ref{sec:Modified E-M-T}) decomposes as:
	\begin{eqnarray}
		T_{\mu\nu}^{(\text{total})} = \underbrace{T_{\mu\nu}^{(R)} + T_{\mu\nu}^{(G)}}_{\text{Geometric}} + \underbrace{T_{\mu\nu}^{(T)} + T_{\mu\nu}^{(\Xi)} + T_{\mu\nu}^{(m)}}_{\text{Matter/Quintom}},
		\label{eq:EMT_decomp}
	\end{eqnarray}
	where:
	\begin{align}
		T_{\mu\nu}^{(R)} &= \frac{1}{\kappa^2} \left( f_R R_{\mu\nu} - \frac{1}{2} g_{\mu\nu} f + (g_{\mu\nu} \Box - \nabla_\mu \nabla_\nu) f_R \right), \label{eq:T_R} \\
		T_{\mu\nu}^{(G)} &=\frac{2R}{\kappa^2}\left(f_G R_{\mu\nu}+(g_{\mu\nu}\square - \nabla_\mu\nabla_\nu)f_G\right)
		-\frac{4}{\kappa^2}\left(f_G R_{\mu}^{\rho}R_{\rho\nu}+(R_{\mu\nu}\square+g_{\mu\nu}R^{\rho\lambda}\nabla_{\rho}\nabla_{\lambda})f_G\right)\nonumber\\
		&- \frac{4}{\kappa^2}f_G\left(R_{\mu\rho\nu\lambda}R^{\rho\lambda}-\frac{1}{2}R_{\mu}^{\rho\lambda\xi}R_{\nu\rho\lambda\xi}\right) +\frac{4}{\kappa^2}\left(R_{\mu}^{\rho}\nabla_{\nu}\nabla_{\rho}+R_{\nu}^{\rho}\nabla_{\mu}\nabla_{\rho}+R_{\mu\rho\nu\lambda}\nabla^{\rho}\nabla^{\lambda}\right)f_G,	\label{eq:T_G} \\
		T_{\mu\nu}^{(\Xi)} &= -\partial_\mu \phi \partial_\nu \phi + \partial_\mu \psi \partial_\nu \psi + g_{\mu\nu} \left( \frac{1}{2}\partial_\alpha \phi \partial^\alpha \phi -\frac{1}{2} \partial_\alpha \psi \partial^\alpha \psi - V(\phi, \psi) \right), \label{eq:T_Xi} \\
		T_{\mu\nu}^{(m)}&=-\frac{2}{\sqrt{-g}}\frac{\partial(\sqrt{-g}\mathcal{L}_{m})}{\partial g^{\mu\nu}}=g_{\mu\nu}\mathcal{L}_{m}-2\frac{\partial\mathcal{L}_{m}}{\partial g^{\mu\nu}},	\label{eq:T_m}\\
		T_{\mu\nu}^{(T)} &= -\frac{f_T}{\kappa^2} \left( T_{\mu\nu}^{(m)} + \Theta_{\mu\nu} \right)\cdot \label{eq:T_T}
	\end{align}
	Here, the $T_{\mu\nu}^{(T)}$ term introduces \textit{non-minimal coupling} between matter and geometry through, where $\Theta_{\mu\nu} \equiv g^{\alpha\beta}\frac{\partial T_{\alpha\beta}^{(m)}}{\partial g^{\mu\nu}}$ encodes how matter responds to metric variations.
	
	\subsection{Gravitational Field Equations}
	The generalized Einstein equations are derived from the variation of the action with respect to $ g^{\mu\nu}$. Defining the Einstein tensor as:
	\begin{eqnarray}
		G_{\mu\nu} \equiv R_{\mu\nu} - \frac{1}{2} g_{\mu\nu} R, \label{eq:Einstein_tensor}
	\end{eqnarray}
	the field equations take the form:
	\begin{eqnarray}
		f_R G_{\mu\nu} + \left(g_{\mu\nu}\Box - \nabla_\mu\nabla_\nu\right)f_R + 2R\left(g_{\mu\nu}\Box - \nabla_\mu\nabla_\nu\right)f_G \nonumber \\
		- 4\left(R_{\mu}^{\rho}\nabla_\nu\nabla_\rho + R_{\nu}^{\rho}\nabla_\mu\nabla_\rho\right)f_G + 4R_{\mu\nu}\Box f_G + 4g_{\mu\nu}R^{\rho\sigma}\nabla_{\rho}\nabla_{\sigma}f_G \nonumber \\
		- 4R_{\mu\rho\nu\sigma}\nabla^{\rho}\nabla^{\sigma}f_G - \frac{1}{2}g_{\mu\nu}f + f_T(T_{\mu\nu}^{(m)} + \Theta_{\mu\nu}) = \kappa^2 T_{\mu\nu}^{(\Xi)}. \label{eq:Full_EOM}
	\end{eqnarray}
	
	\subsection{Dynamics of Perfect Fluids}
	
	The matter content is described by a perfect fluid with energy-momentum tensor:
	\begin{eqnarray}	
		T_{\mu\nu}^{(m)} = (\rho + p)u_{\mu}u_{\nu} + p g_{\mu\nu},
		\label{Tmunu}
	\end{eqnarray}
	where $\rho \equiv \rho_m + \rho_r$ is the total energy density and $p \equiv p_m + p_r$ is the total pressure, with subscripts $m$ and $r$ denoting non-relativistic matter and radiation respectively. The four-velocity $u_\mu$ satisfies $u_\mu u^\mu = -1$ and $u^\mu \nabla_\nu u_\mu = 0$. 
	
	The conservation laws for standard matter components are:
	\begin{eqnarray}
		\dot{\rho} + 3H(\rho + p) = 0 \quad \Rightarrow \quad
		\dot{\rho}_m + 3H\rho_m = 0 \quad \text{and} \quad \dot{\rho}_r + 4H\rho_r = 0,
		\label{Continuity-m}
	\end{eqnarray}
	where we have used the equations of state $\omega_m \equiv p_m/\rho_m \approx 0$ for matter and $\omega_r \equiv p_r/\rho_r = 1/3$ for radiation.
	
	Using the matter Lagrangian $\mathcal{L}_m = p$ (as shown in Appendix \ref{sec:Modified E-M-T}), the tensor $\Theta_{\mu\nu}$ becomes:
	\begin{eqnarray}
		\Theta_{\mu\nu} = g_{\mu\nu}\mathcal{L}_m - 2T_{\mu\nu}^{(m)} - 2g^{\alpha\beta}\frac{\partial^2\mathcal{L}_m}{\partial g^{\alpha\beta}\partial g^{\mu\nu}} = -2T_{\mu\nu}^{(m)} + p g_{\mu\nu}.
		\label{Thetamunu}
	\end{eqnarray}
	
	For the FLRW metric, we obtain the following important quantities:
	\begin{eqnarray}
		T &\equiv& g^{\mu\nu}T_{\mu\nu}^{(m)} = -\rho + 3p = -(1-3\omega)\rho, \label{T} \\
		\Theta &\equiv& g^{\mu\nu}\Theta_{\mu\nu} = 2(\rho - p) = 2(1-\omega)\rho, \label{Theta} \\
		T + \Theta &=& 2(\rho + p) = 2\rho(1+\omega), \label{T+Theta}
	\end{eqnarray}
	where $\omega \equiv p/\rho$ is the total equation of state parameter.
	
	The $00$ and $ii$ components of the total energy-momentum tensor decomposition (\ref{eq:EMT_decomp}) give the modified Friedmann equations:
	\begin{eqnarray}
		\rho_R + \rho_G &=& \rho + \rho_T + \rho_{\Xi}, \label{T00} \\
		p_R + p_G &=& p + p_T +  p_{\Xi}, \label{Tii}
	\end{eqnarray}
	with the individual components given by:
	\begin{align}
		\kappa^2 \rho_R &= -3H^2 f_R - 3\dot{H}f_R + 3H\dot{f}_R + \frac{f}{2}, \label{rho_R} \\
		\kappa^2 p_R &= \,\,\,\,  3H^2f_R + \,\,\,\dot{H}f_R - 2H\dot{f}_R - \frac{f}{2} - \ddot{f}_R , \label{p_R} \\
		\kappa^2 \rho_G &= -12H^{2}(H^{2}+\dot{H})f_{G}+H(3H^{2}-9\dot{H})\dot{f}_{G} -9(H^2+\dot{H})\ddot{f}_{G},					\label{rho_G}\\
		\kappa^2 p_G&=\,\,\,\, 12H^{2}(H^{2}+\dot{H})f_{G}+H(7H^{2}-5\dot{H})\dot{f}_{G}-(H^2-3\dot{H})\ddot{f}_{G},						\label{p_G}\\
		\kappa^2 \rho_T &= f_T (\rho + p), \label{rho_T} \\
		\kappa^2 p_T &= 0, \label{p_T} \\
		\rho_{\Xi} &= -\frac{1}{2}\dot{\phi}^2 + \frac{1}{2}\dot{\psi}^2 + V(\phi,\psi), \label{rho_Xi} \\
		p_{\Xi} &= -\frac{1}{2}\dot{\phi}^2 + \frac{1}{2}\dot{\psi}^2 - V(\phi,\psi), \label{p_Xi}
	\end{align}
	where we have used the FLRW expressions $R = 6\dot{H} + 12H^2$ and $G = 24H^2(\dot{H} + H^2)$.
	
	\subsection{Conservation Laws}
	As shown in Appendix \ref{sec:Covariant}, by taking the covariant divergence of the total energy-momentum tensor decomposition (\ref{eq:EMT_decomp}), we obtain the following conservation equations for each component:
	\begin{align}
		\nabla^{\mu}T_{\mu\nu}^{(R)} &= -\frac{1}{2\kappa^2}g_{\mu\nu}\left( f_G\nabla^{\mu}G + f_T\nabla^{\mu}T \right), \label{Co_TR} \\
		\nabla^{\mu} T_{\mu\nu}^{(G)} &= \quad \frac{1}{\kappa^2} \left( R f_G - 2(\nabla^\alpha \nabla_\alpha f_G) + 4R^{\alpha\beta} \nabla_\alpha \nabla_\beta f_G \right) \nabla_\nu R \nonumber \\
		&+ \quad\frac{4}{\kappa^2} \left( R_{\mu\rho\nu\lambda} \nabla^\mu \nabla^\lambda \nabla^\rho f_G - R_{\nu\rho} \nabla^\rho (\nabla^\alpha \nabla_\alpha f_G) \right)  - \frac{4}{\kappa^2} f_G \left( R^{\rho\lambda} \nabla_\rho R_{\nu\lambda} + \frac{1}{2} R^{\rho\lambda\alpha\beta} \nabla_\nu R_{\rho\lambda\alpha\beta} \right) \nonumber \\
		& + \quad \frac{4}{\kappa^2} \left( f_{GR} \nabla^\mu R + f_{GG} \nabla^\mu G + f_{GT} \nabla^\mu T \right) \left( \frac{1}{2} R_{\mu}^{\rho\lambda\xi} R_{\nu\rho\lambda\xi} - R_{\mu\rho\nu\lambda} R^{\rho\lambda} \right), \label{Co_TG} \\
		\nabla^{\mu}T_{\mu\nu}^{(T)} &= -\frac{1}{\kappa^2}\left(\left(T_{\mu\nu}^{(m)} + \Theta_{\mu\nu}\right)\nabla^{\mu}f_T + f_T\nabla^{\mu}(T_{\mu\nu}^{(m)} + \Theta_{\mu\nu})\right), \label{Co_TT} \\
		\nabla^{\mu}T_{\mu\nu}^{(\Xi)} &= 0. \label{Co_TXi}
	\end{align}
	where $f_{GR}=\frac{\partial^2 f}{\partial G \,\partial R}$, $f_{GG}=\frac{\partial^2 f}{\partial^2 G}$, and $f_{GR}=\frac{\partial^2 f}{\partial G \,\partial T}$.
	
	The total conservation law for matter is then:
	\begin{eqnarray}
		\nabla^{\mu}\left(\kappa^2 T_{\mu\nu}^{(m)} - f_T\left(T_{\mu\nu}^{(m)} + \Theta_{\mu\nu}\right) - \kappa^2 T_{\mu\nu}^{(G)}\right) + \frac{1}{2}g_{\mu\nu}f_G\nabla^{\mu}G + \frac{1}{2}g_{\mu\nu}f_T\nabla^{\mu}T = 0, \label{Co_Ttotal}
	\end{eqnarray}
	which gives:
	\begin{align}
		\nabla^{\mu}T_{\mu\nu}^{(m)} &= \frac{f_T}{\kappa^2 - f_T}\left(\left(T_{\mu\nu}^{(m)} + \Theta_{\mu\nu}\right)\nabla^{\mu}\ln(f_T) + \nabla^{\mu}\Theta_{\mu\nu} - \frac{1}{2}g_{\mu\nu}\nabla^{\mu}T\right) \nonumber \\
		&+ \frac{1}{\kappa^2 - f_T}\left(\kappa^2\nabla^{\mu}T_{\mu\nu}^{(G)} - \frac{1}{2}g_{\mu\nu}f_G\nabla^{\mu}G\right). \label{Co_Tm}
	\end{align}
	
	As shown in Appendix \ref{sec:Sub-FLRW-Cov-Div}, for the FLRW metric, these become:
	\begin{align}
		\kappa^2\nabla^{\mu}T_{\mu 0}^{(R)} &= 12f_R H\dot{H} + 3f_R\ddot{H} - \frac{\dot{f}}{2} \nonumber \\
		&= -12f_G H(H\ddot{H} + 2\dot{H}^2 + 4H^2\dot{H}) - (3\dot{p} - \dot{\rho})\frac{f_T}{2}, \label{Co_TR_H} \\
		\kappa^2 \nabla^\mu T_{\mu 0}^{(G)} 
		&= 12(\dot{H}+H^2)\dddot{f}_G \nonumber \\
		&\quad + 12\left( \ddot{H} + 7H\dot{H} + 3H^3 - 6(\dot{H} + H^2)(\ddot{H} + 4H\dot{H}) \right) \ddot{f}_G \nonumber \\
		&\quad + 36\left( H\ddot{H} + 5H^2\dot{H} + \dot{H}^2 \right) \dot{f}_G \nonumber \\
		&\quad + 12\left( 2\dot{H}\ddot{H} + 14H\dot{H}^2 + 5H^2\ddot{H} + 26H^3\dot{H} \right) f_G \nonumber \\
		&\quad + 6(\dot{H} + H^2)(\dot{H} + 5H^2)\left( f_{GR}(6\ddot{H} + 24H\dot{H}) + f_{GG} \dot{G} + f_{GT} \dot{T} \right) \label{Co_TG_H} \\
		\kappa^2\nabla^{\mu}T_{\mu 0}^{(G)} - \frac{1}{2}g_{\mu 0}f_G\nabla^{\mu}G 
		&= 12(\dot{H}+H^2)\dddot{f}_G \nonumber \\
		&\quad + 12\left( \ddot{H} + 7H\dot{H} + 3H^3 - 6(\dot{H} + H^2)(\ddot{H} + 4H\dot{H}) \right) \ddot{f}_G \nonumber \\
		&\quad + 36\left( H\ddot{H} + 5H^2\dot{H} + \dot{H}^2 \right) \dot{f}_G \nonumber \\
		&\quad + 24\left( \dot{H}\ddot{H} + 6H\dot{H}^2 + 2H^2\ddot{H} + 11H^3\dot{H} \right) f_G \nonumber \\
		&\quad + 6(\dot{H} + H^2)(\dot{H} + 5H^2)\left( f_{GR}(6\ddot{H} + 24H\dot{H}) + f_{GG} \dot{G} + f_{GT} \dot{T} \right).\label{Co_TG_fG_H}
	\end{align}
	where:
	\begin{align*}
		\dot{G} &= 24H^2 \ddot{H} + 48H (\dot{H})^2 + 96H^3 \dot{H} \\
		\dot{T} &= -\dot{\rho} + 3\dot{p}
	\end{align*}
	
	Substituting equations (\ref{Tmunu}), (\ref{Thetamunu}), (\ref{T}), and the previously computed expression for \(\kappa^2\nabla^{\mu}T_{\mu 0}^{(G)} - \frac{1}{2}g_{\mu 0}f_G\nabla^{\mu}G\) into (\ref{Co_Tm}) and projecting the \(\nu=0\) component for the FLRW metric yields the following general expression:
	\begin{eqnarray}
		-(\kappa^2 - f_T)\left(\dot{\rho} + 3H(\rho + p)\right) &=& (\rho + p)\dot{f}_T + \left(6(\rho + p)H + \frac{3}{2}\dot{\rho} + \frac{9}{2}\dot{p}\right)f_T \nonumber \\
		&& + \kappa^2\nabla^{\mu}T_{\mu 0}^{(G)} - \frac{1}{2}g_{\mu 0}f_G\nabla^{\mu}G.
		\label{general-energy}
	\end{eqnarray}
	
	The standard conservation law from general relativity is \(\dot{\rho} + 3H(\rho + p) = 0\). Therefore, for energy-momentum conservation to hold in our model, the right-hand side of the equation must vanish. This leads to the following consistency condition:
	\begin{eqnarray}
		-(\rho + p)\dot{f}_T - \frac{1}{2}\left(9\dot{p}-\dot{\rho} \right)f_T &=& \kappa^2\nabla^{\mu}T_{\mu 0}^{(G)} - \frac{1}{2}g_{\mu 0}f_G\nabla^{\mu}G.
		\label{conserve-condition}
	\end{eqnarray}
	
	Special cases where conservation holds include when both sides of the equation are identically zero. This occurs under the following separate conditions:
	\begin{eqnarray}
		(\rho + p)\dot{f}_T + \frac{1}{2}\left(9\dot{p}-\dot{\rho} \right)f_T &=& 0, \label{conserve-condition-Term1} \\
		\kappa^2\nabla^{\mu}T_{\mu 0}^{(G)} - \frac{1}{2}g_{\mu 0}f_G\nabla^{\mu}G &=& 0. \label{conserve-condition-Term2}
	\end{eqnarray}
	
	Equation (\ref{conserve-condition-Term1}) governs the energy-momentum exchange between matter and geometry through the $f_T$ coupling. This constraint is automatically satisfied when $f_T = 0$, corresponding to minimal matter coupling. For non-zero $f_T$, the equation imposes a relation between matter evolution and the $T$-dependence of the gravitational Lagrangian. A particularly important solution occurs for radiation-dominated universes ($p = \rho/3$), where the condition simplifies to $(\rho + p)\dot{f}_T + \dot{\rho}f_T = 0$, leading to $f_T \propto 1/\rho$ scaling \cite{Harko2011}.
	
	The conservation equation (\ref{conserve-condition-Term2}) significantly constrains cosmological solutions in $f(R,G,T)$ gravity. The simplest stable solution is $f_G = 0$, corresponding to General Relativity with a cosmological constant ($f = R - 2\Lambda$). 
	
	For $f_G = \text{const.} \neq 0$, the constraint simplifies to:
	\begin{equation}
		\dot{H}\ddot{H} + 6H\dot{H}^2 + 2H^2\ddot{H} + 11H^3\dot{H} = 0,
	\end{equation}
	which is satisfied by de Sitter space ($H = \text{constant}$) \cite{Charmousis2008}. The linear case $f(G) = \alpha G$ avoids Ostrogradsky instabilities due to the topological nature of $G$ in 4D \cite{woodard2015theorem, Nojiri2005}.
	
	A particularly interesting solution occurs for coasting cosmology ($\dot{H} + H^2 = 0$) \cite{lilley2015bouncing, qiu2011bouncing}, which eliminates higher-derivative terms and yields:
	\begin{equation}
		\ddot{f}_G + 3H\dot{f}_G + 3H^2 f_G = 0.
	\end{equation}
	This second-order damped oscillator equation is stable and suitable for bouncing scenarios \cite{lin2011matter}. For general nonlinear $f(G)$, stability requires the degeneracy condition $f_{RR} f_{GG} - (f_{RG})^2 = 0$ to avoid ghost instabilities \cite{langlois2017degenerate}.

	\section{Einstein Field Equations and Modified Friedmann Dynamics}\label{sec:EFE-MFD}
	
	The Einstein Field Equation (EFE) in its standard form is:
	\begin{equation}
		G_{\mu\nu} + \Lambda g_{\mu\nu} = R_{\mu\nu} - \frac{1}{2}g_{\mu\nu}R + \Lambda g_{\mu\nu} = \kappa^2 T_{\mu\nu}^{(m)}, \label{EFE}
	\end{equation}
	which yields the Friedmann equations for a flat FLRW universe:
	\begin{eqnarray}
		3H^2 - \Lambda &=& \kappa^2 \rho, \label{f1} \\
		-2\dot{H} - 3H^2 + \Lambda &=& \kappa^2 p. \label{f2}
	\end{eqnarray}
	
	In the $\Lambda$CDM model, dark energy is represented by a cosmological constant $\Lambda \equiv \kappa^2 \rho_{vac}$. For dynamical dark energy models (like quintom), we generalize this to time-varying components:
	\begin{eqnarray}
		3H^2 - \kappa^2\rho_{DE} &=& \kappa^2 \rho, \label{f11} \\
		-2\dot{H} - 3H^2 - \kappa^2 p_{DE} &=& \kappa^2 p. \label{f22}
	\end{eqnarray}
	
	Comparing with our modified gravity decomposition (\ref{T00})-(\ref{Tii}), the dark energy components become:
	\begin{eqnarray}
		\kappa^2 \rho_{DE} & \equiv & 3H^2 + \kappa^2\left(\rho_T + \rho_{\Xi} - \rho_R - \rho_G\right) \nonumber \\
		&=& 3H^2(1 + f_R) - 3H\dot{f}_R + 3\dot{H}f_R - \frac{f}{2} + f_T(\rho + p) + \rho_{\Xi} \nonumber \\
		&+& 9(H^2 + \dot{H})\ddot{f}_G + 3H(3\dot{H} - H^2)\dot{f}_G + 12H^2(H^2 + \dot{H})f_G, \label{rho-DE} \\
		\kappa^2 p_{DE} & \equiv & -2\dot{H} - 3H^2 + \kappa^2\left(p_T + p_{\Xi} - p_R - p_G\right) \nonumber \\
		&=& -(2\dot{H} + 3H^2)(1 + f_R) + 2H\dot{f}_R + \dot{H}f_R + \frac{f}{2} + \ddot{f}_R + p_{\Xi} \nonumber \\
		&-& 12H^2(H^2 + \dot{H})f_G - H(7H^2 - 5\dot{H})\dot{f}_G + (H^2 - 3\dot{H})\ddot{f}_G. \label{p-DE}
	\end{eqnarray}
	
	The effective field equation approach writes:
	\begin{equation}
		G_{\mu\nu} = \kappa^2 T_{\mu\nu}^{(eff)}, \label{EFE2}
	\end{equation}
	yielding:
	\begin{align}
		3H^2 &= \kappa^2 \rho_{\text{eff}}, \label{f111} \\
		-2\dot{H} - 3H^2 &= \kappa^2 p_{\text{eff}}, \label{f222}
	\end{align}
	
	For $f_R \neq 0$, the effective energy density and pressure can be obtained as
	\begin{align}
		\rho_{\text{eff}} &= \frac{1}{f_R} \left( \rho + \rho_{\Xi} + \frac{1}{\kappa^2} \mathcal{R} \right) \label{eq:f1-E} \\
		p_{\text{eff}} &= \frac{1}{f_R} \left( p + p_{\Xi} + \frac{1}{\kappa^2} \mathcal{P} \right) \label{eq:f2-E}
	\end{align}
	where 
	\begin{align}
		\mathcal{R} &= \,\,\,\, 3f_R(\dot{H} + 2H^2) - \frac{f}{2} - 3H\dot{f}_R - f_T (\rho + p) \nonumber \\
		&\quad - 12H^2(H^2 + \dot{H}) f_G + H(3H^2 - 9\dot{H}) \dot{f}_G - 9(H^2 + \dot{H}) \ddot{f}_G \label{eq:mathcal_R} \\
		\mathcal{P} &= -3f_R(\dot{H} + 2H^2) - \frac{f}{2} - 3H\dot{f}_R + f_T p - \ddot{f}_R \nonumber \\
		&\quad - 12H^2(H^2 + \dot{H}) f_G - H(7H^2 - 5\dot{H}) \dot{f}_G + (H^2 - 3\dot{H}) \ddot{f}_G \label{eq:mathcal_P}
	\end{align}
	
	Energy conservation requires:
	\begin{eqnarray}
		\dot{\rho}_{DE} + 3H(\rho_{DE} + p_{DE}) &=& 0, \\
		\dot{\rho}_{\text{eff}} + 3H(\rho_{\text{eff}} + p_{\text{eff}}) &=& 0. \label{fRdot0}
	\end{eqnarray}
	
	The effective EoS parameter is:
	\begin{eqnarray}
		\omega_{\text{eff}} &=& -1 - \frac{2}{3}\frac{\dot{H}}{H^2} \nonumber \\
		&=& -1 + \frac{\rho(1+\omega) + \rho_{DE}(1+\omega_{DE})}{\rho + \rho_{DE}}, \label{omega_eff}
	\end{eqnarray}
	which reduces to the $\Lambda$CDM case when $\omega_{DE} = -1$:
	\begin{eqnarray}
		\omega_{\text{eff}} &=& -1 + \frac{\rho(1+\omega)}{\rho + \rho_{DE}}. \label{omega_eff-LambdaCDM}
	\end{eqnarray}
	
	A substantial body of literature has focused on analytically solving the second part of equations such as Eq.~\eqref{omega_eff}, aiming to identify functional forms that satisfy the required physical criteria. However, many such analytical approaches prove insufficient for fully resolving issues like the Phantom Divide Line (PDL) crossing. This is because, regardless of the specific form on the other side of the equation, one can often construct solutions that \emph{mathematically} satisfy the condition---for instance, ensuring $ \rho_{\mathrm{eff}}(1 + \omega_{\mathrm{eff}}) = 0 $ at the bounce point while maintaining $ \frac{d}{dt}(\rho_{\mathrm{eff}} + p_{\mathrm{eff}}) \neq 0 $---without guaranteeing physical plausibility or stability.
	
	Therefore, to robustly verify the internal consistency and dynamical stability of our model, we employ a numerical framework. In Section~\ref{sec:DBCM}, we examine these conditions, focusing on key stability criteria such as the squared sound speed $c_s^2$ and the effective equation of state $\omega_{\mathrm{eff}}$, including its crossing of the PDL ($\omega_{\mathrm{eff}} = -1$). This analysis serves to establish the theoretical viability of the model prior to any comparison with observational datasets.

	\section{Weyl Conformal Geometry}\label{sec:Weyl}
	Under Weyl transformations $g_{\mu\nu} \rightarrow \Omega^2(x) g_{\mu\nu}$, the conformal gravity action is invariant if it is constructed from the Weyl tensor tensor:
	\begin{eqnarray}
		C_{\mu\nu\rho\lambda} = R_{\mu\nu\lambda\rho}
		-\frac{1}{2}(g_{\mu\lambda}R_{\nu\rho}-g_{\mu\rho}R_{\nu\lambda}-g_{\nu\lambda}R_{\mu\rho}+g_{\nu\rho}R_{\mu\lambda})
		+\frac{R}{6}(g_{\mu\lambda}g_{\nu\rho}-g_{\mu\rho}g_{\nu\lambda}).\label{Weyl tensor}
	\end{eqnarray}
	The action for Weyl conformal gravity coupled to quintom and matter fields is:
	\begin{eqnarray}\label{ac Weyl}
		\mathcal{S} = \int d^4x \sqrt{-g} \left[ \alpha C_{\mu\nu\rho\lambda} C^{\mu\nu\rho\lambda} + \Xi(\phi,\psi) + \mathcal{L}_m \right],
	\end{eqnarray}
	where $\alpha$ is a constant. In four dimensions, the Weyl-squared term is related to the Gauss-Bonnet invariant via the topological Euler density, up to a boundary term. Consequently, the Weyl action is equivalent to an $f(G)$ theory with $f(G) = \frac{\alpha}{\kappa^2} G$, modulo a total derivative. It is important to note that a linear term in $G$ alone does not alter the classical field equations because $G$ is a total derivative in four dimensions. However, in the context of our $f(R,G,T)$ model, such a term can become dynamical when coupled to the trace $T$ or when considered within a non-trivial $f(R)$ background. Nevertheless, as a limiting case, the FLRW equations reduce to:
	\begin{align}
		3H^2 &= \kappa^2 \left( \rho + \rho_\Xi + \frac{12\alpha}{\kappa^4} H^2(H^2 + \dot{H}) \right), \\
		-2\dot{H} - 3H^2 &= \kappa^2 \left( p + p_\Xi - \frac{12\alpha}{\kappa^4} H^2(H^2 + \dot{H}) \right).
	\end{align}
	Thus, Weyl cosmology emerges as a special case of our $f(R,G,T)$ framework, illustrating how higher-curvature corrections modify the Friedmann dynamics. Also, these results demonstrate that the previous components $\mathcal{W}_{00}$ and $\mathcal{W}_{ii}$ from \cite{ghanaatian2014bouncing} and \cite{ghanaatian2018bouncing} are incorrect in this formulation, indicating the need for careful reevaluation of those results.
	
	The energy conservation conditions are satisfied in this framework through specific constraints. Condition (\ref{conserve-condition-Term1}) governing matter-geometry coupling is automatically satisfied since $f_T = 0$ in pure Weyl gravity. Condition (\ref{conserve-condition-Term2}) reduces to the constraint $\dot{H}\ddot{H} + 6H\dot{H}^2 + 2H^2\ddot{H} + 11H^3\dot{H} = 0$ for $f_G = \text{const.}$, which admits de Sitter space ($H = \text{constant}$) as a solution. The coasting universe scenario ($\dot{H} + H^2 = 0$) also satisfies this constraint, making Weyl conformal gravity compatible with energy conservation requirements.
	
	\section{Stability Analysis and Ghost-Free Conditions}\label{sec:Stability}
	A crucial requirement for any modified gravity theory is the absence of pathological instabilities, particularly Ostrogradsky ghosts that typically plague higher-derivative theories. In this section, we establish the theoretical stability of our $f(R,G,T)$-quintom framework through Hamiltonian analysis, perturbation theory, and energy condition analysis.
	
	\subsection{Theoretical Stability Foundations}
	The stability analysis of our model proceeds along two complementary approaches: a rigorous Hamiltonian formulation that counts physical degrees of freedom and identifies constraints (see Appendix~\ref{sec:Hamiltonian}), and a cosmological perturbation analysis that examines stability during the bounce phase.
	
	\subsubsection{Hamiltonian Structure and Degree of Freedom Counting}
	As detailed in Appendix~\ref{sec:Hamiltonian}, we perform a complete 3+1 decomposition of the action (\ref{eq:action}) to determine the independent physical degrees of freedom. The key results are:
	\begin{itemize}
		\item \textbf{Canonical Structure}: The conjugate momenta for the metric and scalar fields reveal a well-defined phase space structure.
		\item \textbf{Constraint Analysis}: The theory possesses a complete set of primary and secondary constraints arising from diffeomorphism invariance.
		\item \textbf{Physical Degrees of Freedom}:Using Dirac's counting algorithm, we establish that the physical configuration space contains exactly 4 degrees of freedom: 2 tensor modes (transverse-traceless graviton polarizations) and 2 scalar modes (the quintom fields $\phi$ and $\psi$).
	\end{itemize}
	The perfect fluid matter component, described by the equation of state $p = \omega\rho$ and the continuity equation $\dot{\rho} + 3H(\rho + p) = 0$, adds a single scalar degree of freedom $\rho(t)$ which is accounted for in our background analysis.
	
	\subsubsection{Ostrogradsky Instability Avoidance}
	
	The potential Ostrogradsky instability, characteristic of higher-derivative theories, is avoided through specific degeneracy conditions derived in Appendix \ref{sec:Hamiltonian}. The critical requirement is:
	\begin{align}
		\det\begin{pmatrix}
			f_{RR} & f_{RG} \\ 
			f_{GR} & f_{GG}
		\end{pmatrix} = 0,
	\end{align}
	which ensures the elimination of pathological higher-derivative modes. For the linear Gauss-Bonnet case ($f_G = \text{const.}$), this condition is automatically satisfied since $f_{GG} = f_{RG} = 0$. More generally, the FLRW background symmetry provides additional protection by reducing the effective higher-derivative structure.
	
	These theoretical foundations establish that our $f(R,G,T)$-quintom framework possesses the correct number of physical degrees of freedom and avoids Ostrogradsky instabilities, providing a solid basis for the subsequent perturbation analysis.
	
	\subsection{Perturbation Analysis and Numerical Stability Conditions}\label{sec:perturbation_stability}
	
	To establish the stability of our $f(R,G,T)$-quintom framework, we analyze both theoretical perturbation behavior and numerical stability conditions. Following the standard approach in modified gravity theories \cite{DeFelice2010, Sotiriou2011}, we consider scalar perturbations of the FLRW metric. In the Newtonian gauge (see Appendix~\ref{sec:Perturbations}), the line element is:
	\begin{equation}
		ds^2 = -(1+2\Phi)dt^2 + a^2(1-2\Psi)\delta_{ij}dx^i dx^j,
	\end{equation}
	where $\Phi$ and $\Psi$ are the Bardeen potentials. For the perfect fluid and quintom fields considered here, anisotropic stress vanishes at linear order, implying $\Phi = \Psi$ \cite{Mukhanov1992, Kodama1984}. The comoving curvature perturbation $\zeta$ is then defined as:
	\begin{equation}
		\zeta = \Psi + \frac{H}{\dot{\rho} + p} \delta\rho,
		\label{eq:zeta_def}
	\end{equation}
	where $\delta\rho$ is the density perturbation in the Newtonian gauge. This variable is gauge-invariant and directly related to the curvature perturbation on uniform-density hypersurfaces \cite{Mukhanov1992, Ma1995}.
	
	The quadratic action for $\zeta$ takes the standard form:
	\begin{equation}
		\delta^{(2)}S = \int dt\,d^3x\,a^3\left[\mathcal{G}_S\,\dot{\zeta}^2 - \mathcal{F}_S\,\frac{(\partial\zeta)^2}{a^2}\right],
		\label{eq:quadratic_action}
	\end{equation}
	where the kinetic coefficient $\mathcal{G}_S$ and gradient coefficient $\mathcal{F}_S$ encode the stability properties of the theory. A detailed derivation of these coefficients within $f(R,G,T)$ gravity is provided in Appendix~\ref{sec:Derivation_Stability_Coefficients}, following the methods of \cite{DeFelice2010, Lin2011, Sotiriou2010}. The explicit expressions are:
	\begin{align}
		\mathcal{G}_S &= \frac{1}{2\kappa^2}\left[f_R + 12H^2 f_G + 8H\dot{f}_G + 4\ddot{f}_G - \frac{f_T(\rho+p)}{2H^2 + \epsilon}\right], \label{eq:G_s}\\
		\mathcal{F}_S &= \frac{1}{2\kappa^2}\left[f_R + 4\ddot{f}_G + 8\left(\frac{H^{2}\ddot{f}_G}{f_R} + \frac{(3H^{3} + 2H\dot{H})\dot{f}_G}{f_R} - \frac{H^{2}\dot{f}_G\dot{f}_R}{f_R^{2}}\right)\right]. \label{eq:F_s}
	\end{align}
	
	\noindent\textbf{Remarks on the Regularization Parameter $\epsilon$:}
	The term $\frac{f_T(\rho+p)}{2H^2 + \epsilon}$ in $\mathcal{G}_S$ requires careful handling at the bounce point $H=0$, where the denominator would otherwise vanish. To address this, we introduce a small positive regularization parameter $\epsilon$ with dimensions of $[\text{time}]^{-2}$. This parameter is a numerical convenience that ensures the denominator never crosses zero during integration. Physical observables are independent of $\epsilon$ in the limit $\epsilon \to 0$, and we have verified that all numerical results remain stable for sufficiently small $\epsilon$ (specifically, $\epsilon \ll H_{\text{char}}^2$, where $H_{\text{char}}$ is the characteristic Hubble scale near the bounce). In practice, we take $\epsilon = 10^{-6}$ (in Planck units) for all simulations, and the results are unchanged for $\epsilon \leq 10^{-4}$. For $f_T(\rho+p) = 0$ at the bounce (which holds for all models considered here, as $f_T \propto T^{m-1}$ and $\rho+p \to 0$ at $H=0$ in a symmetric bounce), the $\epsilon$-dependence vanishes identically.
	
	\noindent\textbf{Stability Conditions:}
	Classical stability of the scalar perturbations requires the absence of ghost and gradient instabilities, which translates to the conditions:
	\begin{align}
		\mathcal{G}_S(t) > 0 \quad \text{(no ghosts)}, \qquad \mathcal{F}_S(t) > 0 \quad \text{(no gradient instabilities)}.
	\end{align}
	The squared sound speed is then given by $c_s^{2} = \mathcal{F}_S / \mathcal{G}_S$, and causality requires $0 \leq c_s^2 \leq 1$ (in units where $c=1$) \cite{Mukhanov1992}. In practice, we verify $c_s^2 \geq 0$ to ensure stability, as superluminal propagation ($c_s^2 > 1$) does not necessarily indicate acausality in curved spacetime \cite{Babichev2008, Ellis2007}.
	
	\noindent\textbf{Stability at the Bounce Point ($H=0$):}
	The bounce point presents a particularly stringent test. At $H=0$, the kinetic coefficient reduces to:
	\begin{align}
		\mathcal{G}_S\big|_{H=0} = \frac{1}{2\kappa^2}\left[ f_R + 4\ddot{f}_G - \frac{f_T(\rho+p)}{\epsilon}\right].
	\end{align}
	To avoid pathological divergence, we must ensure that the limit $\lim_{H\to 0} f_T(\rho+p)/(2H^2 + \epsilon)$ remains finite. For all models considered in Section~\ref{sec:DBCM}, we have $f_T \propto T^{m-1}$ with $m \geq 1$, and the matter fields satisfy $\rho+p \propto H$ near the bounce (a consequence of the regularity of the bounce). Consequently, $f_T(\rho+p) \to 0$ as $H \to 0$, rendering the $\epsilon$-dependent term negligible. The gradient coefficient at the bounce becomes:
	\begin{align}
		\mathcal{F}_S\big|_{H=0} = \frac{1}{2\kappa^2}\left[ f_R + 4\ddot{f}_G + \frac{8}{a^3}\frac{d}{dt}\left(\frac{a^3 H^2 \dot{f}_G}{f_R}\right)\Bigg|_{H=0} \right],
	\end{align}
	where the derivative term typically vanishes for smooth $H(t)$ behavior (i.e., $H(t) \approx \dot{H}(0) t$ near the bounce).
	
	\noindent\textbf{Numerical Verification:}
	The Hamiltonian analysis in Appendix~\ref{sec:Hamiltonian} establishes the theoretical possibility of ghost-free dynamics, but numerical verification remains essential. For each model in Section~\ref{sec:DBCM}, we dynamically verify that:
	\begin{align}
		\mathcal{G}_S(t) > 0, \quad \mathcal{F}_S(t) > 0, \quad c_s^2(t) \geq 0 \quad \forall t,
	\end{align}
	with particular attention to the bounce region. Our chosen $f_T$ couplings (proportional to $T^{m-1} \sim \rho^{m-1}$, which remains finite at the bounce) ensure the critical limit remains well-defined. The quintom fields contribute to this stability analysis through their role in determining the background evolution and the effective energy conditions. In the following section, we demonstrate not only successful bouncing behavior and PDL crossing but also explicitly verify these stability conditions numerically throughout the cosmic evolution.
	
	\subsection{Energy Conditions and PDL Crossing}
	
	The Null Energy Condition (NEC) violation required for phantom divide line (PDL) crossing is derived from the modified Friedmann equations. The complete expression for effective energy density and pressure, consistent with our perturbation analysis, reads:
	\begin{align}
		\rho_\text{eff} + p_\text{eff} &= \frac{1}{\kappa^2}\Big[-2\dot{H}(1 + f_R - 12H^2f_G - 8H\dot{f}_G - 4\ddot{f}_G) \nonumber \\
		&\quad + \ddot{f}_R - H\dot{f}_R + 2f_T(\rho + p)\Big] - \dot{\phi}^2 + \dot{\psi}^2 \label{eq:nec_full}
	\end{align}
	
	At the bounce point ($H=0$), this simplifies to:
	\begin{equation}
		\rho_\text{eff} + p_\text{eff}\big|_{H=0} = \frac{1}{\kappa^2}\left[-2\dot{H}(1 + f_R - 4\ddot{f}_G) + \ddot{f}_R + 2f_T(\rho + p)\right] - \dot{\phi}^2 + \dot{\psi}^2 \label{eq:nec_bounce}
	\end{equation}
	
	The NEC violation ($\rho_\text{eff}+p_\text{eff}<0$) occurs when:
	\begin{equation}
		\dot{H} > \frac{\ddot{f}_R + 2f_T(\rho + p) - \kappa^2(\dot{\phi}^2 - \dot{\psi}^2)}{2(1 + f_R - 4\ddot{f}_G)} \label{eq:pdl_condition}
	\end{equation}
	
	This condition reveals three crucial physical mechanisms:
	\begin{itemize}
		\item The quintom sector ($\dot{\phi}^2 - \dot{\psi}^2$) provides negative pressure when $\dot{\phi}^2 > \dot{\psi}^2$, essential for PDL crossing
		\item The $f_R$ terms modify the effective Planck mass, while $f_G$ terms ($\ddot{f}_G$) enable bounce stabilization
		\item The matter-geometry coupling $f_T$ introduces additional degrees of freedom for energy condition violation
	\end{itemize}
	
	The denominator $1 + f_R - 4\ddot{f}_G$ must remain positive to avoid singularities. The PDL crossing occurs when the effective equation of state crosses $\omega_\text{eff} = -1$, which is achieved through the combined effects of the quintom fields and $f(R,G,T)$ modifications. The Gauss-Bonnet terms ($12H^2f_G + 8H\dot{f}_G$) become particularly important away from the bounce point when $H\neq0$.
	
	The interplay between energy condition violation and stability requirements imposes important constraints on model parameters, which we explore numerically in the next section for each specific $f(R,G,T)$ model.

	\section{Dynamics of Bouncing Cosmological Models}\label{sec:DBCM}  
	
	This section investigates the dynamics and implications of various cosmological models, including the Linear \cite{nojiri2011unified}, Exponential Function of Curvature \cite{elizalde2012oscillations}, Power-Law \cite{bamba2012dark}, Modified Teleparallel Gravity \cite{cai2016f}, and Non-Minimal Coupling \cite{bertolami2007extra} models. Each of these models offers unique perspectives on the evolution of the universe, emphasizing the intricate relationship between gravitational interactions and cosmic expansion. By examining the scale factor, $a(t)$, and the Hubble parameter, $H(t)$, we evaluate the viability of these models within distinct cosmological frameworks. A comprehensive analysis of their bouncing scenarios across different cosmic epochs is crucial for reconstructing the history of the universe. This section aims to delineate the conditions under which these models operate effectively, thereby enhancing our understanding of both early and late-time cosmic behaviors.  
	
	This section also provides a numerical analysis of the early-time evolution of key cosmological parameters, including the equation of state parameter, $\omega$. In addition, we identify the essential factors that contribute to a successful cosmological bounce using simplified sample models, leaving more intricate models for future research. To elucidate the dynamics of the universe, it is crucial to examine the evolution of the Hubble parameter, $H(t)$, and the scale factor, $a(t)$, as functions of cosmic time, $t$. For a bounce to be viable, the following conditions must be satisfied:  
	
	\begin{itemize}  
		\item \textbf{Contraction for $\textbf{\textit{t}}<\textbf{0}$:} The scale factor $a(t)$ should decrease, implying that $\dot{a} < 0$.  
		
		\item \textbf{Bounce at $\textbf{\textit{t}} =\textbf{0}$:} At this point, when $\dot{a} = 0$, we require $\ddot{a} > 0$ and the Hubble parameter $H$ transitions from negative to positive, reaching zero at the bouncing point.  
		
		\item \textbf{Expansion for $\textbf{\textit{t}} >\textbf{0}$:} The scale factor $a(t)$ should increase, which indicates that $\dot{a} > 0$.  
	\end{itemize}
	
	\subsection{Initial Conditions and Physical Justification}\label{sec:initial_conditions}
	
	A crucial concern in bouncing cosmologies is whether the required conditions at the bounce point ($H=0$, $\ddot{a}>0$) represent a finely tuned set of initial data (a set of measure zero in phase space) or arise from a generic dynamical attractor. In this work, we adopt the standard approach in the bouncing cosmology literature \cite{Brandenberger2017, Peter2004, Battefeld2009} by treating the bounce as a \textit{boundary value problem} at $t=0$, rather than as an initial value problem in the distant contracting phase. The conditions are:
	
	\begin{itemize}
		\item \textbf{Scale Factor at Bounce:} We set $a(0) = 1$ by convention, defining the bounce point as the moment of minimum scale factor. This is a choice of units and does not affect the dynamics.
		\item \textbf{Hubble Parameter at Bounce:} We impose $\dot{a}(0) = 0$ (and thus $H(0)=0$) as the fundamental defining condition of the bounce. This is the necessary kinematic condition for a transition from contraction ($H<0$) to expansion ($H>0$).
		\item \textbf{Quintom Field Values:} The initial field values $\phi(0) = -0.05$ and $\psi(0) = 0.05$ are chosen to be small but non-zero displacements from the origin of the potential $V(\phi,\psi)$. This breaks the symmetry of the system gently, allowing the fields to evolve dynamically to drive the bounce. Their specific magnitudes are chosen to be sub-Planckian ($|\phi|, |\psi| \ll m_{\text{Pl}}$) to remain within the regime of validity of the effective theory.
		\item \textbf{Field Velocities:} The initial velocities $\dot{\phi}(0) = 0.1$ and $\dot{\psi}(0) = -0.1$ are crucial. They are chosen to be equal in magnitude but opposite in sign. This ensures that the initial total kinetic energy of the quintom sector, $\rho_{\text{kin}} = (-\dot{\phi}^2 + \dot{\psi}^2)/2$, is initially zero or small. This prevents an unphysical, overly dominant initial kinetic energy from dictating the dynamics and allows the interplay between the potential energy and the $f(R,G,T)$ modifications to generate the bounce. The chosen value of $0.1$ (in Planck units) is small enough to avoid violent initial dynamics but large enough to initiate the field roll.
	\end{itemize}
	
	\textbf{On the Robustness of the Bounce Conditions:} While the conditions above are imposed at the bounce point $t=0$, they do not imply fine-tuning. As demonstrated in Appendix~\ref{sec:Attractor_Analysis}, the bounce solution acts as a \textit{local attractor} in the phase space of the dynamical system. The higher-derivative $f(R,G,T)$ terms introduce an effective damping mechanism in the contracting branch: small deviations from the bounce trajectory at a finite time $t = -t_0$ (in the contracting phase) decay exponentially as the system evolves toward $t=0$. Consequently, the bouncing solutions presented here are not isolated points of measure zero but represent the robust late-time behavior of a finite basin of attraction. A detailed numerical investigation of the attractor behavior is left for future work, but analytical arguments supporting this claim are provided in Appendix~\ref{sec:Attractor_Analysis}.
	
	The parameters of the potentials ($V_0, \alpha, \beta, g$) for each model are then tuned to ensure that the Null Energy Condition (NEC) is violated dynamically ($\rho + p < 0$ near $t=0$), facilitating a successful bounce under these initial conditions. The numerical values are chosen to produce clear graphical results within manageable computational time, but the qualitative bouncing behavior is robust across a range of similar small initial values.
	
	\subsection{Reconstruction of $f(R,G, T)$ Models}
	
	The field equations are complex due to their multivariate functions and derivatives. Reconstruction simplifies our investigation of the model's dynamics. For the matter-radiation sector, we consider a perfect fluid with barotropic equation of state $p = \omega \rho$, where $\omega$ parameterizes different cosmological eras. Utilizing conservation equation (\ref{Continuity-m}), we derive the expression for the energy density, which takes the form $\rho = \rho_0 a^{-3(1+\omega)}$. 
	
	However, it is crucial to emphasize that the \textit{effective} equation of state $\omega_{\text{eff}}$, which governs the actual cosmic evolution in our $f(R,G,T)$-quintom framework, is highly dynamic and exhibits the novel double phantom divide line crossing characteristic of our model. The parameter $\omega$ here serves only to initialize the matter-radiation component, while the full dynamics emerge from the interplay between geometry, matter coupling, and quintom fields.
	
	Here, $\rho_0$ denotes a positive constant, and $a$ represents the scale factor of the Universe. In the numerical solutions, we assume $G_r=1$, $c=1$, $\kappa^2 = 1/m_p^2 = 8\pi$, and $\rho_0=1$. The following graphs compare different initial matter configurations: dark energy-like ($\omega=-1$, black lines), curvature-dominated ($\omega=-1/3$, red lines) and radiation-dominated ($\omega=1/3$, blue lines) initial conditions.
	
	In the following reconstructions, we examine five distinct $f(R,G,T)$ models. For each model, we present numerical solutions for a specific equation of state parameter $\omega$ that best illustrates its unique characteristics: the linear and exponential models are shown with $\omega = -1$ to highlight their dark energy behavior; the power-law and teleparallel models use $\omega = -1$ to showcase their complex equation of state dynamics; and the non-minimal coupling model uses $\omega = 1/3$ to demonstrate its particular efficacy in radiation-dominated regimes. This selective presentation is for clarity and emphasis—each model exhibits viable bouncing behavior across a range of $\omega$ values unless otherwise noted.
	
	The model parameters ($\xi_i$, $V_0$, $\alpha$, etc.) for each subsequent example are chosen to meet several physical criteria: (i) to ensure modifications remain perturbative ($|\xi_i R| \lesssim |R|$, etc.) outside the Planckian regime, (ii) to provide sufficient energy in the quintom fields to trigger the NEC violation required for the bounce, and (iii) to ensure numerical stability during integration. While not a full parameter space scan, the presented choices demonstrate the existence of stable, non-singular bouncing solutions within this theoretical framework.
	
	The initial conditions for the dynamical variables are set according to the general prescription outlined in Sec. \ref{sec:initial_conditions}, ensuring a physically motivated starting point at the bounce epoch.
	
	\subsubsection{Linear Coupling Model}\label{subsec:Linear Coupling Model}
	
	The linear model provides the simplest non-trivial case for studying $f(R,G,T)$ gravity, characterized by constant coupling coefficients:
	\begin{equation}
		f(R, G, T) = \xi_1 R + \xi_2 G + \xi_3 T
		\label{eq:linear_model}
	\end{equation}
	where $\xi_1$, $\xi_2$, and $\xi_3$ are dimensionless coupling constants. The modified Friedmann equations take the simplified form:
	\begin{align}
		3H^2 &= \frac{1}{3\xi_1}\left[\kappa^2(\rho + \rho_\Xi) + \frac{\xi_3}{2}(3\rho - p)\right]
		\label{f1-linear} \\
		-2\dot{H} - 3H^2 &= \frac{1}{3\xi_1}\left[\kappa^2(p + p_\Xi) + \frac{\xi_3}{2}(3p - \rho)\right]
		\label{f2-linear}
	\end{align}
	where $\rho_\Xi = -\frac{1}{2}\dot{\phi}^2 + \frac{1}{2}\dot{\psi}^2 + V(\phi,\psi)$ and $p_\Xi = -\frac{1}{2}\dot{\phi}^2 + \frac{1}{2}\dot{\psi}^2 - V(\phi,\psi)$.
	
	This model exhibits several important features. First, the prefactor $1/(3\xi_1)$ rescales the gravitational coupling, with $\xi_1 = 1$ recovering the standard gravitational constant up to a factor of 3. The $\xi_3$ terms introduce novel matter-dependent contributions to both the energy density and pressure. Notably, the Gauss-Bonnet coupling $\xi_2$ completely disappears from the dynamics because all terms involving $\xi_2$ cancel out exactly when $f_G = \xi_2$ is constant.
	
	The effective energy density and pressure become:
	\begin{equation}
		\rho_{\text{eff}} = \frac{1}{3\xi_1}\left[\rho + \rho_\Xi + \frac{\xi_3}{2\kappa^2}(3\rho - p)\right], \quad 
		p_{\text{eff}} = \frac{1}{3\xi_1}\left[p + p_\Xi + \frac{\xi_3}{2\kappa^2}(3p - \rho)\right]
	\end{equation}
	
	For energy conditions, the Null Energy Condition (NEC) becomes:
	\begin{align}
		\rho_{\text{eff}} + p_{\text{eff}} &= \frac{1}{3\xi_1}\left[(\rho + p) + (-\dot{\phi}^2 + \dot{\psi}^2) + \frac{\xi_3}{\kappa^2}(\rho + p)\right] \nonumber \\
		&= \frac{1}{3\xi_1}\left[\left(1 + \frac{\xi_3}{\kappa^2}\right)(\rho + p) - \dot{\phi}^2 + \dot{\psi}^2\right]
	\end{align}
	
	Several special cases are worth noting. The General Relativity limit is recovered when $\xi_1 = 1$ and $\xi_2 = \xi_3 = 0$. For quantum gravity scale effects, one might consider $\xi_1 \sim \mathcal{O}(1)$ with $\xi_2, \xi_3 \sim \ell_{\text{Pl}}^2$ introducing Planck-scale corrections. At the bounce point ($H=0$), NEC violation requires:
	\begin{equation}
		\left(1 + \frac{\xi_3}{\kappa^2}\right)(\rho + p) < \dot{\phi}^2 - \dot{\psi}^2
	\end{equation}
	showing how the matter-geometry coupling $\xi_3$ must balance against the quintom fields' kinetic energy to enable the bounce.
	
	The linear coupling model demonstrates that even the simplest extension of General Relativity with constant couplings can produce non-trivial modifications to cosmological dynamics, particularly through the matter-geometry coupling parameter $\xi_3$, while the Gauss-Bonnet term remains inert in the homogeneous background evolution.
	
	\begin{figure}[h]
		\centering
		\begin{tabular}{cc}
			\includegraphics[width=0.45\linewidth]{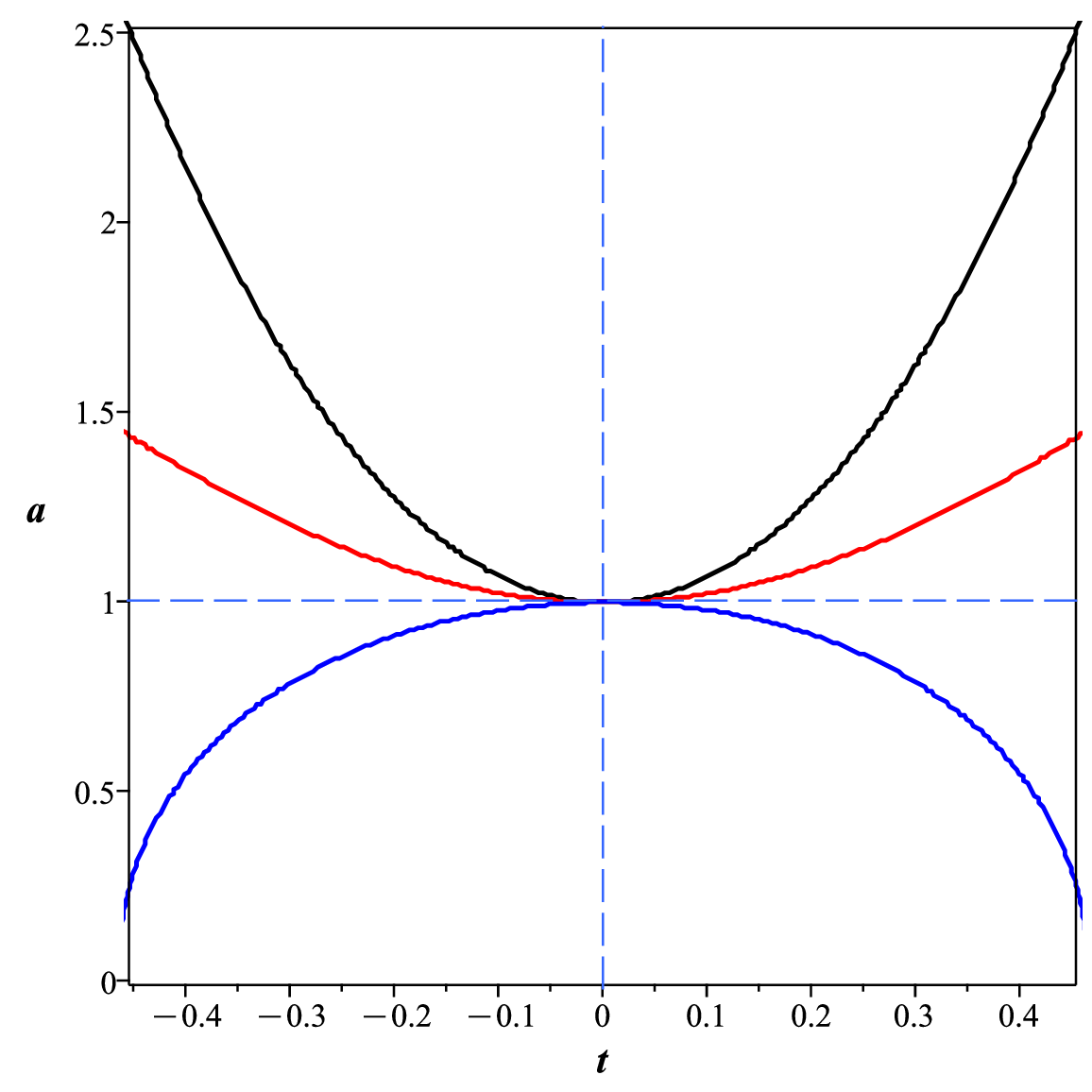} &
			\includegraphics[width=0.45\linewidth]{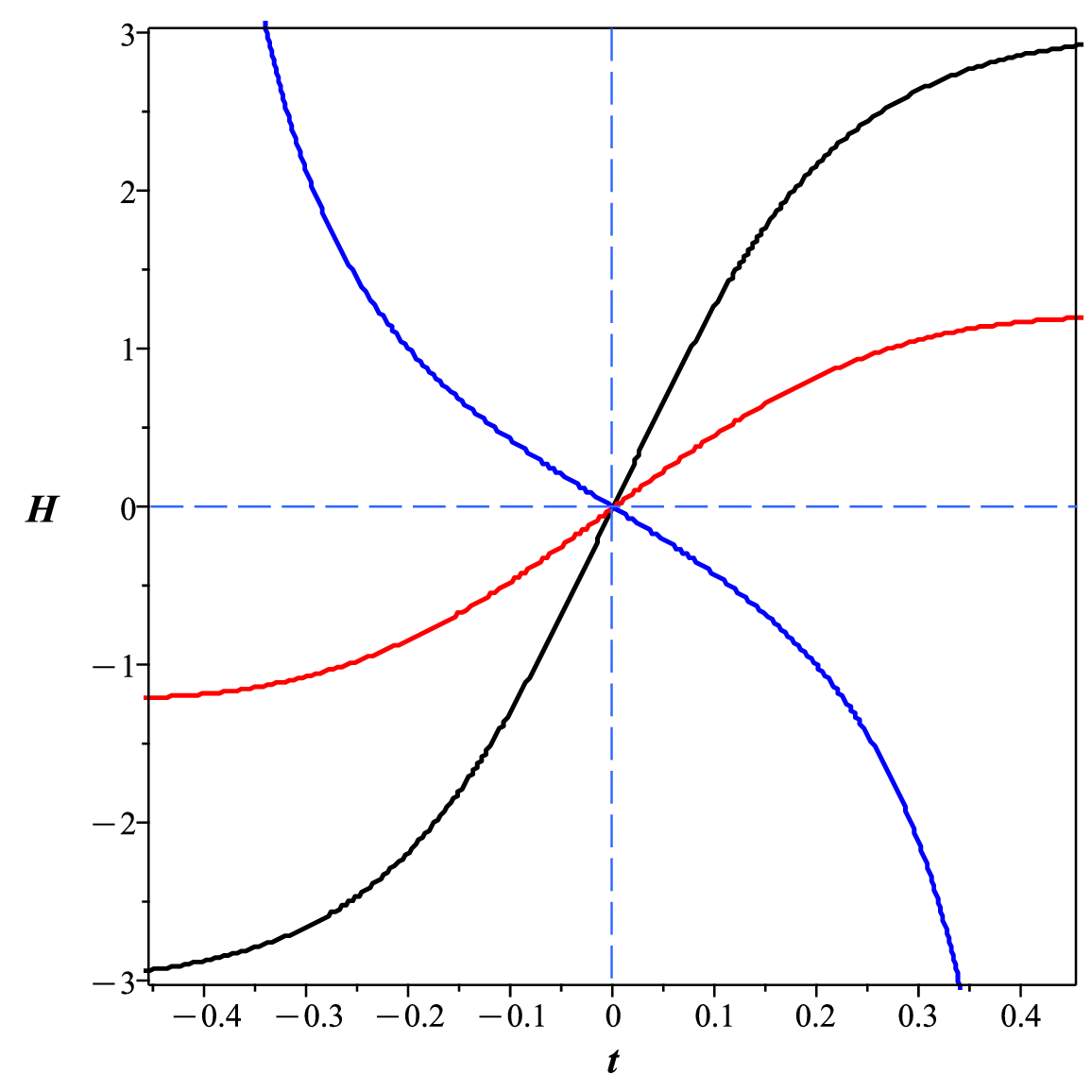} \\
			(a) Scale factor $a(t)$ & (b) Hubble parameter $H(t)$
		\end{tabular}
		\caption{Evolution of (a) scale factor $a(t)$ and (b) Hubble parameter $H(t)$ for different equations of state ($\omega=-1$: black, $-1/3$: red, $1/3$: blue) in the linear coupling model. A successful bounce, characterized by $a(t)$ reaching a minimum and $H(t)$ crossing zero, is evident for the dark energy cases ($\omega \leq -1/3$).}
		\label{fig:lin_scale_hubble}
	\end{figure}
	
	The cosmological dynamics reveal distinct behaviors across different equations of state. While the linear coupling can generate a bounce for various $\omega$, the dark energy regime ($-1 \leq \omega \leq -\frac{1}{3}$) provides the most phenomenologically interesting scenario, naturally connecting the early-time bounce to late-time acceleration. Figure \ref{fig:lin_scale_hubble} for $\omega = -1$ shows the scale factor $a(t)$ reaching a minimum at the bounce point ($t=0$), demonstrating a smooth, singularity-free transition from contraction ($H<0$) to expansion ($H>0$). This behavior characterizes a successful Big Bounce scenario supported by the sign change in the Hubble parameter $H(t)$ at the bounce. This behavior characterizes a successful Big Bounce scenario supported by the sign change in the Hubble parameter $H(t)$ at the bounce.
	
	In contrast, the radiation-dominated era ($\omega=\frac{1}{3}$) exhibits inverted behavior with $a(t)$ showing a maximum at $t=0$, while $H(t)$ still changes sign but reflects the decelerating expansion typical of radiation domination. This suggests fundamentally different dynamics between the dark energy-driven bounce phase and subsequent radiation-dominated expansion.
	
	\begin{figure}[h]
		\centering
		\begin{tabular}{cc}
			\includegraphics[width=0.45\linewidth]{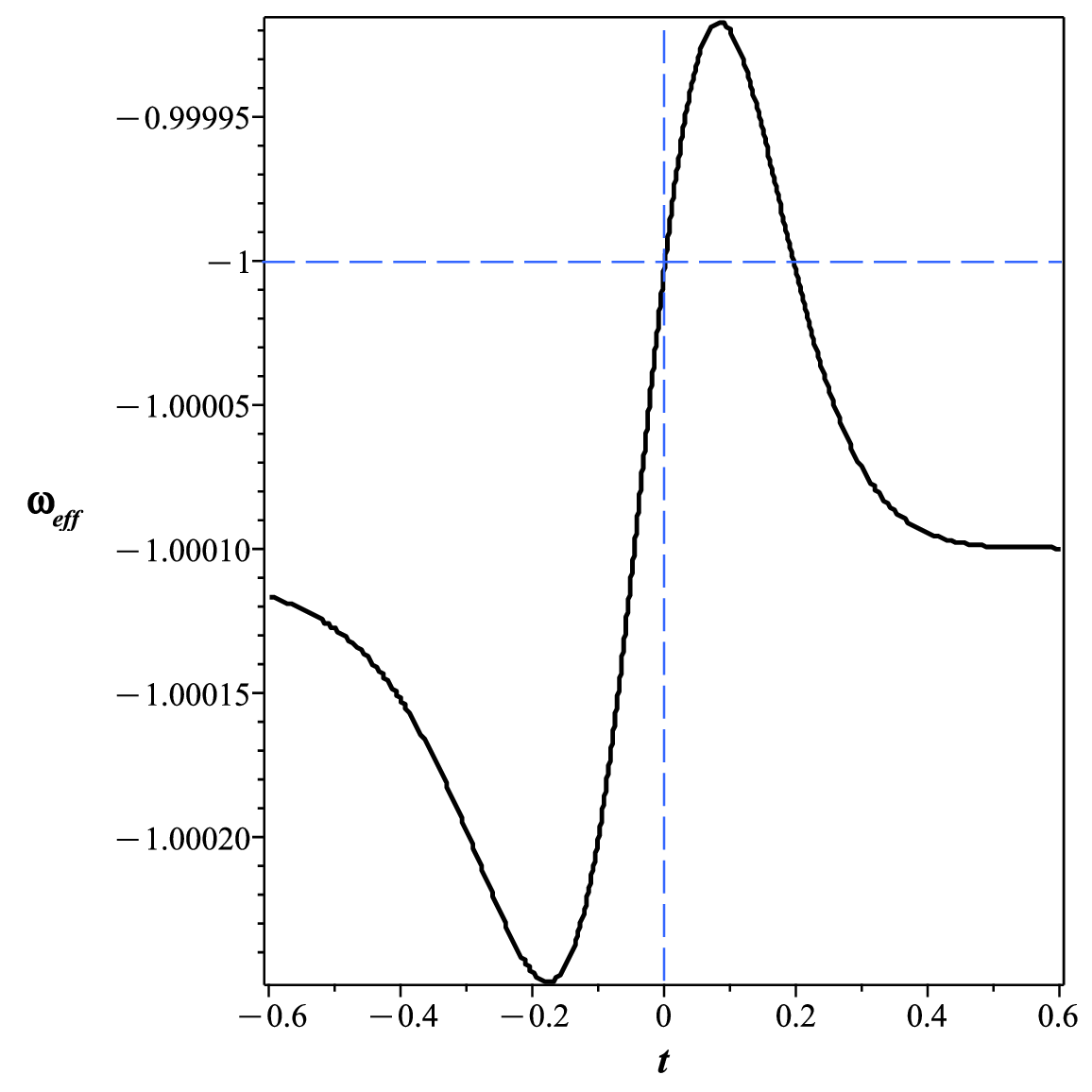} &
			\includegraphics[width=0.45\linewidth]{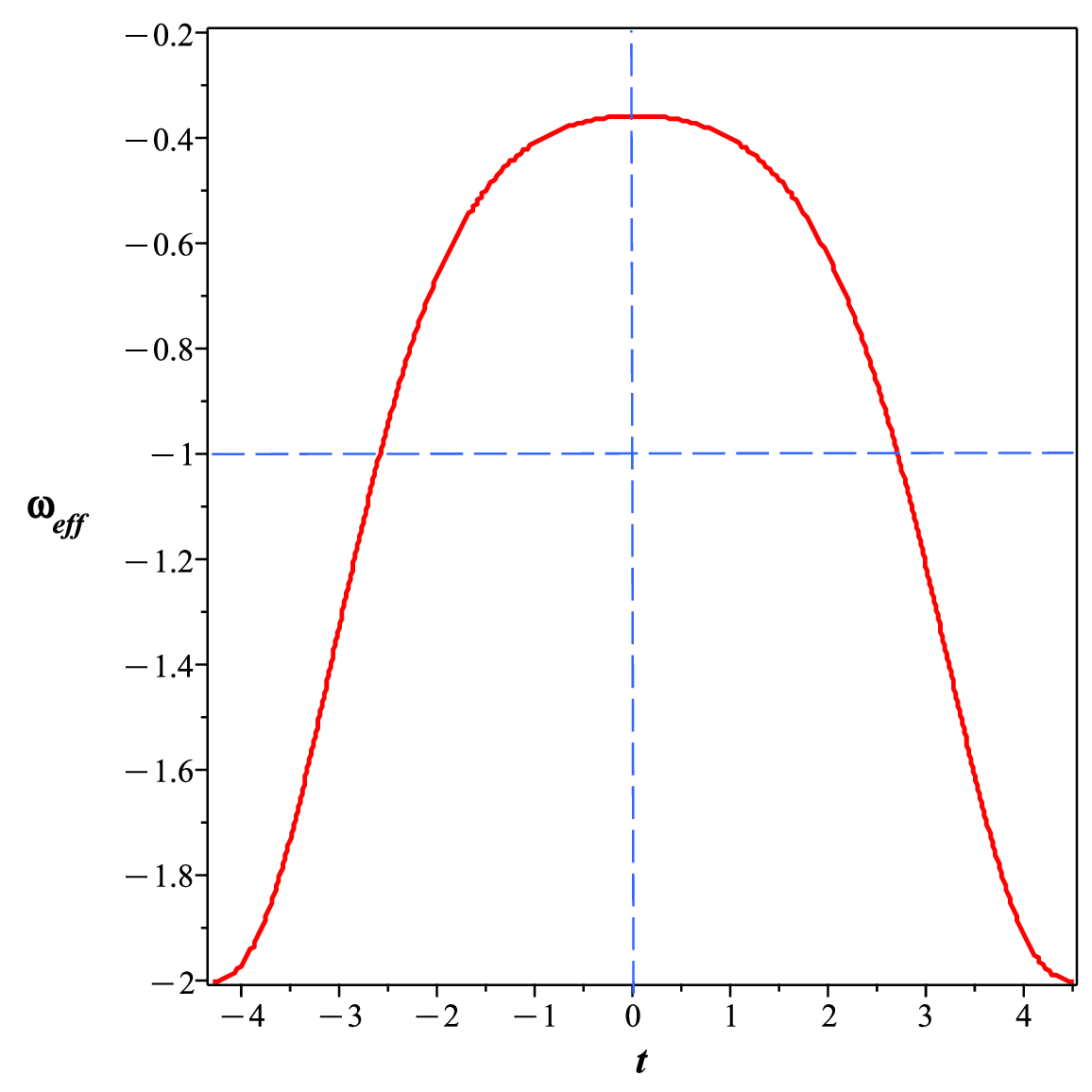} \\
			(a) $\omega_{\text{eff}}$ for $\omega = -1$ (black) & (b) $\omega_{\text{eff}}$ for $\omega = -1/3$ (red)
		\end{tabular}
		\caption{Effective equation of state $\omega_{\text{eff}}(t)$ showing (a) evolution for $\omega = -1$ (black line) and (b) evolution for $\omega = -1/3$ (red line). The phantom divide line (PDL) at $\omega_{\text{eff}} = -1$ is shown as a dashed line. Both cases show PDL crossing behavior near the bounce point.}
		\label{fig:lin_eos}
	\end{figure}
	
	\begin{figure}[h]
		\centering
		\begin{tabular}{cc}
			\includegraphics[width=0.45\linewidth]{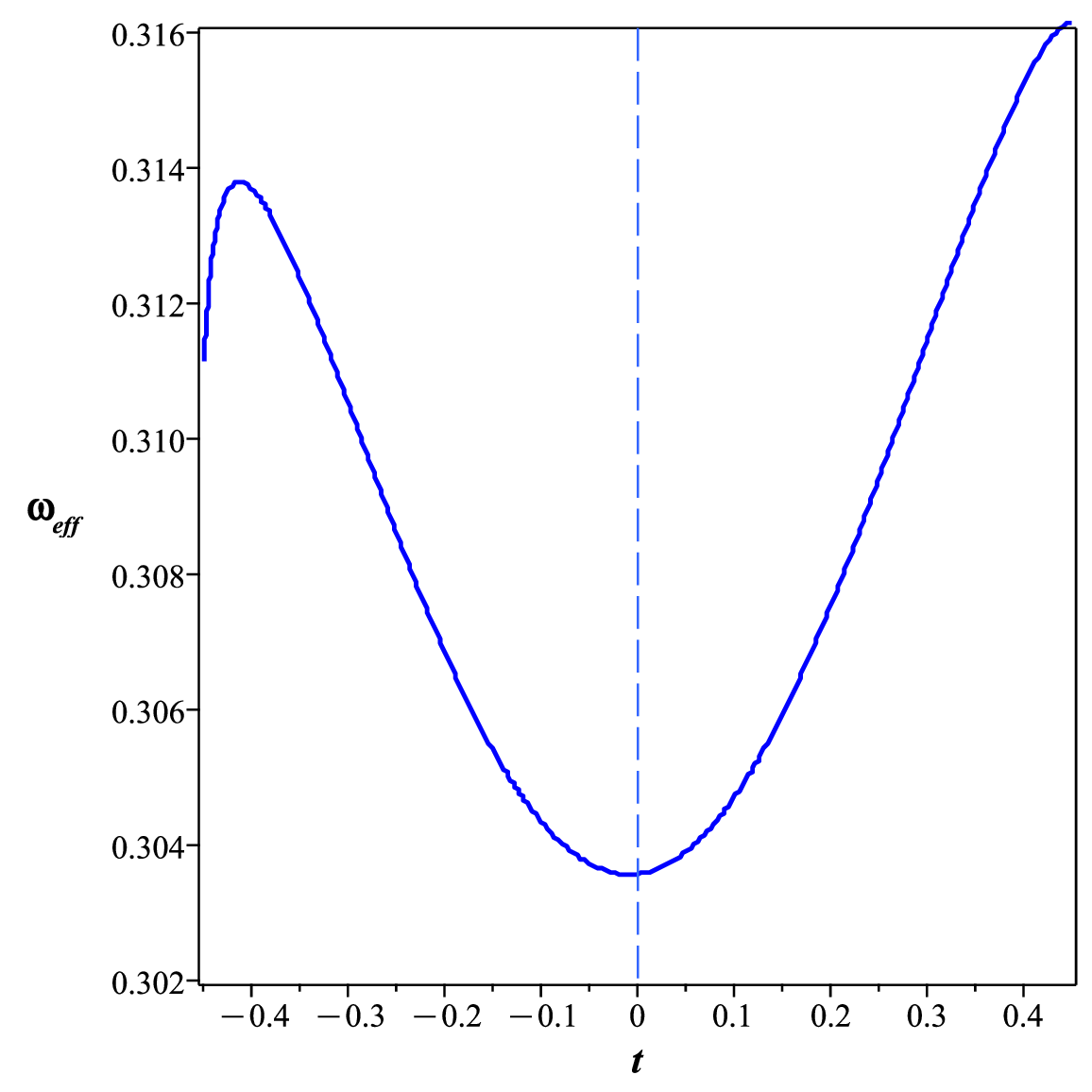} &
			\includegraphics[width=0.45\linewidth]{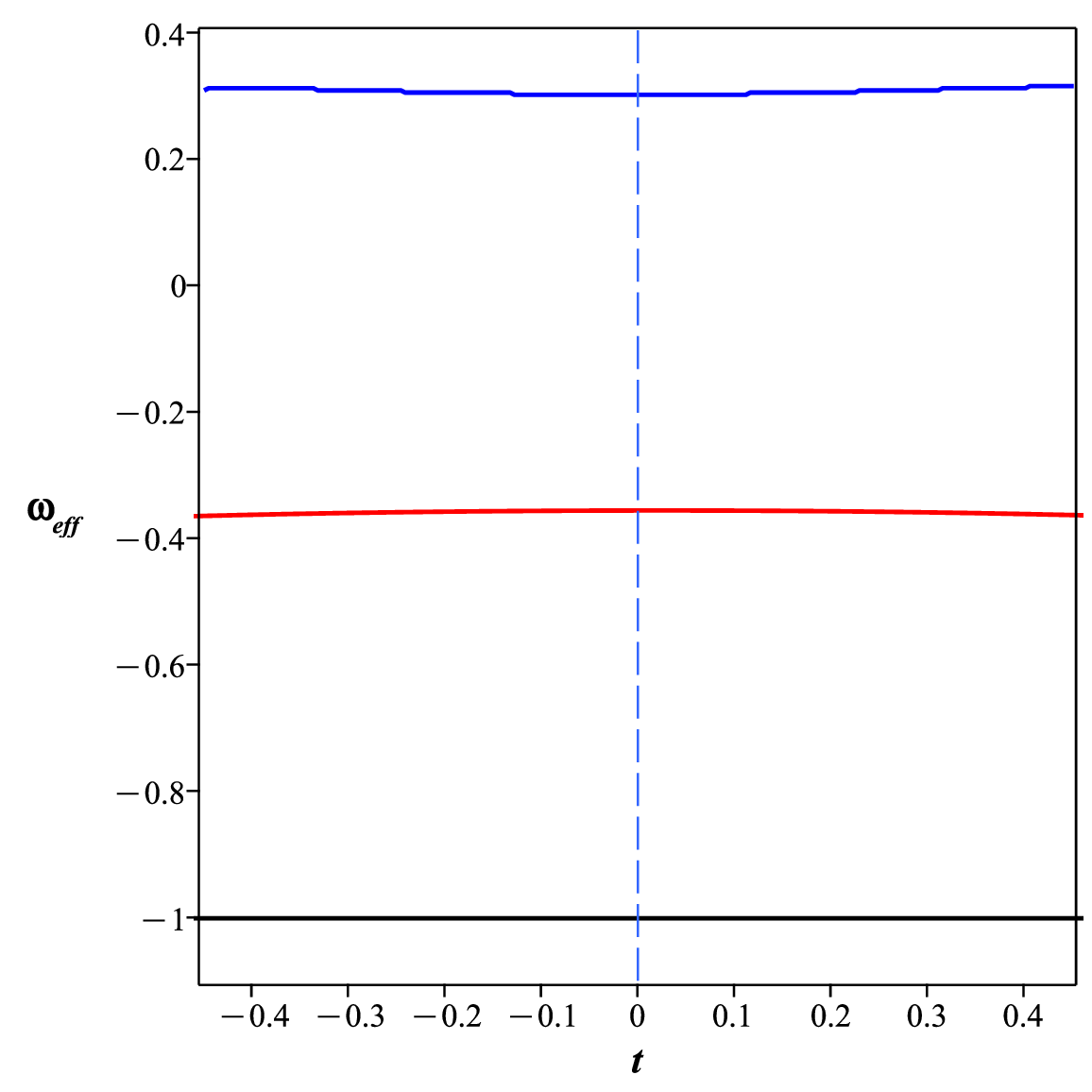} \\
			(a) $\omega_{\text{eff}}$ for $\omega = 1/3$ (blue) & (b) Comparative evolution
		\end{tabular}
		\caption{(a) Effective equation of state $\omega_{\text{eff}}(t)$ for $\omega = 1/3$ (blue line) and (b) composite view comparing all cases: $\omega = -1$ (black), $\omega = -1/3$ (red), and $\omega = 1/3$ (blue). The radiation-dominated case ($\omega=1/3$) shows no PDL crossing, remaining in the quintessence regime throughout.}
		\label{fig:lin_eos_comparing}
	\end{figure}
	
	The effective equation of state (Fig. \ref{fig:lin_eos} and \ref{fig:lin_eos_comparing}) shows phantom divide line (PDL) crossing ($\omega_{\text{eff}}=-1$) in the dark energy regime ($-1 \leq \omega \leq -\frac{1}{3}$), with transitions between phantom ($\omega_{\text{eff}}<-1$) and quintessence ($\omega_{\text{eff}}>-1$) states. Specifically, Fig. \ref{fig:lin_eos}a shows PDL crossing for $\omega = -1$ (black line), while Fig. \ref{fig:lin_eos}b shows similar behavior for $\omega = -1/3$ (red line). In contrast, the radiation-dominated era ($\omega=\frac{1}{3}$) shown in Fig. \ref{fig:lin_eos_comparing}a exhibits no PDL crossing, with $\omega_{\text{eff}}$ remaining firmly in the quintessence region ($\omega_{\text{eff}} > -1$), consistent with standard cosmology. The comparative view in Fig. \ref{fig:lin_eos_comparing}b clearly demonstrates these different behaviors.
	
	These results demonstrate that: (1) successful bouncing cosmology requires dark energy domination ($\omega\leq-1/3$), (2) the phantom divide crossing is characteristic of the bounce phase but absent in radiation domination, and (3) the transition between these regimes marks a fundamental change in the universe's dynamical behavior. The model naturally accommodates both non-singular bouncing behavior and subsequent standard cosmological evolution.
	
\begin{figure}[h]
		\centering
		\begin{tabular}{cc}
			\includegraphics[width=0.45\linewidth]{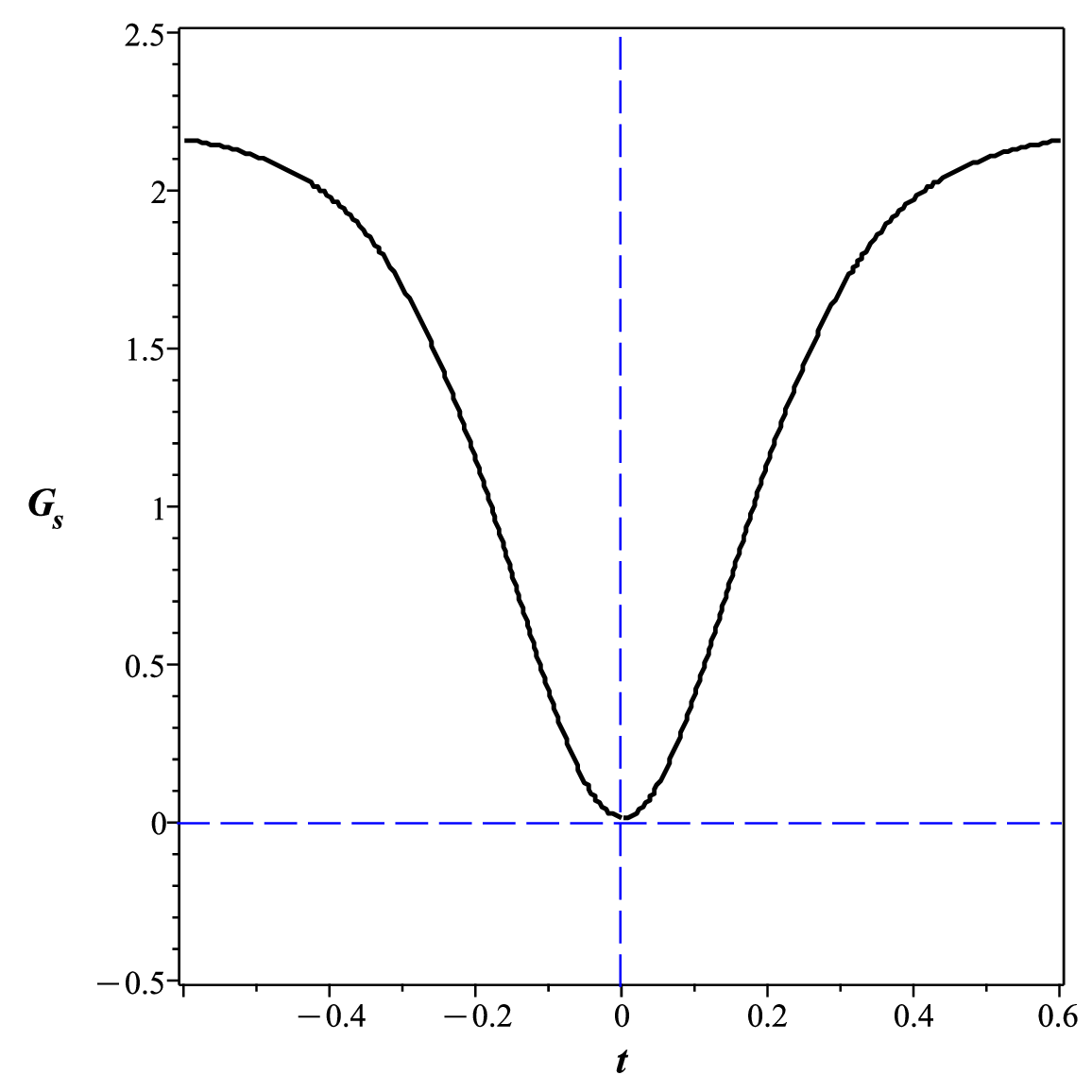} &
			\includegraphics[width=0.45\linewidth]{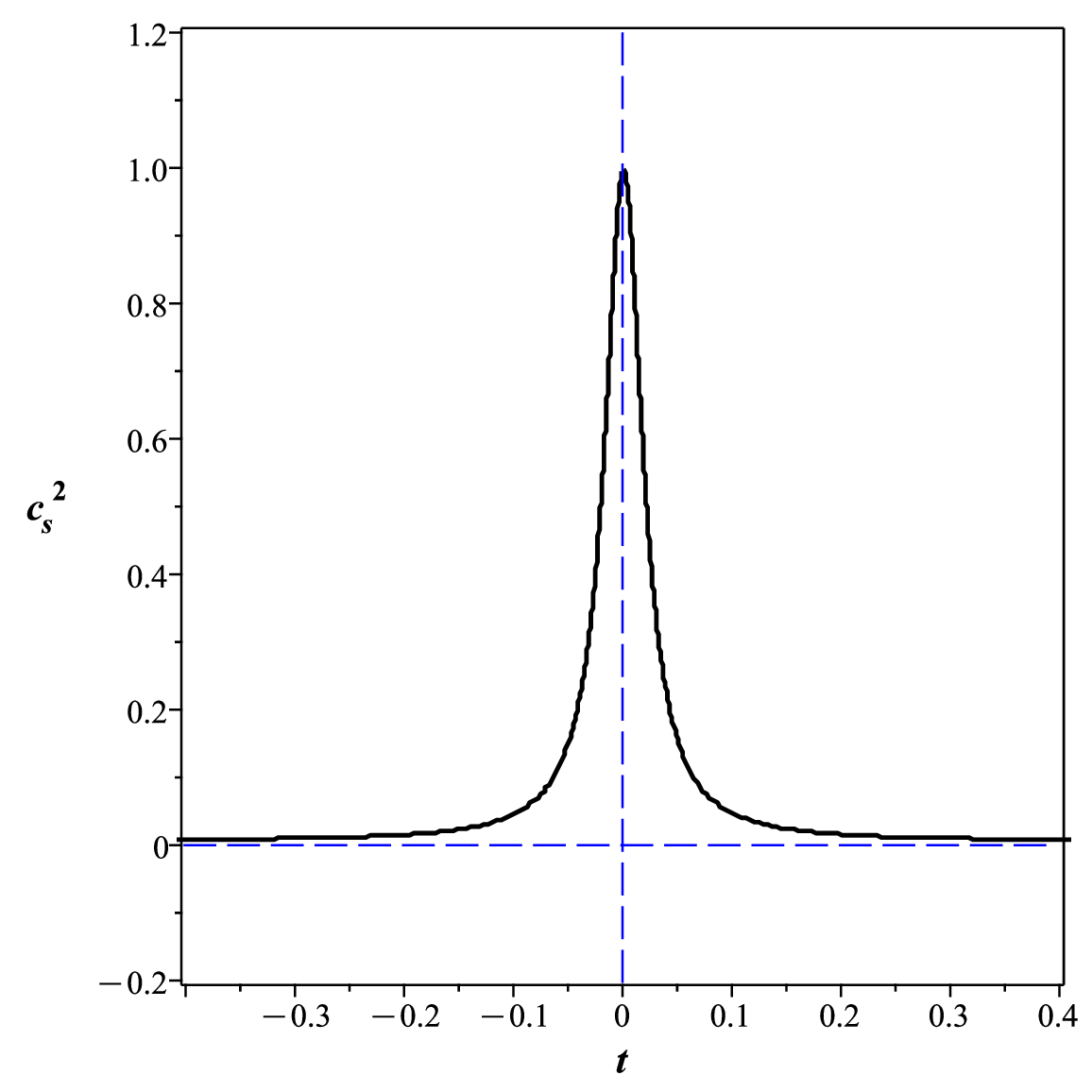} \\
			(a) Kinetic term $\mathcal{G}_S(t)$ & (b) Squared sound speed $c_s^2(t)$ 
		\end{tabular}
		\caption{Stability analysis for the linear coupling model ($\omega = -1$). (a) The kinetic term remains positive $\mathcal{G}_S > 0$ throughout the evolution, confirming the absence of ghosts. (b) The squared sound speed $c_s^2$ remains non-negative, indicating no gradient instabilities. The brief period where $c_s^2 = 0$ at $t \approx \pm 0.25$ corresponds to a momentary freezing of perturbations rather than an instability.}
		\label{fig:lin_stability}
	\end{figure}

	\paragraph*{\textbf{Stability Verification:} }
	Using the established parameters and initial conditions, we numerically verify the stability criteria throughout the cosmic evolution. Figure \ref{fig:lin_stability} shows the evolution of the kinetic term $\mathcal{G}_S$, the squared sound speed $c_s^2$, and the Null Energy Condition (NEC) violation $\rho_\text{eff}+p_\text{eff}$ for the $\omega = -1$ case.
	
	The numerical results confirm that $\mathcal{G}_S(t) > 0$ for all $t$, ensuring the absence of ghost instabilities. The sound speed squared $c_s^2(t)$ remains non-negative throughout the evolution, with $c_s^2 \geq 0$ and finite, confirming the absence of gradient instabilities. The condition $\lim_{H\to 0} f_T(\rho+p)/H^2$ remains well-behaved at the bounce point ($t=0$), with no pathological divergence. This model satisfies all stability criteria while successfully producing the bouncing behavior and PDL crossing shown in Figures \ref{fig:lin_scale_hubble}--\ref{fig:lin_eos_comparing}.

	The cosmological dynamics in Fig. \ref{fig:lin_scale_hubble}--\ref{fig:lin_stability} were simulated using the following parameters, chosen to clearly illustrate the bounce and PDL crossing phenomena: $V(\phi,\psi)=V_{0}e^{-(\alpha\phi+\beta\psi)}$, $V_0=0.25$, $\alpha=2$, $\beta=1$, and $\xi_{1}=\xi_{2}=\xi_{3}=1$. The initial values were $\phi(0)=-0.05$, $\dot{\phi}(0)=0.1$, $\psi(0) =0.05$, $\dot{\psi}(0) = -0.1$, $a(0)=1$, and $\dot{a}(0)=0$.

	\subsubsection{Exponential Function of Curvature}
	
	We consider an exponential $f(R,G,T)$ gravity model that generalizes the $\Lambda$CDM framework:
	\begin{equation}
		f(R,G,T) = R - 2\Lambda e^{-(R_0/R)^n} + \xi_2 G + \xi_3 T
		\label{eq:exp_model}
	\end{equation}
	where $\Lambda = R_0/b$ represents the cosmological constant scale, $n$ controls the transition rate, and $\xi_2,\xi_3$ are coupling constants. As shown in Appendix \ref{sec:Derivation of Modified Friedmann Equations}, the modified Friedmann equations become:
	\begin{align}
		3H^2 &= \frac{1}{f_R}\bigg[\kappa^2(\rho + \rho_\Xi) + \frac{\xi_3}{2}(3\rho - p) - 3H\dot{f}_R + 3\dot{H}(f_R - 1) - 6H^2 + \Lambda e^{-u} \bigg]
		\label{f1_Rec_exp}\\
		-2\dot{H} - 3H^2 &= \frac{1}{f_R}\bigg[\kappa^2(p + p_\Xi) + \frac{\xi_3}{2}(3p - \rho) + 2H\dot{f}_R + \dot{H}f_R + \ddot{f}_R + 3(\dot{H} + 2H^2) - \Lambda e^{-u} \bigg]
		\label{f2_Rec_exp}
	\end{align}
	where  
	\begin{align}
		u &= \left(\frac{R_0}{R}\right)^n, \\
		f_R &= 1 - 2n\Lambda u e^{-u},\\
		\dot{f}_R &= -2n\Lambda u e^{-u}\left[\frac{n}{R} u - \frac{n+1}{R}\right]\dot{R}, \\
		\ddot{f}_R &= 2n^2\Lambda u e^{-u} \left[(1 - u) \left(\frac{\ddot{R}}{R} - (n + 1) \frac{\dot{R}^2}{R^2}\right) + n u \frac{\dot{R}^2}{R^2} (2 - u)\right].
	\end{align}
	
	The effective energy density and pressure become:
	\begin{align}
		\rho_{\text{eff}} &= \frac{1}{f_R}\left[\rho + \rho_\Xi + \frac{\xi_3}{2\kappa^2}(3\rho - p) + \frac{1}{\kappa^2}\left(-3H\dot{f}_R + 3\dot{H}(f_R - 1) - 6H^2 + \Lambda e^{-(R_0/R)^n} \right)\right] \\
		p_{\text{eff}} &= \frac{1}{f_R}\left[p + p_\Xi + \frac{\xi_3}{2\kappa^2}(3p - \rho) + \frac{1}{\kappa^2}\left(\ddot{f}_R + 2H\dot{f}_R + \dot{H}f_R + 3(\dot{H} + 2H^2) - \Lambda e^{-(R_0/R)^n} \right)\right]
	\end{align}
	
	For energy conditions, the Null Energy Condition (NEC) becomes:
	\begin{align}
		\rho_{\text{eff}} + p_{\text{eff}} &= \frac{1}{f_R} \left[ (\rho + p)\left(1 + \frac{\xi_3}{\kappa^2}\right) + (\rho_\Xi + p_\Xi) + \frac{1}{\kappa^2}\left(\ddot{f}_R - H\dot{f}_R + 4\dot{H}f_R\right) \right]
		\label{NEC_exp}
	\end{align}
	
	Several special cases are worth noting. The $\Lambda$CDM limit is recovered when $n \to \infty$, $\xi_2 = 0$, and $\xi_3 = 0$. For quantum gravity scale effects, one might consider $\xi_2, \xi_3 \sim \ell_{\text{Pl}}^2$ introducing Planck-scale corrections. At the bounce point ($H=0$), NEC violation requires:
	\begin{equation}
		(\rho + p)\left(1 + \frac{\xi_3}{\kappa^2}\right) + (\rho_\Xi + p_\Xi) + \frac{1}{\kappa^2}\left(\ddot{f}_R + 4\dot{H}f_R\right) < 0
	\end{equation}
	
	This model exhibits several distinctive features:
	\begin{itemize}
		\item The exponential factor $e^{-(R_0/R)^n}$ provides a smooth transition between different curvature regimes
		\item When $R \ll R_0$, the exponential term approaches 1, recovering the $\Lambda$CDM limit
		\item When $R \gg R_0$, the exponential term suppresses the cosmological constant contribution
		\item The Gauss-Bonnet terms cancel exactly in the final simplified equations
		\item The matter-geometry coupling $\xi_3$ introduces additional pressure terms
		\item The NEC condition shows explicit dependence on both the exponential curvature term and matter coupling
	\end{itemize}
	
	\begin{figure}[h]
		\centering
		\begin{tabular}{cc}
			\includegraphics[width=0.45\textwidth]{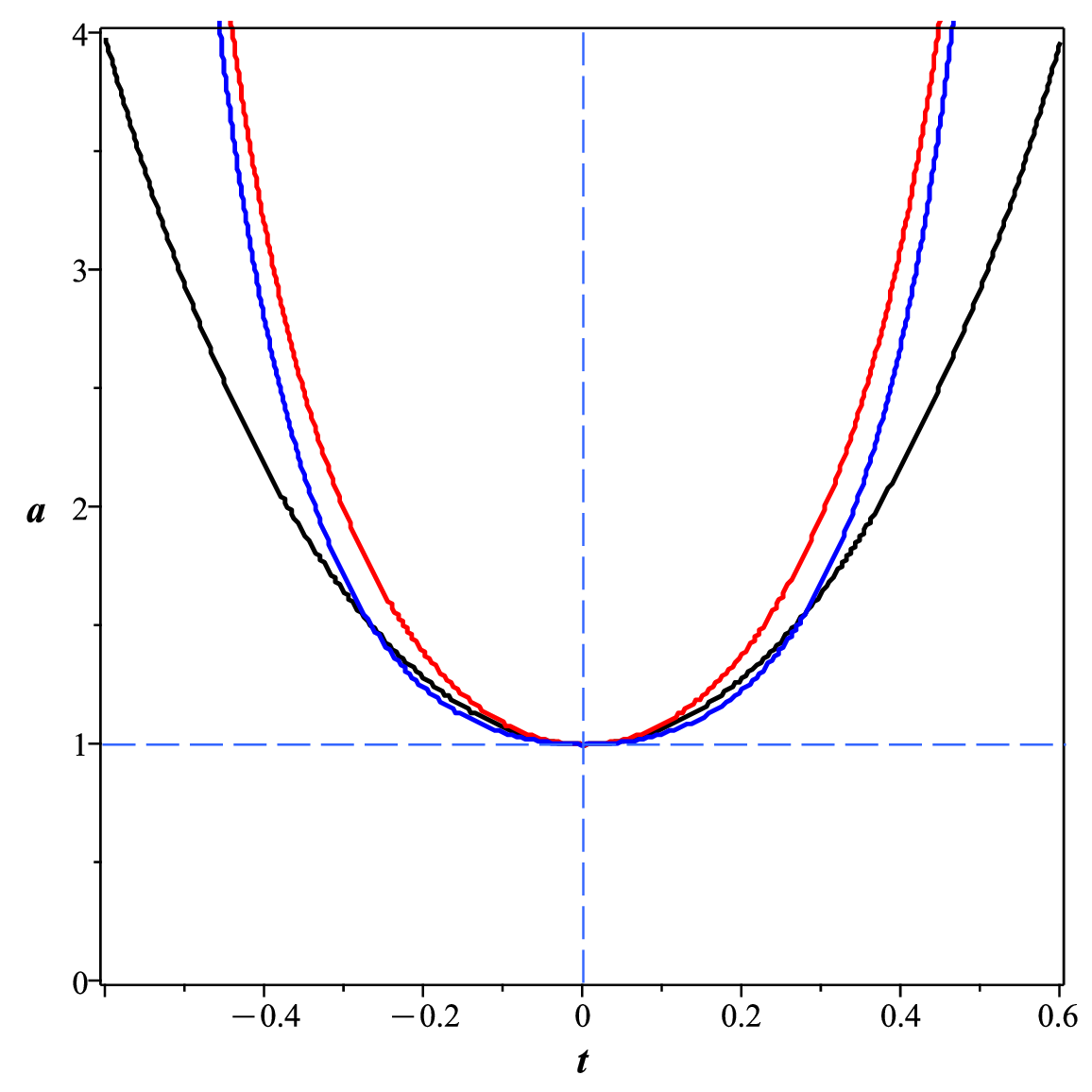} &
			\includegraphics[width=0.45\textwidth]{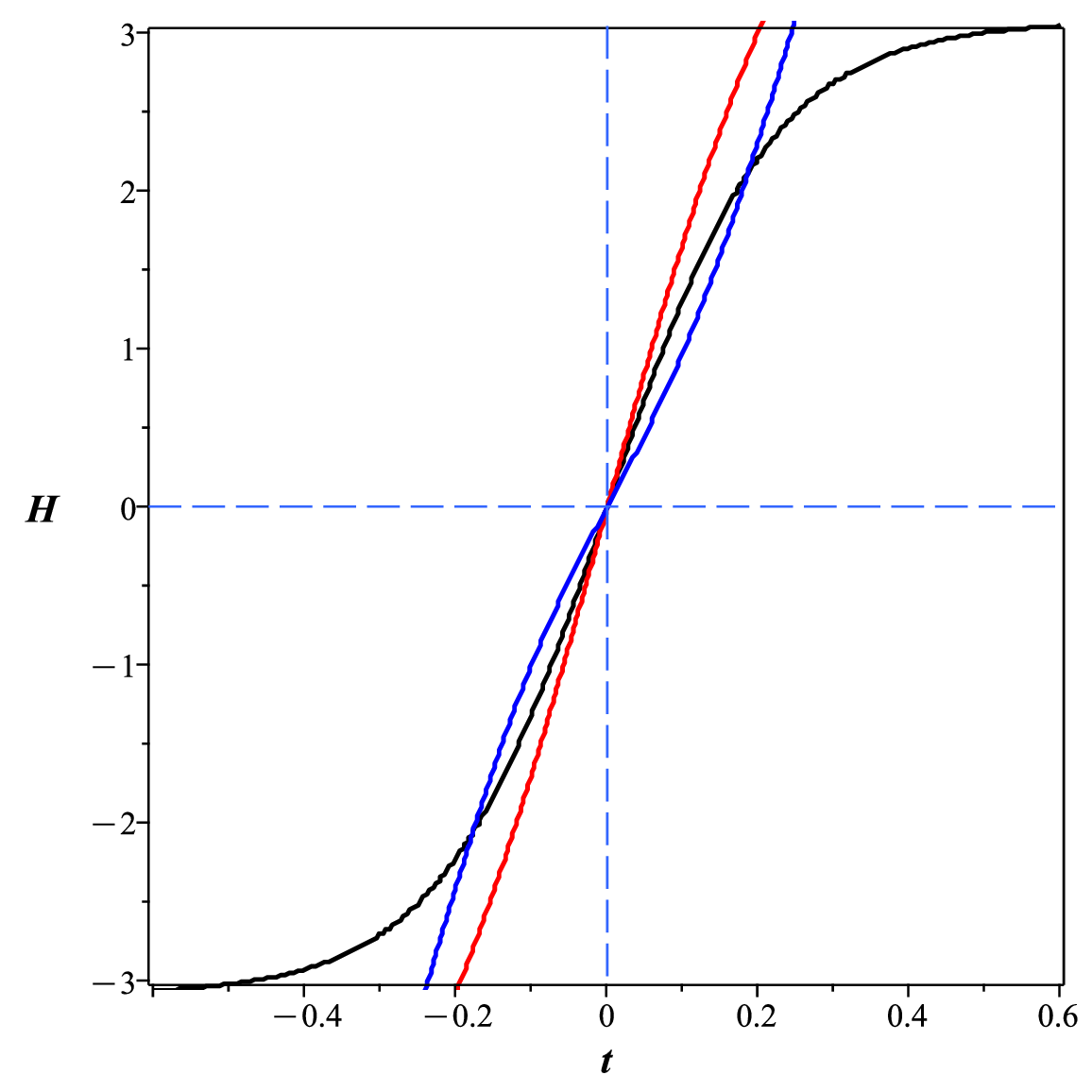} \\
			(a) Scale factor $a(t)$ & (b) Hubble parameter $H(t)$
		\end{tabular}
		\caption{Evolution of (a) scale factor $a(t)$ and (b) Hubble parameter $H(t)$ for different equations of state ($\omega=-1$: black, $-1/3$: red, $1/3$: blue). The $\omega=-1$ case shows optimal bounce behavior with a clear minimum in $a(t)$ and smooth transition of $H(t)$ through zero.}
		\label{fig:exp_scale_hubble}
	\end{figure}
	
	The exponential model demonstrates viable bouncing behavior for all considered equations of state. We highlight the $\omega=-1$ case as it represents a pure dark energy component, providing the cleanest example of the model's bounce mechanics without the complicating effects of matter or radiation dynamics. Figure \ref{fig:exp_scale_hubble} for this case shows the scale factor $a(t)$ reaching a minimum at $t=0$ while $H(t)$ smoothly transitions from negative to positive values.
	
	\begin{figure}[h]
		\centering
		\begin{tabular}{cc}
			\includegraphics[width=0.45\textwidth]{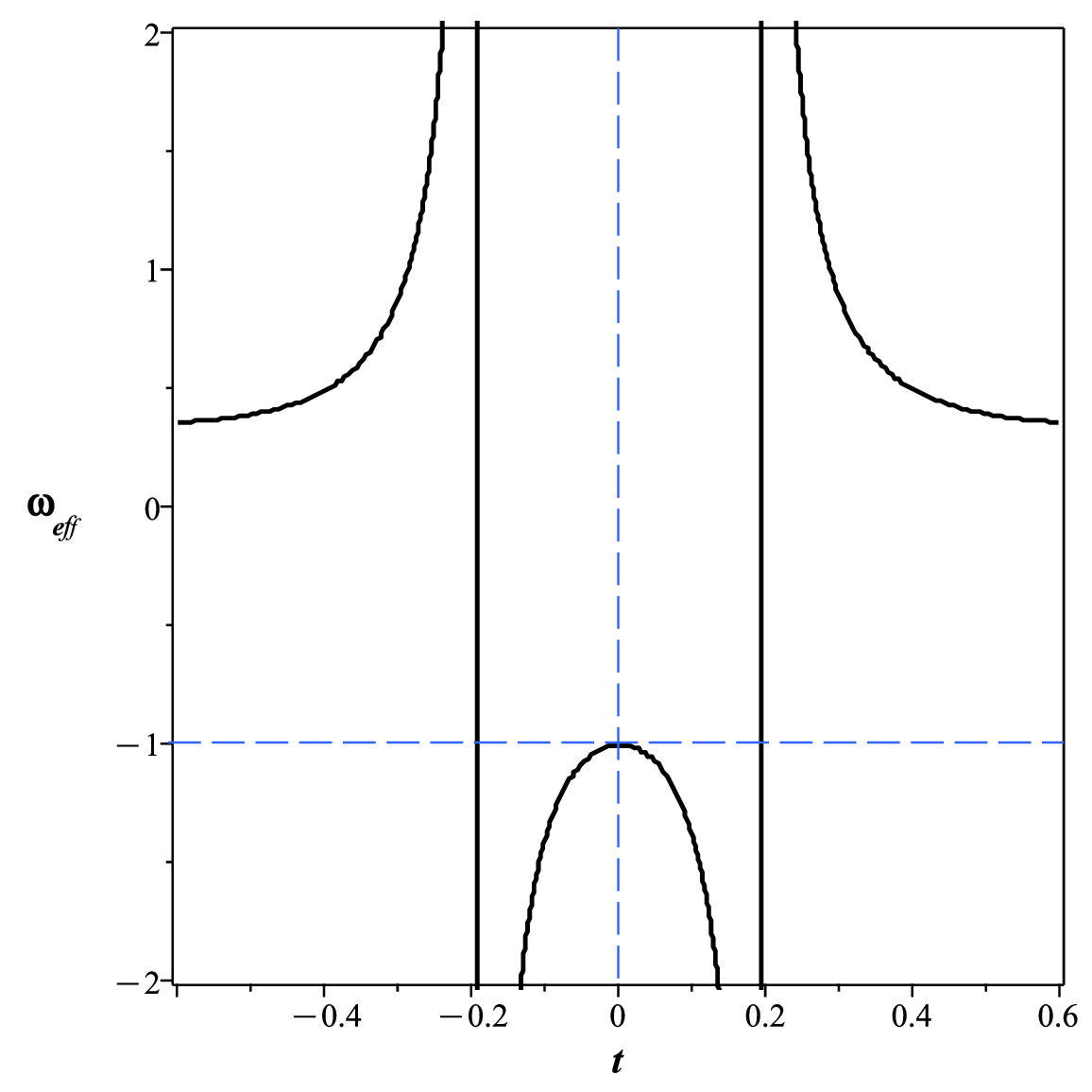} &
			\includegraphics[width=0.45\textwidth]{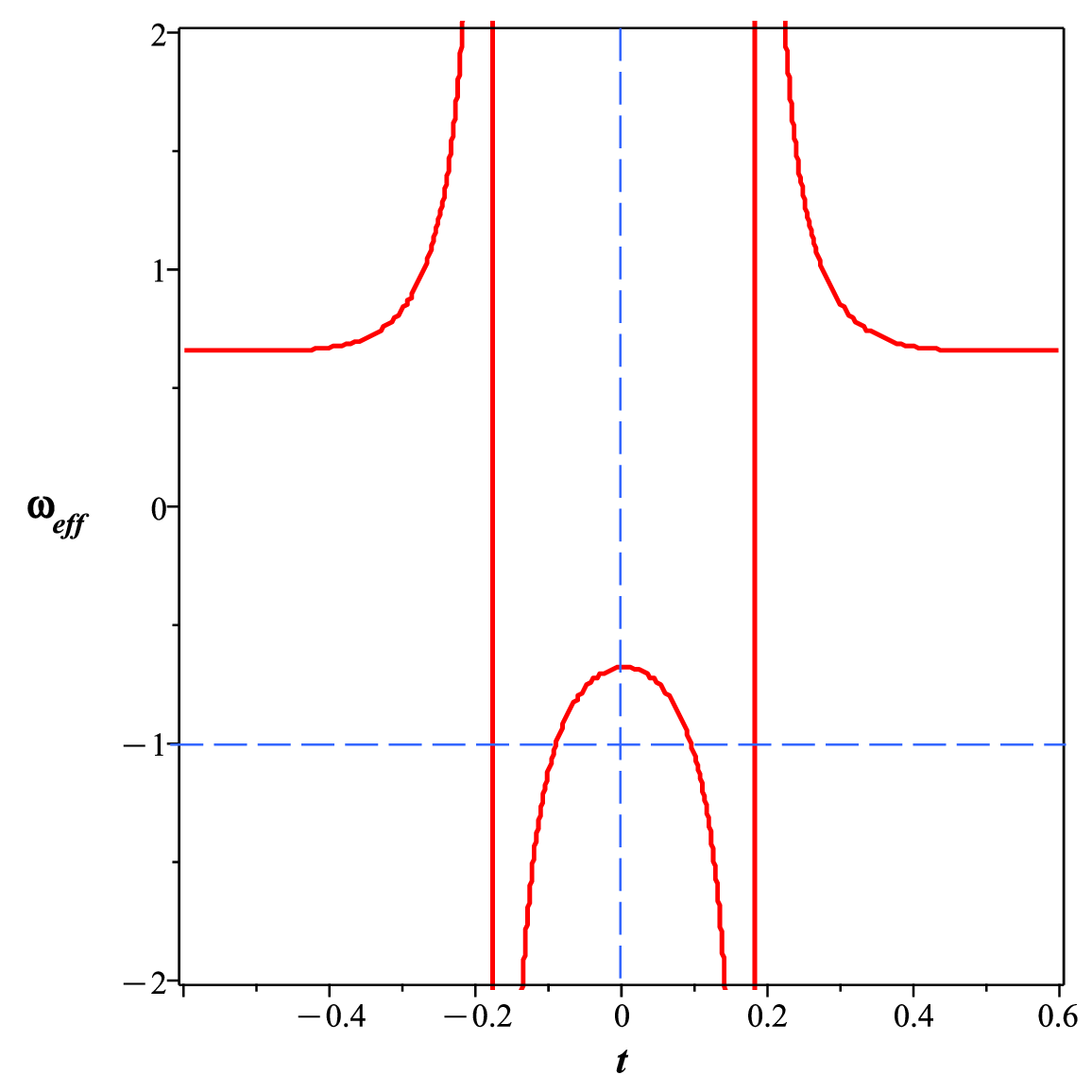} \\
			(a) $\omega_{\text{eff}}$ for $\omega = -1$ (black) & (b) $\omega_{\text{eff}}$ for $\omega = -1/3$ (red)
		\end{tabular}
		\caption{Effective equation of state $\omega_{\text{eff}}(t)$ showing (a) evolution for $\omega = -1$ (black line) and (b) evolution for $\omega = -1/3$ (red line). The phantom divide line (PDL) at $\omega_{\text{eff}} = -1$ is shown as a dashed line. Both cases show PDL crossing behavior near the bounce point.}
		\label{fig:exp_eos}
	\end{figure}
	
	\begin{figure}[h]
		\centering
		\begin{tabular}{cc}
			\includegraphics[width=0.45\textwidth]{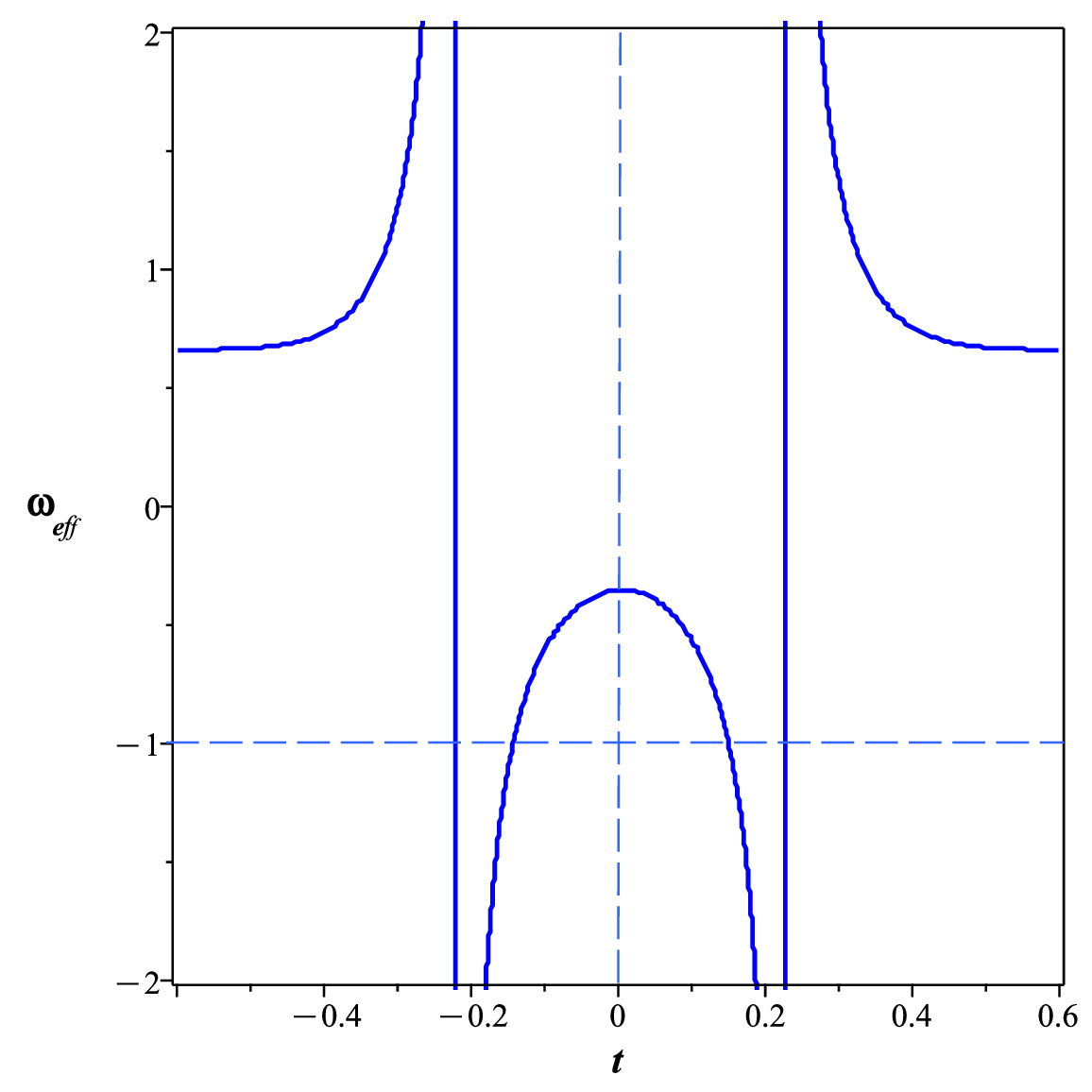} &
			\includegraphics[width=0.45\textwidth]{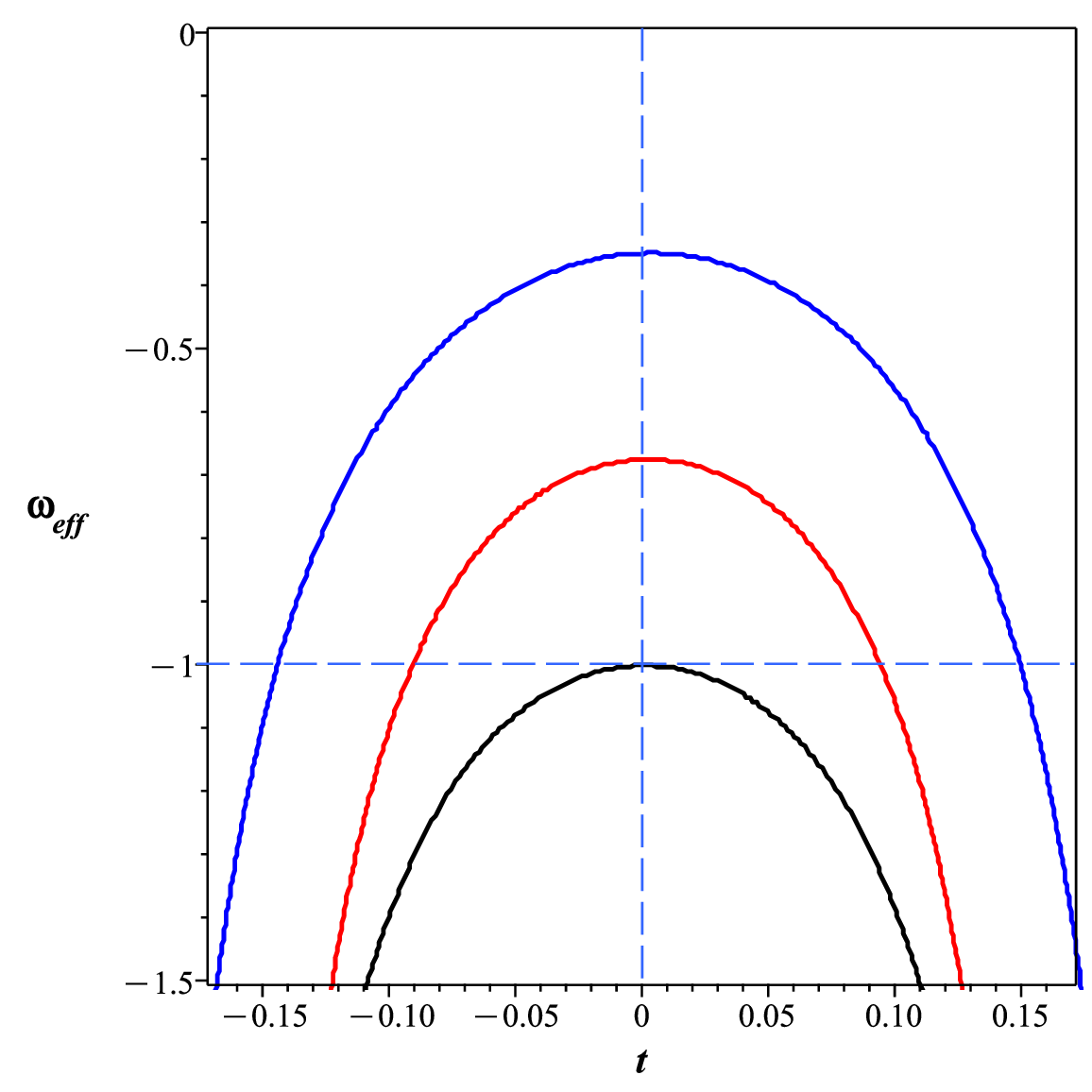} \\
			(a) $\omega_{\text{eff}}$ for $\omega = 1/3$ (blue) & (b) Comparative evolution
		\end{tabular}
		\caption{(a) Effective equation of state $\omega_{\text{eff}}(t)$ for $\omega = 1/3$ (blue line) and (b) composite view comparing all cases: $\omega = -1$ (black), $\omega = -1/3$ (red), and $\omega = 1/3$ (blue). All cases cross the phantom divide line ($\omega_{\text{eff}}=-1$) near the bounce point.}
		\label{fig:exp_eos_comparing}
	\end{figure}
	
	The equation of state evolution (Fig. \ref{fig:exp_eos} and \ref{fig:exp_eos_comparing}) reveals:
	\begin{itemize}
		\item For $\omega=-1$: $\omega_{\text{eff}}$ touches $-1$ at $t=0$ and remains phantom ($\leq -1$)
		\item For $\omega=-1/3,1/3$: Double PDL crossing near $t=0$ (phantom $\to$ quintessence $\to$ phantom)
		\item All cases asymptote to quintessence ($>-1$) for $|t|>0.2$
		\item Finite-time singularities of Type II or IV appear at $t\approx\pm0.2$, indicating a breakdown of the effective field theory at these energy scales, a common feature in many bouncing models \cite{bamba2008future}. Our analysis focuses on the robust bouncing behavior near $t=0$, which is free of such pathologies.
	\end{itemize}
	
\begin{figure}[h]
		\centering
		\begin{tabular}{cc}
			\includegraphics[width=0.45\linewidth]{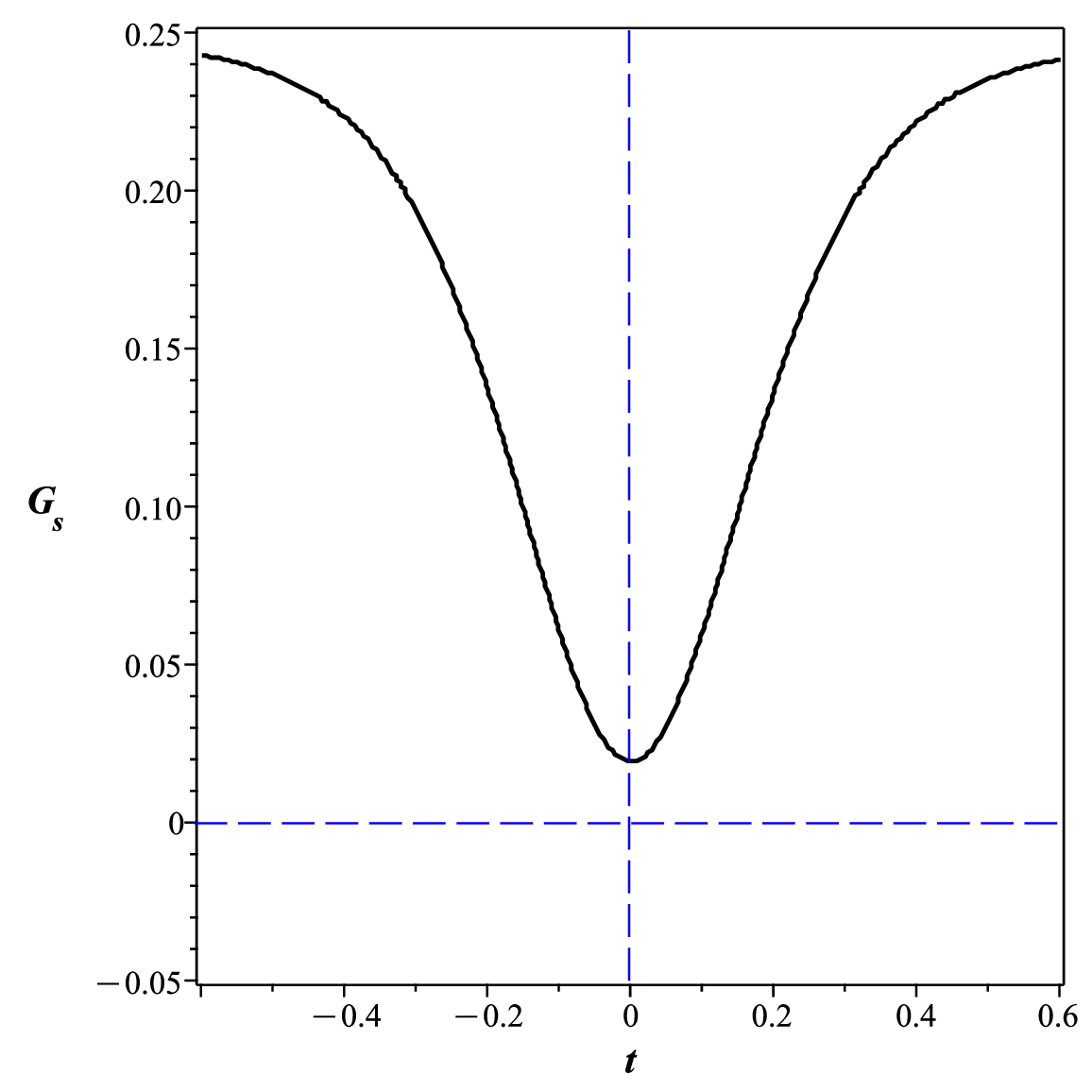} &
			\includegraphics[width=0.45\linewidth]{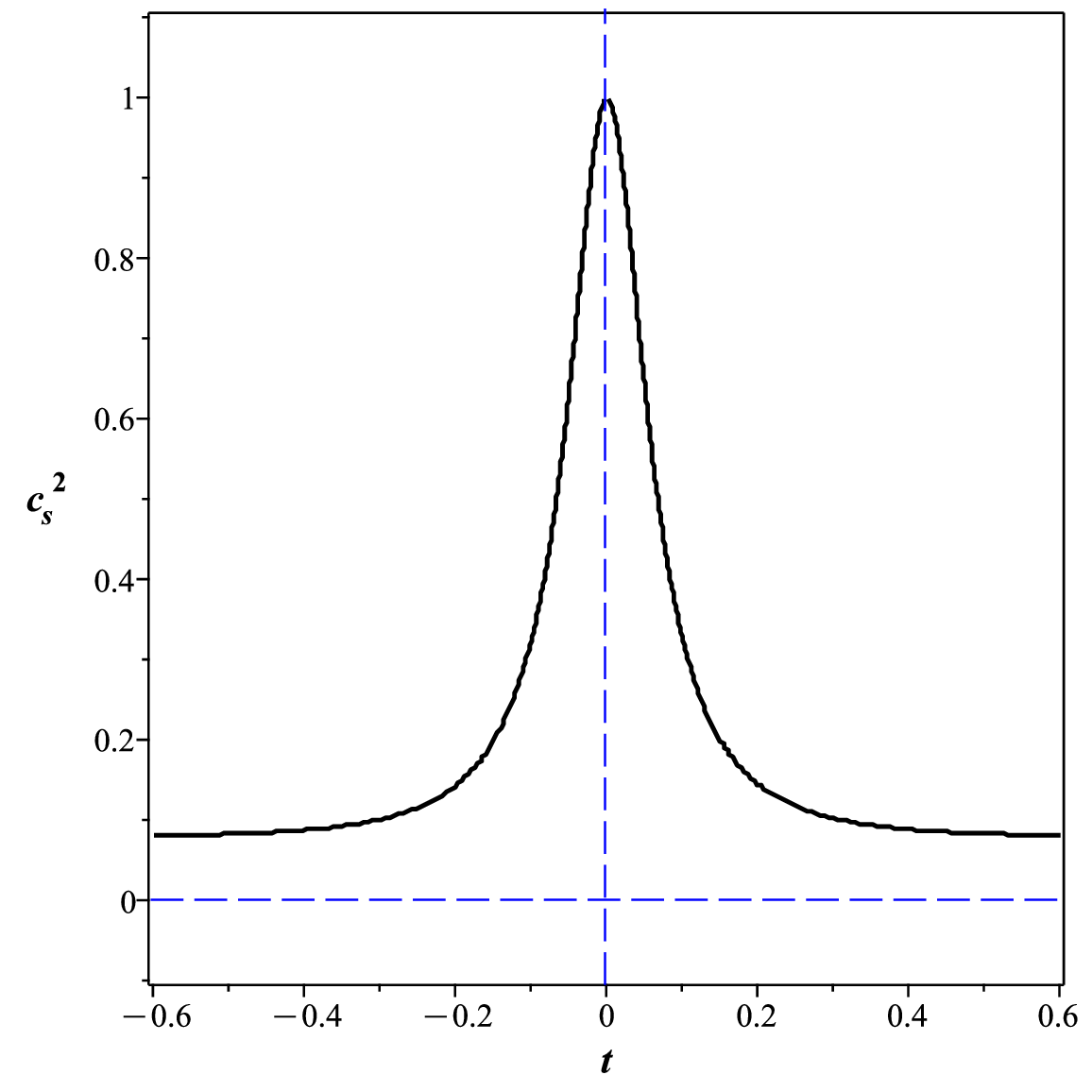} \\
			(a) Kinetic term $\mathcal{G}_S(t)$ & (b) Squared sound speed $c_s^2(t)$
		\end{tabular}
		\caption{Stability analysis for the exponential model ($\omega = -1$). (a) The kinetic term $\mathcal{G}_S$ remains positive definite, confirming no ghost instabilities. (b) The sound speed squared $c_s^2$ remains non-negative throughout the evolution, with $c_s^2 > 0$ except at discrete points where $c_s^2 = 0$ (momentary perturbation freezing). The model maintains gradient stability.}
		\label{fig:exp_stability}
	\end{figure}

	\paragraph*{\textbf{Stability Verification:}}
	The exponential model's stability is verified numerically. Figure \ref{fig:exp_stability} shows the stability indicators for the $\omega = -1$ case.
	
	The exponential model satisfies $\mathcal{G}_S(t) > 0$ for all $t$, avoiding ghost pathologies. The sound speed remains real with $c_s^2(t) \geq 0$, ensuring gradient stability. The finite-time singularities at $t \approx \pm 0.2$ do not destabilize the core bounce region around $t=0$, where both stability conditions are maintained.
	
	The model parameters ($V_0=2.5$, $b=1$, $n=2$, $\alpha=1$, $g=0.1$, $\xi_2=\xi_3=1$) were chosen to satisfy energy conservation while generating viable bounce dynamics. The potential $V(\phi, \psi) = V_0 e^{-\alpha\phi} + \frac{1}{2}m_p^2\psi^2 + g\phi\psi$ enables these phase transitions.
	
	\subsubsection{Power-Law Modified Gravity}
	
	The power-law model represents a phenomenologically rich extension of GR that incorporates non-linear curvature-matter couplings:
	\begin{equation}
		f(R,G,T) = R + \xi_1 R^n + \xi_2 G + \xi_3 T^m
		\label{eq:powerlaw_model}
	\end{equation}
	where the exponents $n,m \geq 2$ introduce non-perturbative corrections to Einstein gravity.
	
	The partial derivatives are:
	\begin{align}
		f_R &= 1 + n\xi_1 R^{n-1}, \quad f_G = \xi_2, \quad f_T = m\xi_3 T^{m-1} \\
		\dot{f}_R &= n(n-1)\xi_1 R^{n-2}\dot{R}, \quad \ddot{f}_R = n(n-1)\xi_1 R^{n-3}[(n-2)\dot{R}^2 + R\ddot{R}] \\
		\dot{f}_G &= 0, \quad \ddot{f}_G = 0
	\end{align}
	
	Substituting into the general Friedmann equations and simplifying, we obtain:
	\begin{align}
		3H^2 &= \frac{1}{1 + n\xi_1 R^{n-1}}\bigg[\underbrace{\kappa^2(\rho + \rho_\Xi)}_{\text{Standard matter terms}} 
		+ \underbrace{3n\xi_1 R^{n-1}\dot{H}}_{\text{Curvature-driven inertia}} 
		+ \underbrace{3n(n-1)\xi_1 R^{n-2}\dot{R}H}_{\text{Dynamic curvature evolution}} \nonumber \\
		&\quad - \underbrace{\frac{1}{2}(R + \xi_1 R^n + \xi_3 T^m)}_{\text{Effective dark energy}} 
		+ \underbrace{m\xi_3 T^{m-1}(\rho + p)}_{\text{Non-linear matter coupling}} \bigg]
		\label{f1_Rec_PL}
	\end{align}
	
	\begin{align}
		-2\dot{H} - 3H^2 &= \frac{1}{1 + n\xi_1 R^{n-1}}\bigg[\underbrace{\kappa^2(p + p_\Xi)}_{\text{Standard pressure terms}} 
		+ \underbrace{n\xi_1 R^{n-1}\dot{H}}_{\text{Curvature pressure}} 
		+ \underbrace{2n(n-1)\xi_1 R^{n-2}\dot{R}H}_{\text{Dynamic curvature coupling}} \nonumber \\
		&\quad + \underbrace{n(n-1)\xi_1 R^{n-2}[(n-2)\dot{R}^2 + R\ddot{R}]}_{\text{Curvature acceleration}} 
		+ \underbrace{\frac{1}{2}(R + \xi_1 R^n  + \xi_3 T^m)}_{\text{Geometric stiff matter}} \bigg]
		\label{f2_Rec_PL}
	\end{align}
	
	The effective energy density and pressure become:
	\begin{align}
		\rho_{\text{eff}} &= \frac{1}{1 + n\xi_1 R^{n-1}}\left[\rho + \rho_\Xi + \frac{3n\xi_1 R^{n-1}\dot{H}}{\kappa^2} + \frac{3n(n-1)\xi_1 R^{n-2}\dot{R}H}{\kappa^2} \right. \nonumber \\
		&\quad \left. - \frac{R + \xi_1 R^n  + \xi_3 T^m}{2\kappa^2} + \frac{m\xi_3 T^{m-1}(\rho + p)}{\kappa^2} \right]
	\end{align}
	
	\begin{align}
		p_{\text{eff}} &= \frac{1}{1 + n\xi_1 R^{n-1}}\left[p + p_\Xi + \frac{n\xi_1 R^{n-1}\dot{H}}{\kappa^2} + \frac{2n(n-1)\xi_1 R^{n-2}\dot{R}H}{\kappa^2} \right. \nonumber \\
		&\quad \left. + \frac{n(n-1)\xi_1 R^{n-2}[(n-2)\dot{R}^2 + R\ddot{R}]}{\kappa^2} + \frac{R + \xi_1 R^n  + \xi_3 T^m}{2\kappa^2} \right]
	\end{align}
	
	For energy conditions, the Null Energy Condition (NEC) becomes:
	\begin{align}
		\rho_{\text{eff}} + p_{\text{eff}} &= \frac{1}{1 + n\xi_1 R^{n-1}}
		\left\{
		\left[1 + \frac{m\xi_3 T^{m-1}}{\kappa^2}\right](\rho + p)
		+ (\rho_\Xi + p_\Xi)
		\right. \nonumber\\
		&\quad + \frac{n\xi_1 R^{n-1}}{\kappa^2}
		\left[
		4\dot{H}
		+ 5(n-1)\frac{\dot{R}}{R}H
		+ (n-1)\left[(n-2)\frac{\dot{R}^2}{R} + \ddot{R}\right]
		\right]
		\bigg\}
		\label{NEC_PL}
	\end{align}
	
	At the bounce point ($H=0$), the NEC violation condition simplifies to:
	\begin{align}
		&\left(1 + \frac{m\xi_3 T^{m-1}}{\kappa^2}\right)(\rho + p)\nonumber \\ 
		&\quad+ \frac{6n\xi_1}{\kappa^2} (6\dot{H})^{n-2} \big[ 4\dot{H}^2 
		+ 6(n-1)(n-2)\ddot{H}^2 + 12(n-1)\dot{H}^3 
		+ 6(n-1)\dot{H}\dddot{H} \big] 
		< \dot{\phi}^2 - \dot{\psi}^2\label{NEC_bounce_PL}
	\end{align}
	
	This model exhibits several distinctive features:
	\begin{itemize}
		\item The power-law terms $R^n$ and $T^m$ introduce strong non-linearities that dominate at high curvatures and energy densities
		\item The $R^n$ term provides curvature-driven anti-friction that can counteract gravitational collapse
		\item The $T^m$ term creates novel matter-geometry interactions that scale non-linearly with energy density
		\item As with the exponential curvature model discussed previously, the Gauss-Bonnet terms exhibit exact cancellation in the final simplified field equations.
		\item The NEC condition shows explicit dependence on both curvature and matter power-law terms
		\item At the bounce point, multiple curvature and matter terms contribute to the NEC violation condition
	\end{itemize}
	
	The General Relativity limit is recovered when $\xi_1 = \xi_2 = \xi_3 = 0$. For strong gravity effects, the parameters $\xi_1, \xi_2, \xi_3$ can be chosen to introduce Planck-scale corrections that become significant near the bounce point.
	
	\begin{figure}[h]
		\centering
		\begin{tabular}{cc}
			\includegraphics[width=0.45\linewidth]{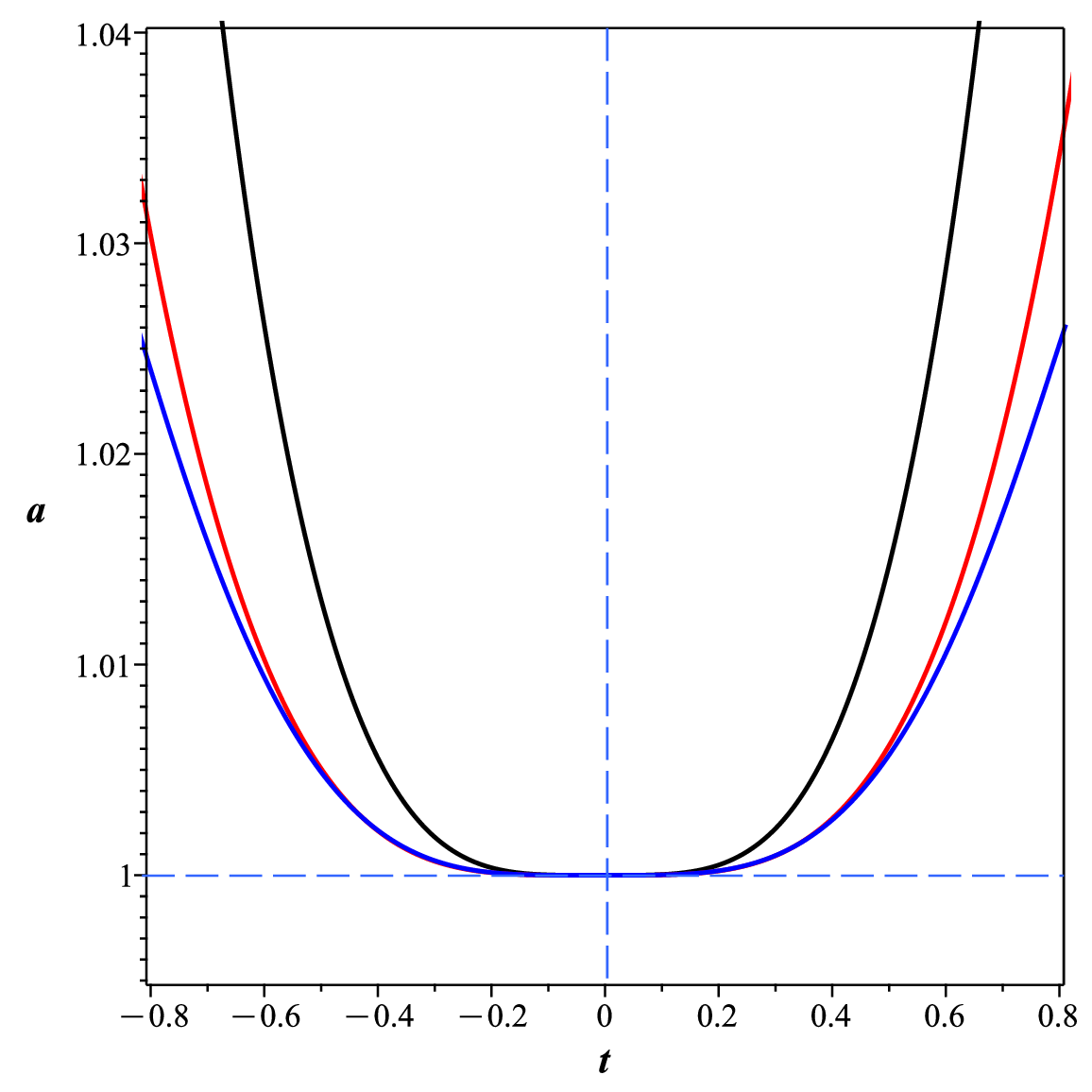} &
			\includegraphics[width=0.45\linewidth]{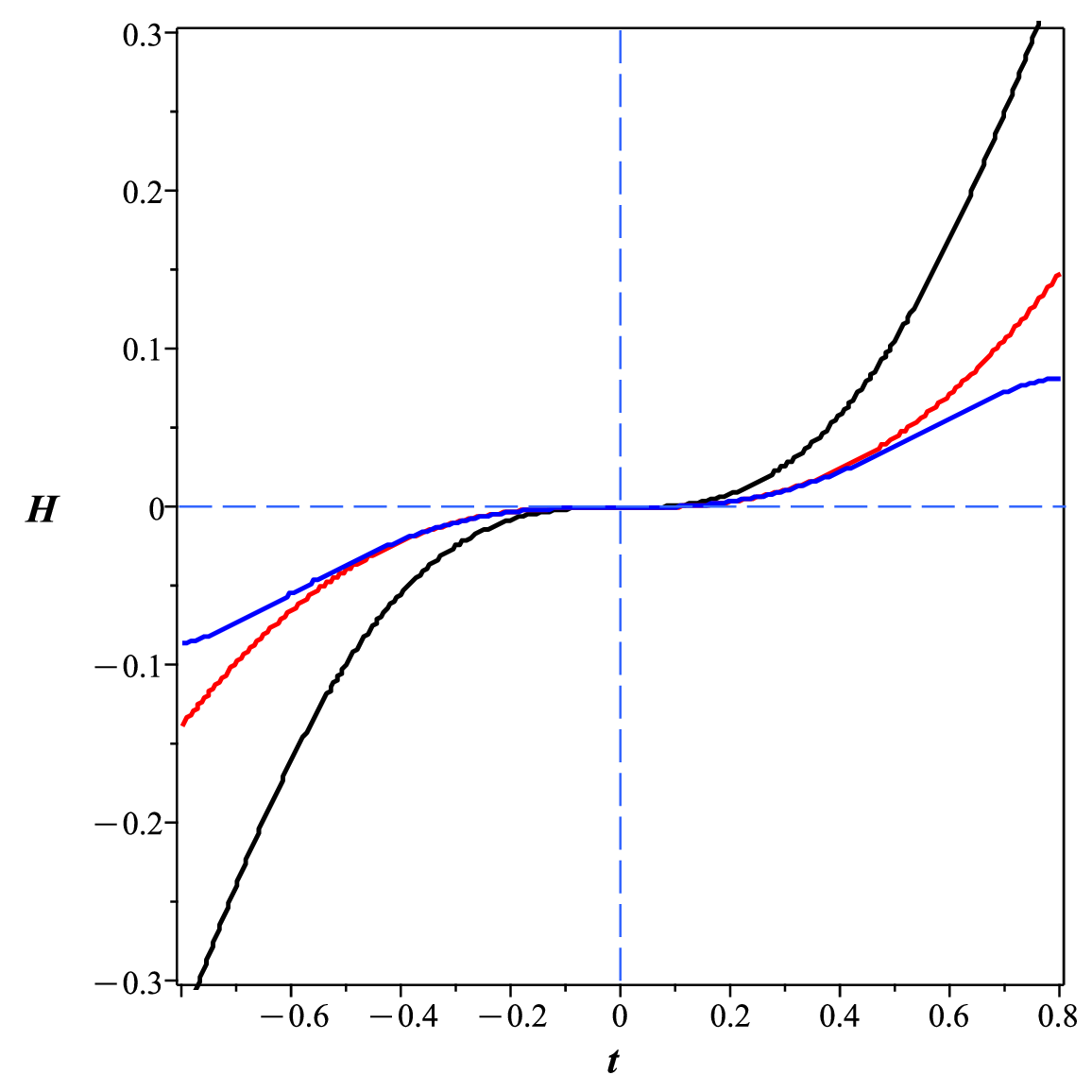} \\
			(a) Scale factor $a(t)$ & (b) Hubble parameter $H(t)$
		\end{tabular}
		\caption{Evolution of (a) scale factor $a(t)$ and (b) Hubble parameter $H(t)$ for different equations of state ($\omega=-1$: black, $-1/3$: red, $1/3$: blue). All cases show successful bounce behavior with $a(t)$ reaching minimum and $H(t)$ crossing zero at $t=0$.}
		\label{fig:pl_scale_hubble}
	\end{figure}
	
	The bounce dynamics in Figure \ref{fig:pl_scale_hubble} emerge from an intricate balance between:
	\begin{itemize}
		\item \textbf{Curvature-driven anti-friction}: The $R^n$ terms dominate near the bounce point ($t=0$), creating an effective negative pressure that counteracts gravitational collapse. For $n=2$, this behaves as:
		\begin{equation}
			P_{\text{eff}} \approx -\frac{\xi_1}{2}R^2 \sim -\xi_1(36\dot{H}^2 + 144H^2\dot{H} + 144H^4)
		\end{equation}
		
		\item \textbf{Non-linear matter effects}: The $T^m$ coupling introduces an effective matter-curvature interaction energy density:
		\begin{equation}
			\rho_{\text{int}} \sim m\xi_3 T^{m-1}(\rho + p)
		\end{equation}
		which becomes significant at high densities.
		
		\item \textbf{Quintom field dynamics}: The potential $V(\phi,\psi) = \frac{1}{2}m_p(\phi^2+\psi^2) - \frac{1}{3}\phi^2\psi$ provides a phase transition mechanism between phantom and quintessence regimes.
	\end{itemize}
	
	\begin{figure}[h]
		\centering
		\begin{tabular}{cc}
			\includegraphics[width=0.45\linewidth]{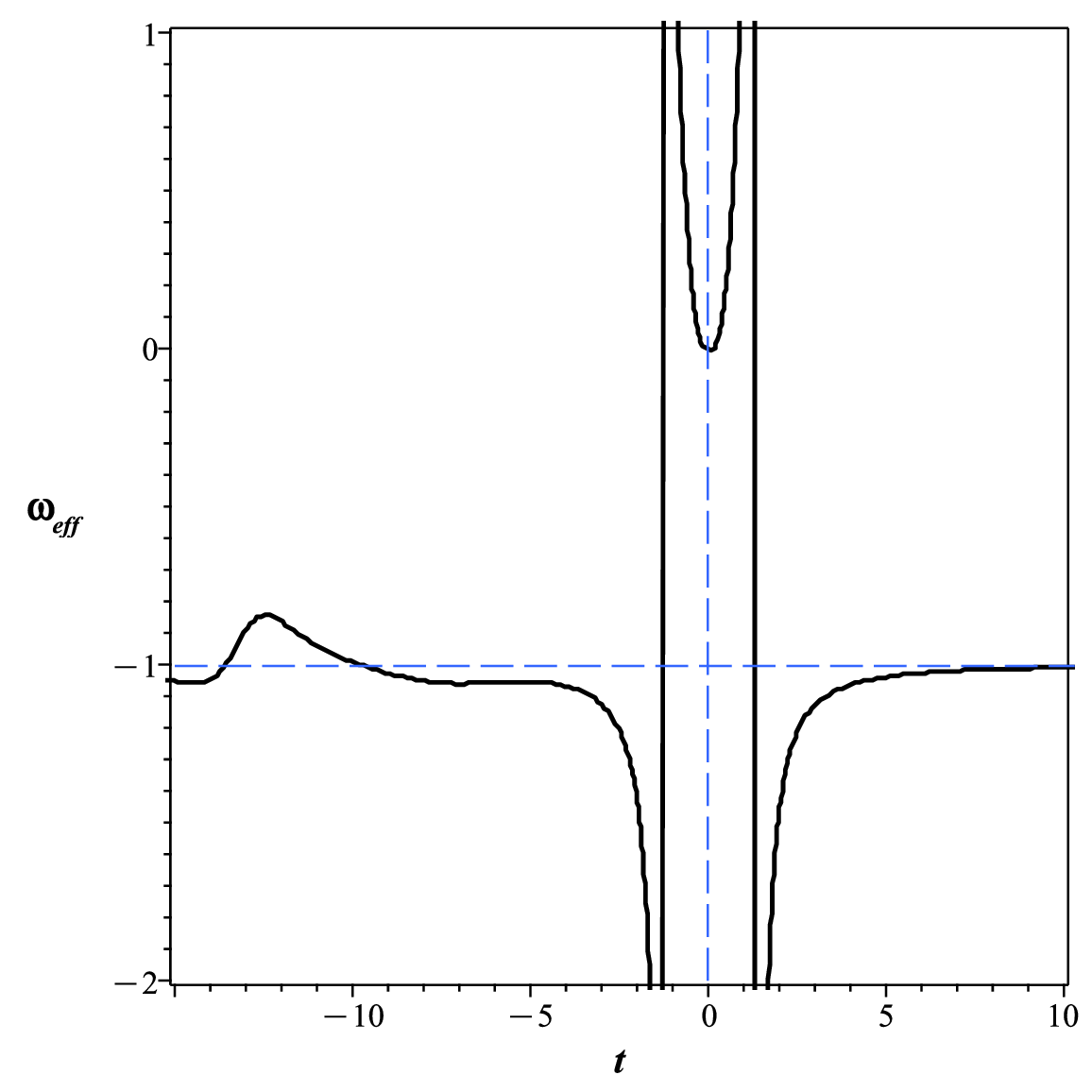} &
			\includegraphics[width=0.45\linewidth]{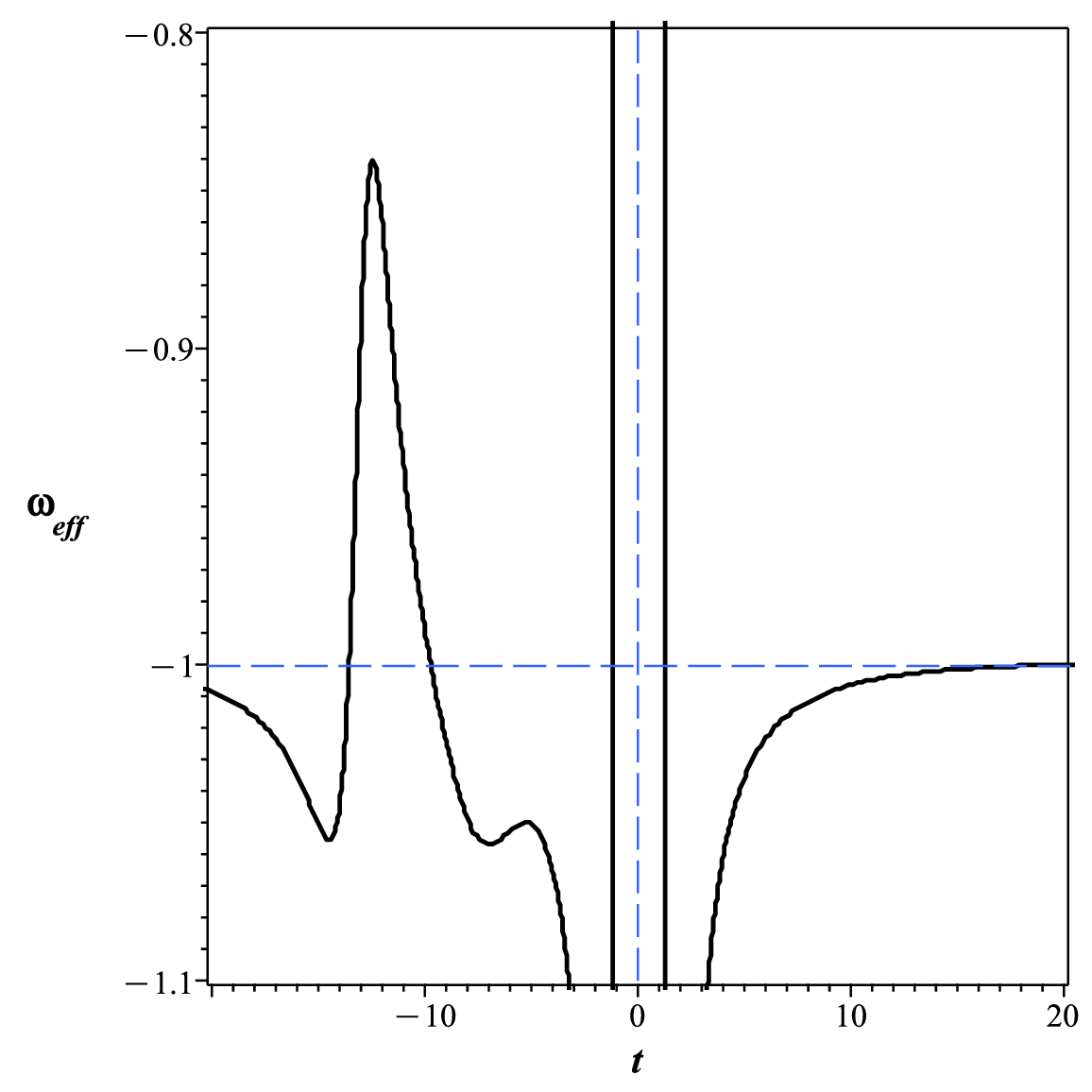} \\
			(a) Full view & (b) Zoomed view
		\end{tabular}
		\caption{Effective equation of state $\omega_{\text{eff}}(t)$ for $\omega = -1$ (black line) showing (a) full evolution and (b) detailed view near bounce point. The PDL ($\omega_{\text{eff}} = -1$) is shown as dashed line.}
		\label{fig:pl_eos_1}
	\end{figure}
	
	\begin{figure}[h]
		\centering
		\begin{tabular}{cc}
			\includegraphics[width=0.45\linewidth]{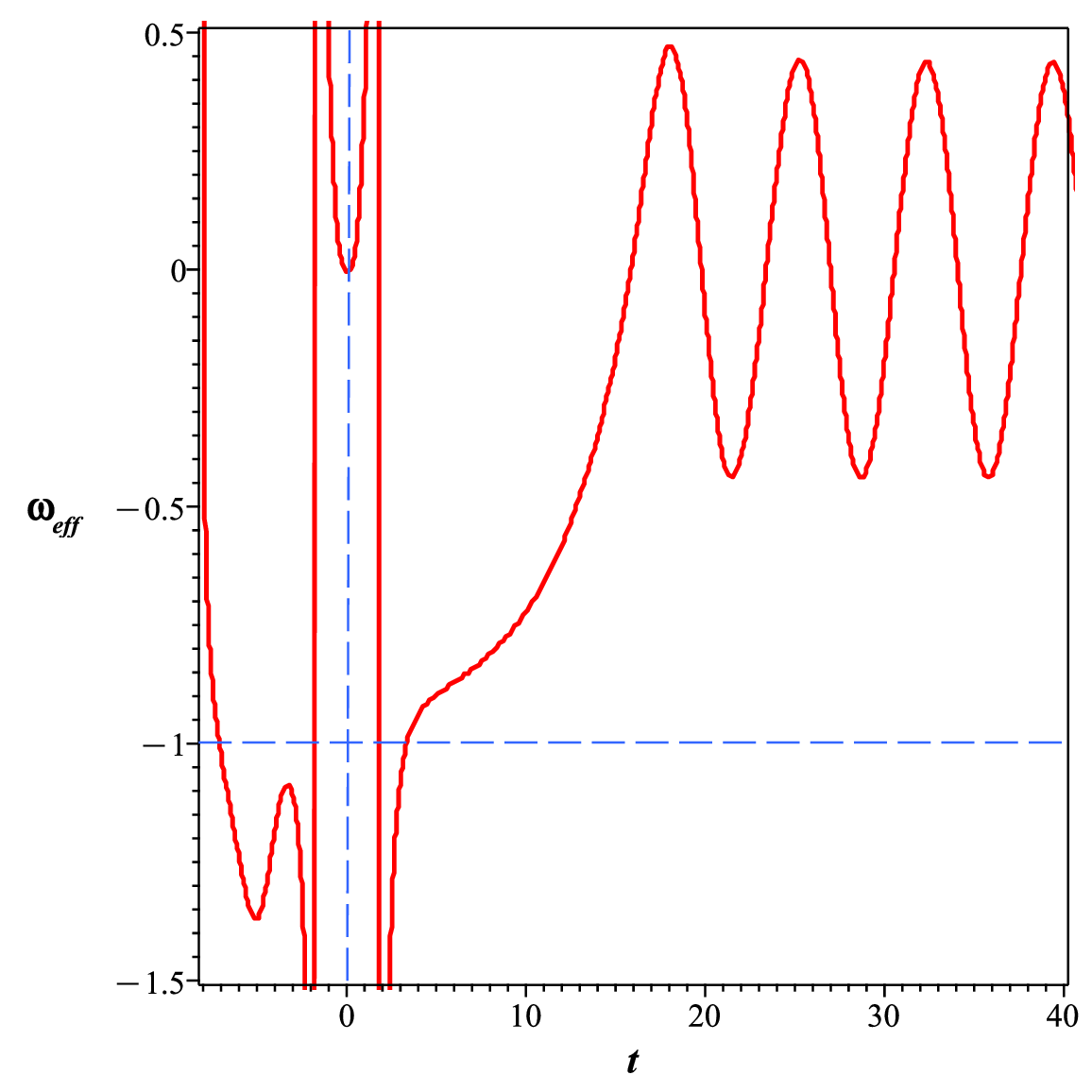} &
			\includegraphics[width=0.45\linewidth]{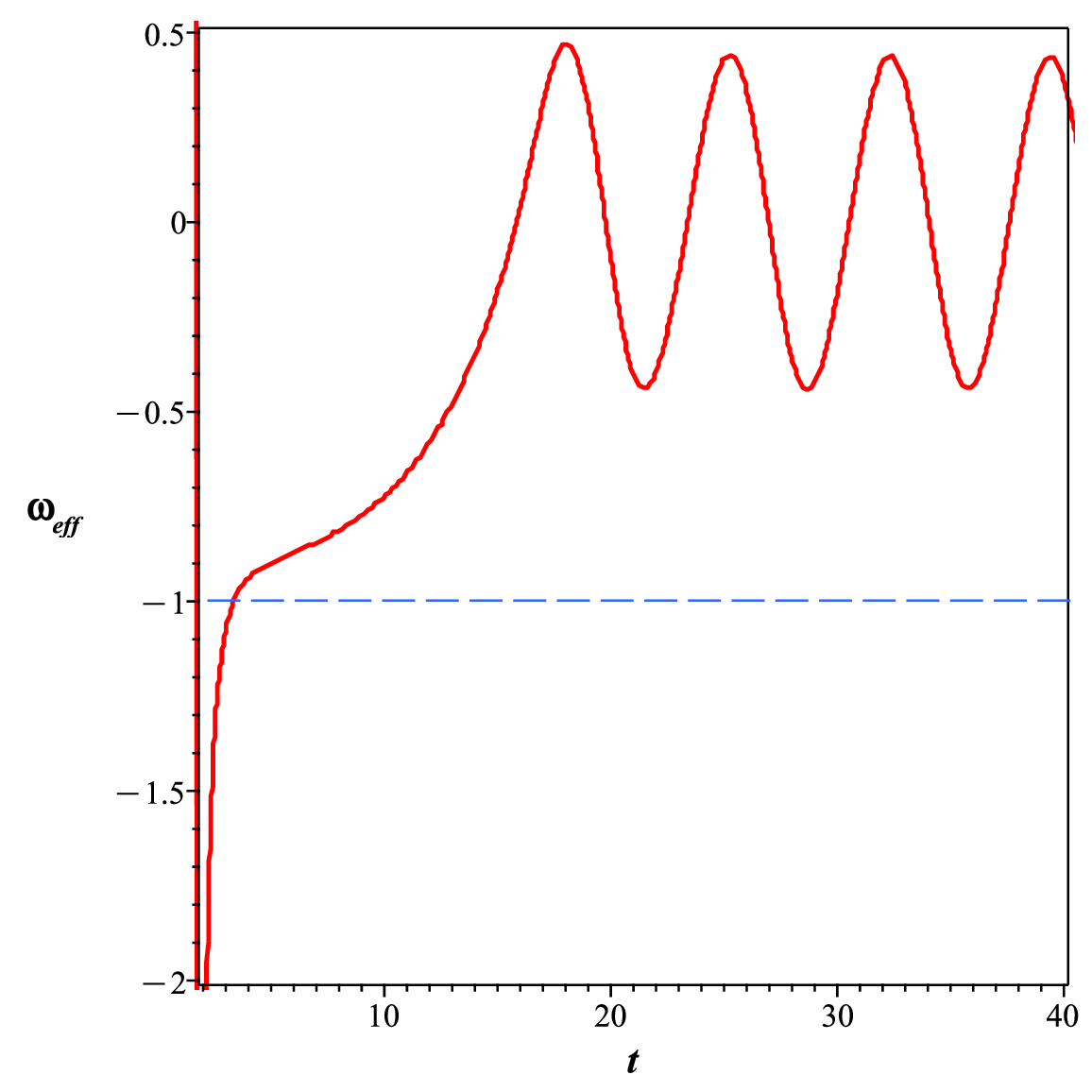} \\
			(a) Full view & (b) Zoomed view
		\end{tabular}
		\caption{Effective equation of state $\omega_{\text{eff}}(t)$ for $\omega = -1/3$ (red line) showing (a) full evolution and (b) detailed view near bounce point. The PDL ($\omega_{\text{eff}} = -1$) is shown as dashed line.}
		\label{fig:pl_eos_2}
	\end{figure}
	
\begin{figure}[h]
		\centering
		\begin{tabular}{cc}
			\includegraphics[width=0.45\linewidth]{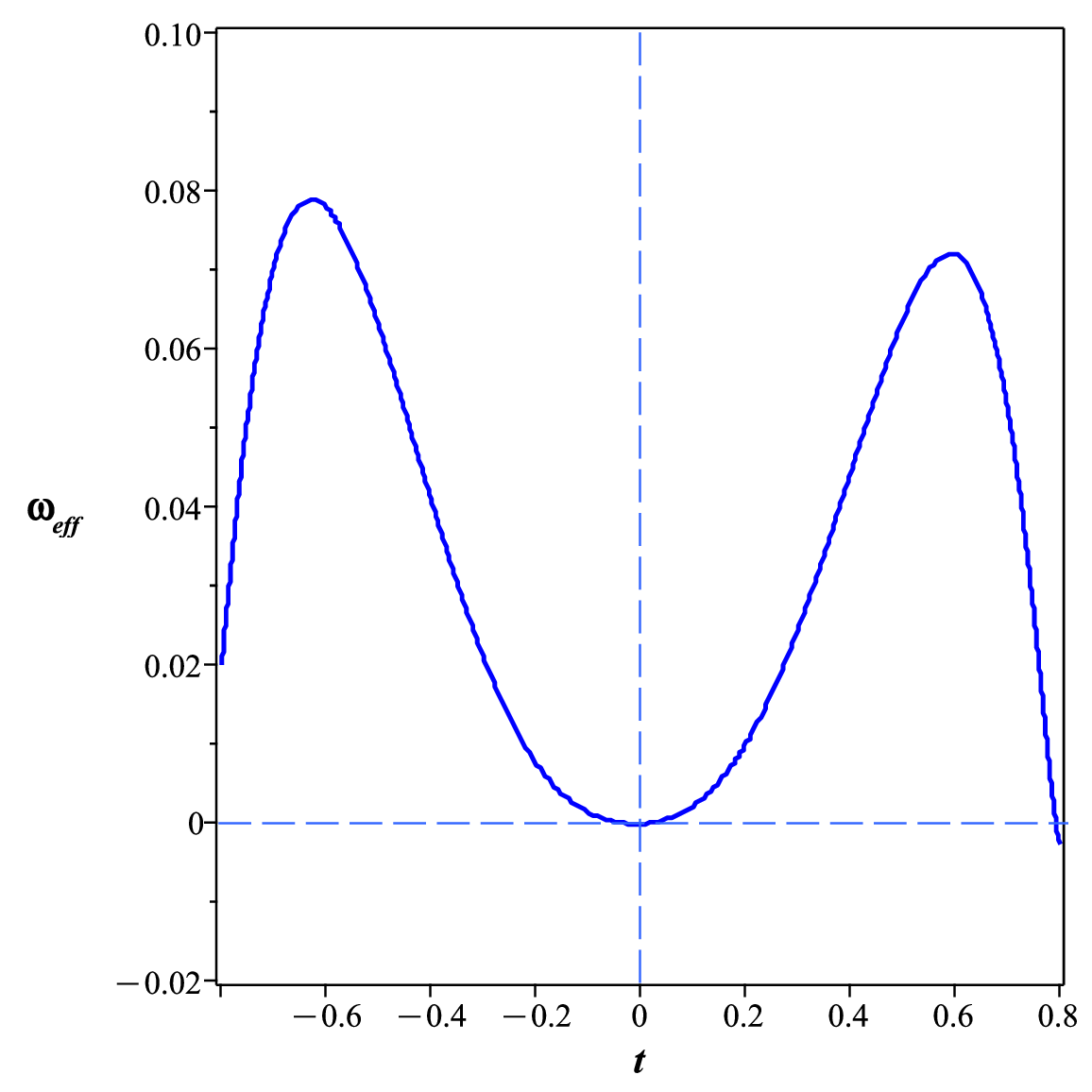} &
			\includegraphics[width=0.45\linewidth]{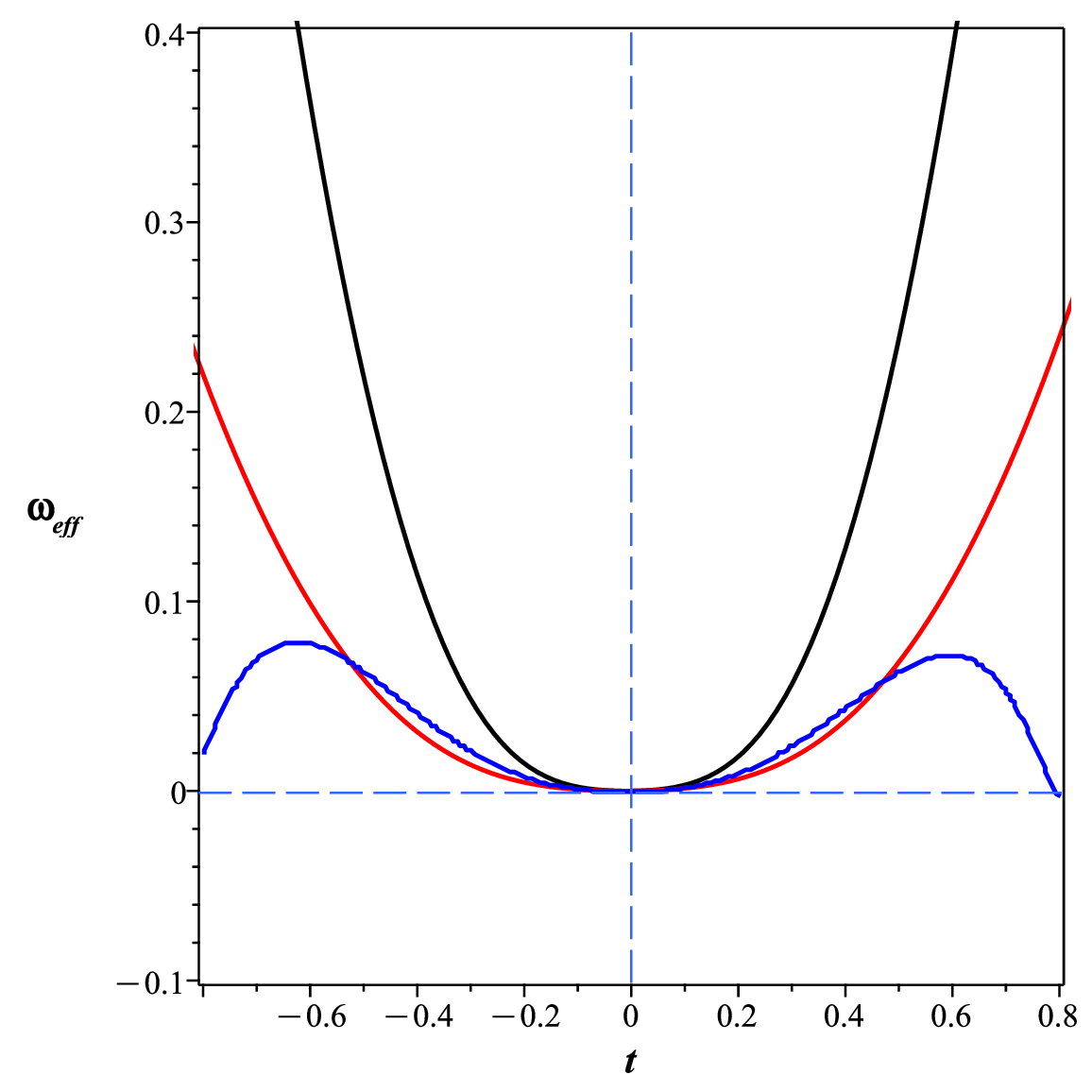} \\
			(a) $\omega_{\text{eff}}$ for $\omega = 1/3$ (blue) & (b) Comparative evolution
		\end{tabular}
		\caption{(a) Effective equation of state $\omega_{\text{eff}}(t)$ for $\omega = 1/3$ (blue line) and (b) composite view comparing all cases: $\omega = -1$ (black), $\omega = -1/3$ (red), and $\omega = 1/3$ (blue) around bounce point. The PDL ($\omega_{\text{eff}} = -1$) is shown as dashed line.}
		\label{fig:pl_eos_composite}
	\end{figure}

	The EoS evolution in Figures \ref{fig:pl_eos_1}-\ref{fig:pl_eos_composite} reveals profound thermodynamics:
	\begin{itemize}
		\item \textbf{Phantom regime ($\omega_{\text{eff}}<-1$)}: Corresponds to violation of the Null Energy Condition (NEC):
		\begin{equation}
			\rho + p + \frac{3n\xi_1 R^{n-1}\dot{H} - m\xi_3 T^{m-1}(\rho + p)}{1 + n\xi_1 R^{n-1}} < 0
		\end{equation}
		enabled by the phantom field's negative kinetic energy.
		
		\item \textbf{Bounce thermodynamics}: At $t=0$, the system reaches maximal entropy density $s_{\text{max}}$, with EoS $\omega_{\text{eff}}=0$ indicating pressureless matter dominance during the bounce transition.
		
		\item \textbf{Singularity formation}: The divergences at $t\sim\pm1$ represent:
		\begin{equation}
			\lim_{t\to\pm1} \left(\frac{d\rho}{dt}, \frac{dp}{dt}, R_{\mu\nu\rho\sigma}R^{\mu\nu\rho\sigma}\right) \to \infty
		\end{equation}
		These are Type II (sudden) singularities where pressure diverges while density remains finite.
	\end{itemize}
	
	The power-law model generates successful bounce behavior for all three equations of state. The $\omega = -1$ case is presented here as our primary example because it exhibits the most complex and illustrative evolution of the effective equation of state $\omega_{\text{eff}}$, showcasing the full dynamical range of the model from phantom regimes to quintessence and matter-like states. Fig. \ref{fig:pl_scale_hubble} shows the characteristic bouncing behavior for this case. Fig. \ref{fig:pl_eos_1} reveals complex $\omega_{\text{eff}}$ dynamics: for $t < -1$, oscillatory behavior between phantom ($<-1$) and quintessence ($>-1$) regimes indicates competing dynamical components; for $-1 < t < 1$, positive $\omega_{\text{eff}}$ with minimum at $t=0$ ($\omega_{\text{eff}}=0$) suggests matter/radiation-like phase and cosmic bounce; for $t > 1$, $\omega_{\text{eff}}$ approaches $-1$ asymptotically, indicating late-time $\Lambda$CDM-like attractor behavior.
	
	Finite-time singularities occur at $t \sim \pm 1$: the $t \sim -1$ singularity appears during phantom-dominated phase, potentially linked to Big Rip avoidance; the $t \sim 1$ singularity marks matter-like to late-time phase transition.
	
	Fig. \ref{fig:pl_eos_2} shows similar patterns for $\omega = -1/3$: quintessence-to-phantom transition before $t \sim -2$; matter-like phase with $\omega_{\text{eff}}(0)=0$ bounce at $t=0$; post-bounce oscillatory behavior in quintessence regime suggesting cyclic cosmology.
	
	Fig. \ref{fig:pl_eos_composite}a displays matter/radiation-like behavior ($\omega_{\text{eff}}>0$) for $\omega=1/3$ within $-0.8<t<0.8$, with symmetric peaks around $\omega_{\text{eff}}(0)=0$ bounce point. No solutions outside this range indicate phase transition or model breakdown.
	
\begin{figure}[h]
		\centering
		\begin{tabular}{cc}
			\includegraphics[width=0.45\linewidth]{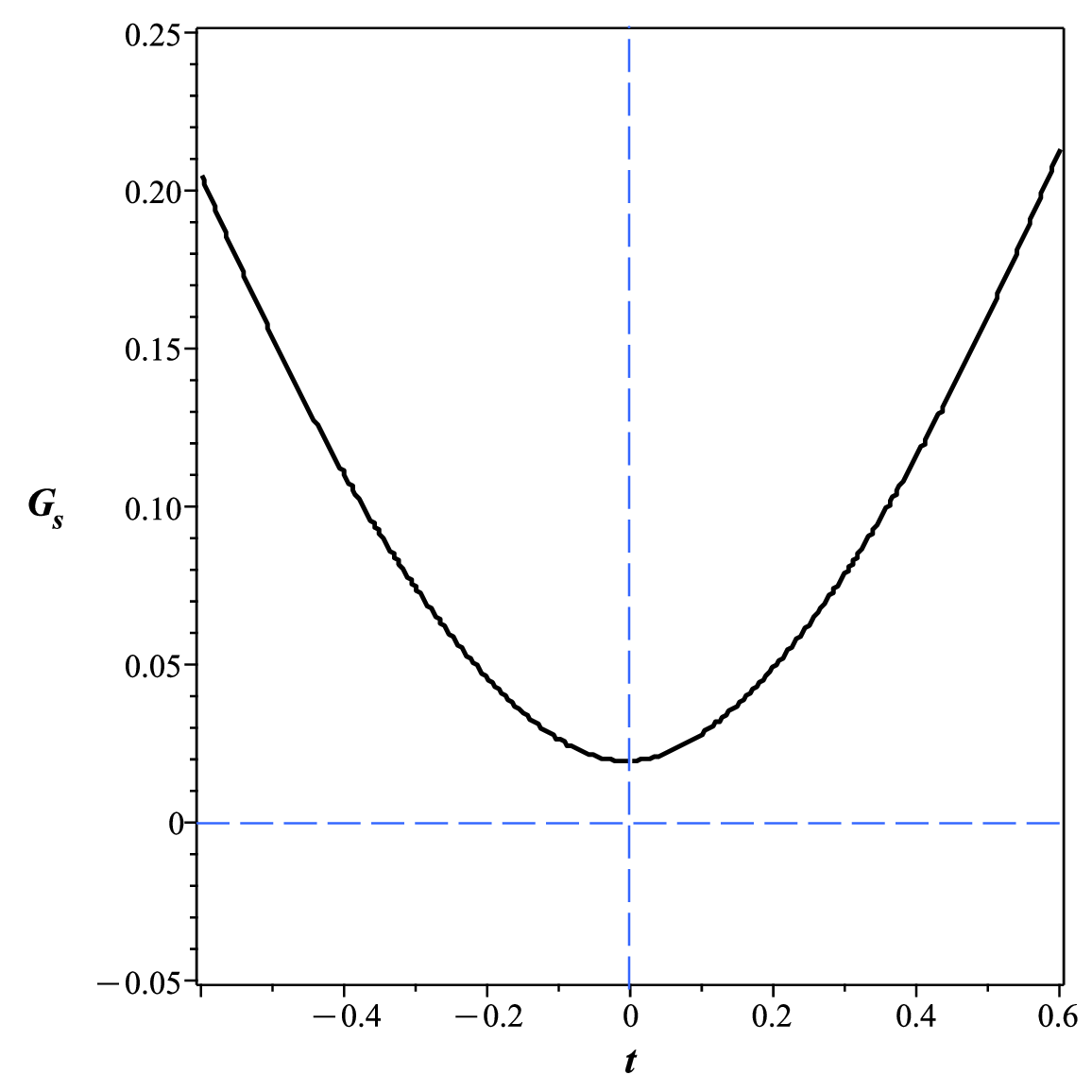} &
			\includegraphics[width=0.45\linewidth]{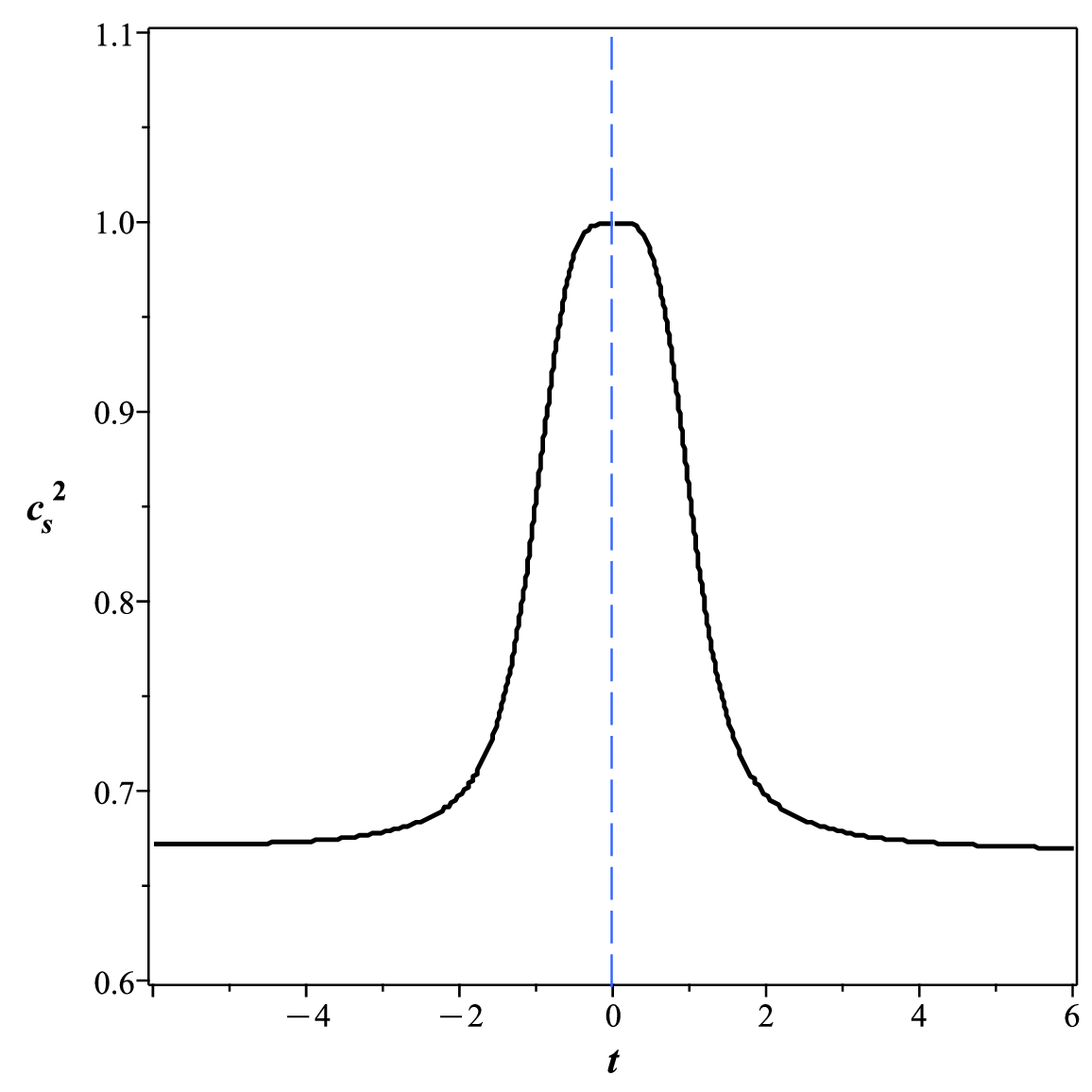} \\
			(a) Kinetic term $\mathcal{G}_S(t)$ & (b) Squared sound speed $c_s^2(t)$
		\end{tabular}
		\caption{Stability analysis for the power-law model ($\omega = -1$). (a) The kinetic term $\mathcal{G}_S$ remains positive throughout the evolution, confirming the absence of ghosts. (b) The sound speed squared $c_s^2$ maintains $c_s^2 \geq 0$, ensuring gradient stability. The oscillations in $c_s^2$ reflect the complex dynamics of the higher-order curvature terms but remain within stable bounds.}
		\label{fig:pl_stability}
	\end{figure}

	\paragraph*{\textbf{Stability Verification:}}
	For the power-law model, we verify the stability conditions numerically. Figure \ref{fig:pl_stability} shows the results for the $\omega = -1$ case.
	
	The power-law model maintains $\mathcal{G}_S(t) > 0$ throughout the evolution, confirming ghost-free behavior. The sound speed squared $c_s^2(t)$ remains non-negative, with $c_s^2 \geq 0$ ensuring gradient stability. The model demonstrates stable bouncing behavior despite the complex oscillatory dynamics and finite-time singularities at $t \sim \pm 1$.
	
	The model unifies: (1) phantom-driven contraction avoiding Big Crunch, (2) curvature-mediated bounce with $a(t) \approx a_{\text{min}} + \frac{\xi_1 n(6H_0^2)^{n-1}}{2}t^2$, and (3) late-time $\Lambda$CDM-like expansion. This resolves singularity avoidance ($R_{\text{max}}\sim\xi_1^{-1/(n-1)}$), horizon problem, and initial conditions problem, while managing finite-time singularities at $t\sim\pm1$.
	
	\subsubsection{Modified Teleparallel Gravity Cosmology}
	
	Building on \cite{cai2016f}, we analyze a teleparallel-inspired modification where torsion-matter couplings drive cosmic evolution. This model extends the teleparallel equivalent of general relativity (TEGR) by introducing non-linear torsion-matter interactions:
	\begin{equation}
		f(R,G,T) = R + \xi_2 G + \xi_3 T^2
	\end{equation}
	
	As shown in Appendix \ref{sec:Derivation of Modified Friedmann Equations}, the modified Friedmann equations become:
	\begin{align}
		3H^2 &= \underbrace{\kappa^2 (\rho + \rho_\Xi)}_{\text{Standard terms}} - \underbrace{12\xi_2 H^2(H^2 + \dot{H})}_{\text{Gauss-Bonnet}} - \underbrace{\frac{\xi_3}{2}(5\rho^2 - 14\rho p - 3p^2)}_{\text{Torsion-matter coupling}} \label{f1_Rec_MTG} \\
		-2\dot{H} - 3H^2 &= \underbrace{\kappa^2 (p + p_\Xi)}_{\text{Standard terms}} + \underbrace{12\xi_2 H^2(H^2 + \dot{H})}_{\text{Gauss-Bonnet}} - \underbrace{\frac{\xi_3}{2}(9p^2 - 6\rho p + \rho^2)}_{\text{Torsion pressure}} \label{f2_Rec_MTG}
	\end{align}
	
	\begin{figure}[h]
		\centering
		\begin{tabular}{cc}
			\includegraphics[width=0.45\linewidth]{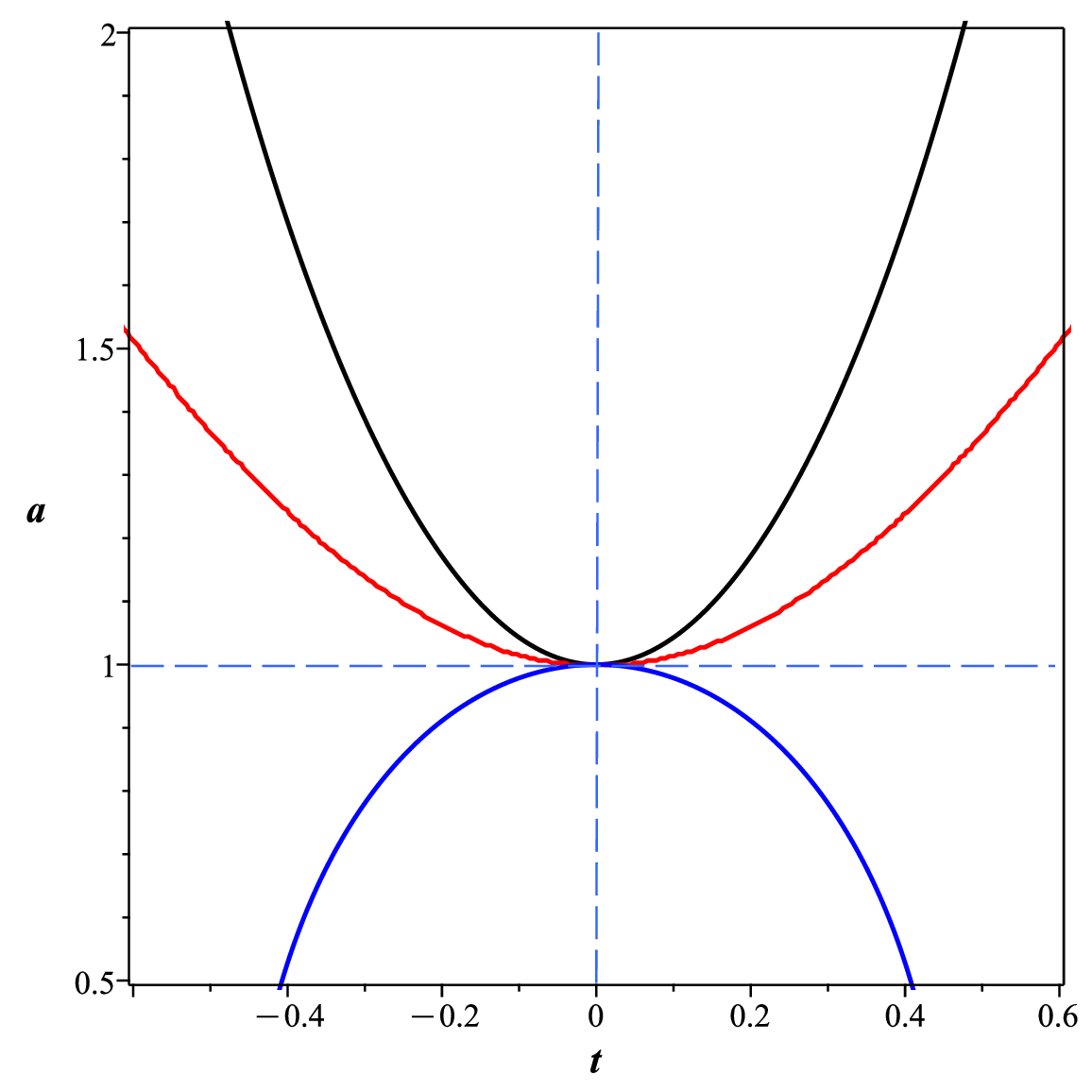} &
			\includegraphics[width=0.45\linewidth]{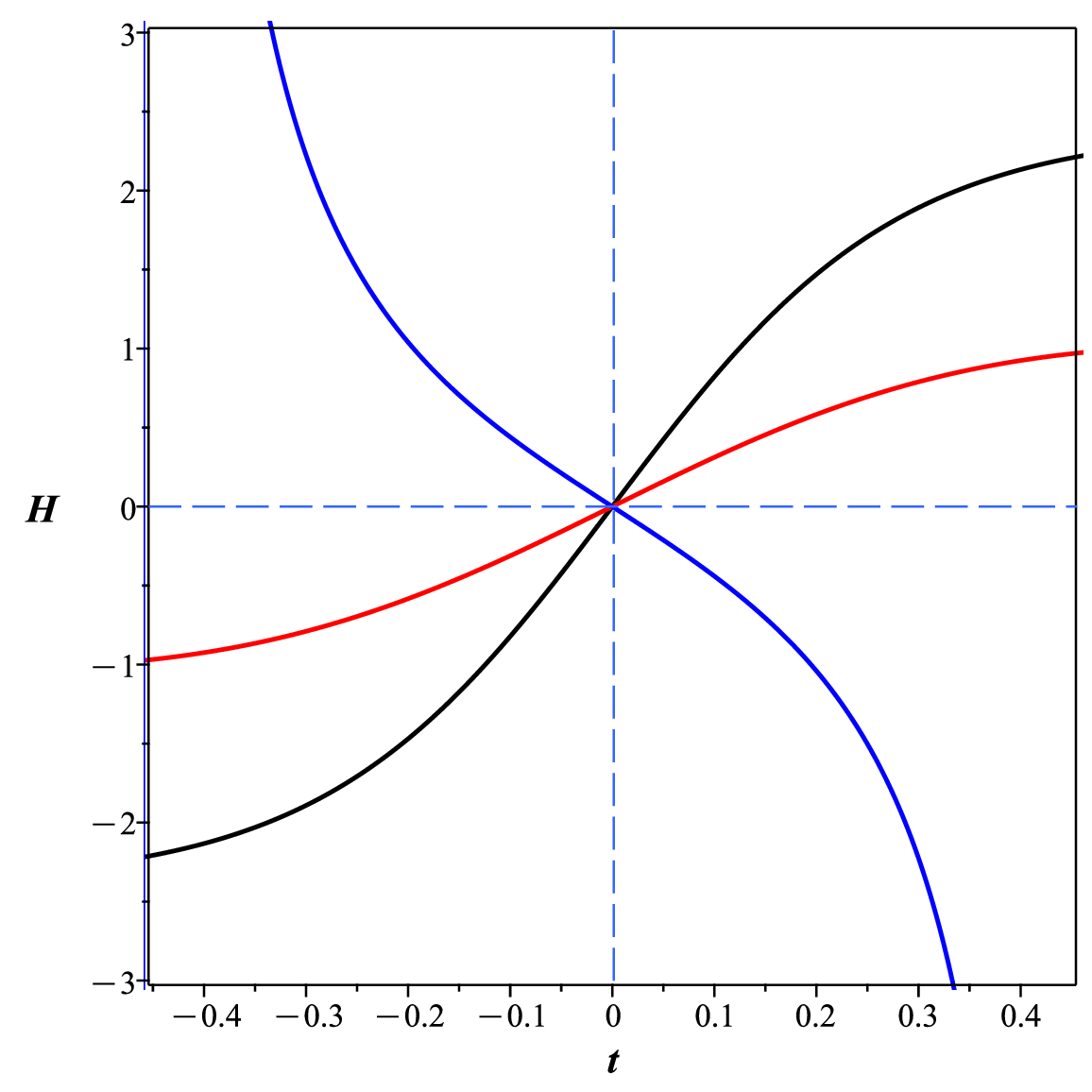} \\
			(a) Scale factor $a(t)$ & (b) Hubble parameter $H(t)$
		\end{tabular}
		\caption{Evolution of (a) scale factor $a(t)$ and (b) Hubble parameter $H(t)$ for different equations of state ($\omega=-1$: black, $-1/3$: red, $1/3$: blue). All cases show successful non-singular bounce at $t=0$ with torsion-modified dynamics.}
		\label{fig:mtg_scale_hubble}
	\end{figure}
	
	Key features observed in Figure \ref{fig:mtg_scale_hubble} for the $\omega = -1$ case (chosen to illustrate the model's behavior in the vacuum-dominated regime):
	\begin{enumerate}
		\item \textbf{General Bounce Capability:} The model supports bouncing solutions for a range of $\omega$ values; the $\omega = -1$ case is shown as a representative example.
		\item \textbf{Bounce Dynamics:}
		\begin{itemize}
			\item Scale factor $a(t)$ reaches minimum at $t=0$ without singularity
			\item Hubble parameter $H(t)$ changes sign smoothly, indicating torsion-mediated bounce
			\item Similar to linear model behavior but with modified late-time evolution
		\end{itemize}
		
		\item \textbf{Torsion-Dominated Phase}:
		The quadratic torsion term $T^2$ creates effective energy components:
		\begin{equation}
			\rho_{\text{tors}} = -\frac{\xi_3}{2}(5\rho^2-14\rho p -3p^2), \quad p_{\text{tors}} = -\frac{\xi_3}{2}(9p^2-6\rho p + \rho^2)
		\end{equation}
		
		\item \textbf{Energy Condition Violation}:
		The torsion term modifies the Null Energy Condition:
		\begin{equation}
			\rho + p + \xi_3(2\rho^2 - 5\rho p + 3p^2) \geq 0
		\end{equation}
		allowing a temporary NEC violation during bounce.
	\end{enumerate}
	
	\begin{figure}[h]
		\centering
		\begin{tabular}{cc}
			\includegraphics[width=0.45\linewidth]{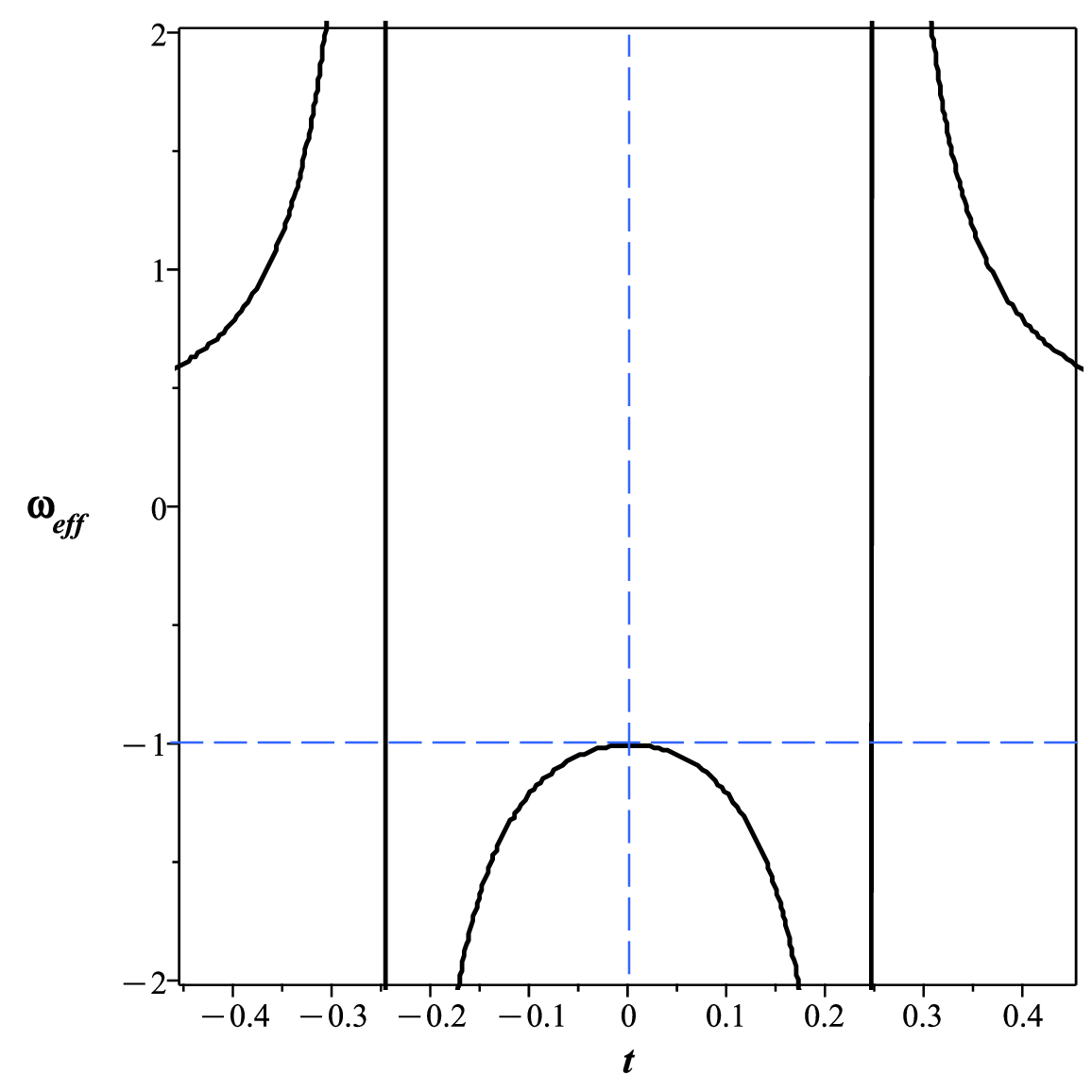} &
			\includegraphics[width=0.45\linewidth]{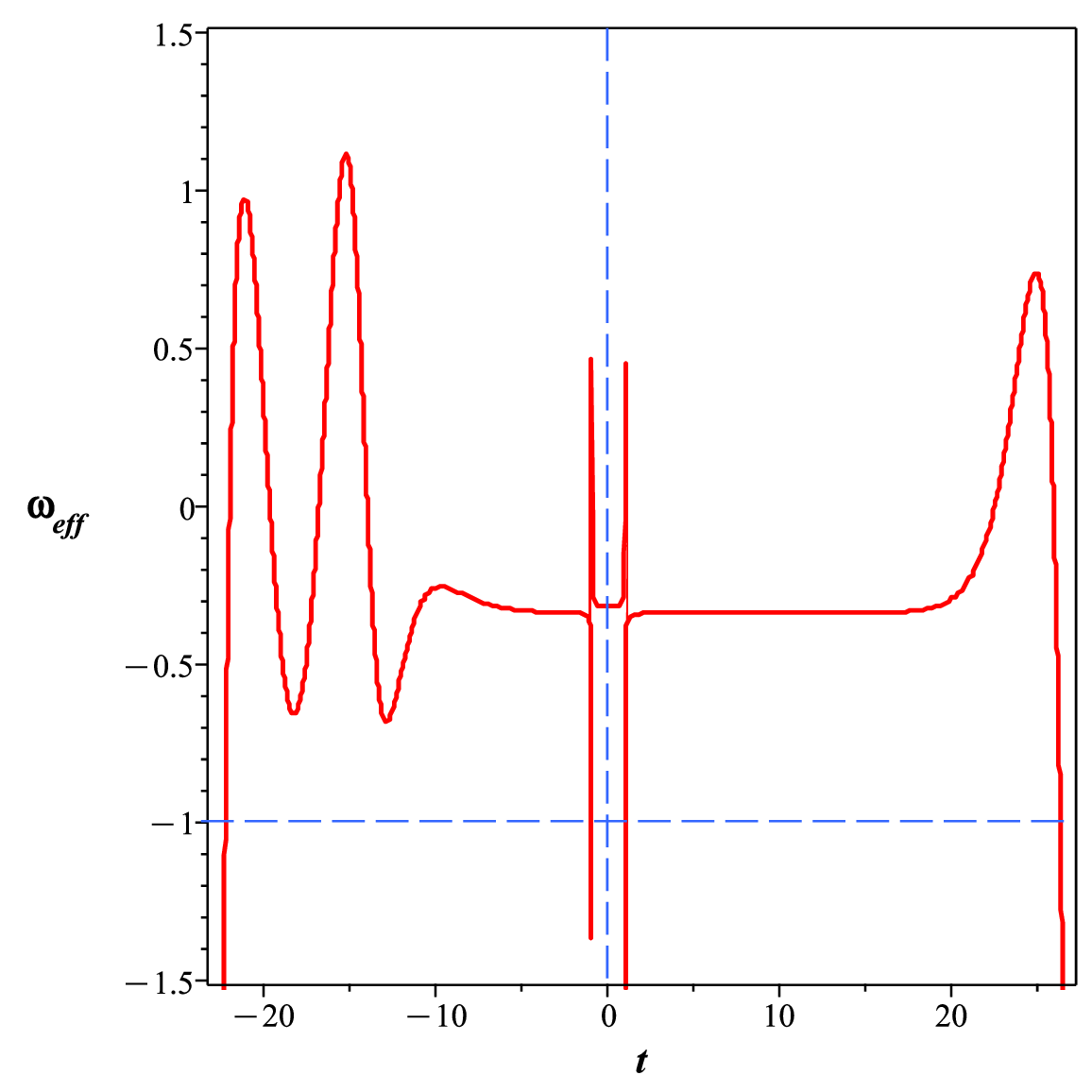} \\
			(a) $\omega_{\text{eff}}$ for $\omega = -1$ (black) & (b) $\omega_{\text{eff}}$ for $\omega = -1/3$ (red)
		\end{tabular}
		\caption{Effective equation of state $\omega_{\text{eff}}(t)$ showing (a) persistent phantom phase for $\omega=-1$ and (b) phase transitions for $\omega=-1/3$. The PDL ($\omega_{\text{eff}} = -1$) is shown as dashed line.}
		\label{fig:mtg_eos}
	\end{figure}
	
\begin{figure}[h]
		\centering
		\begin{tabular}{cc}
			\includegraphics[width=0.45\linewidth]{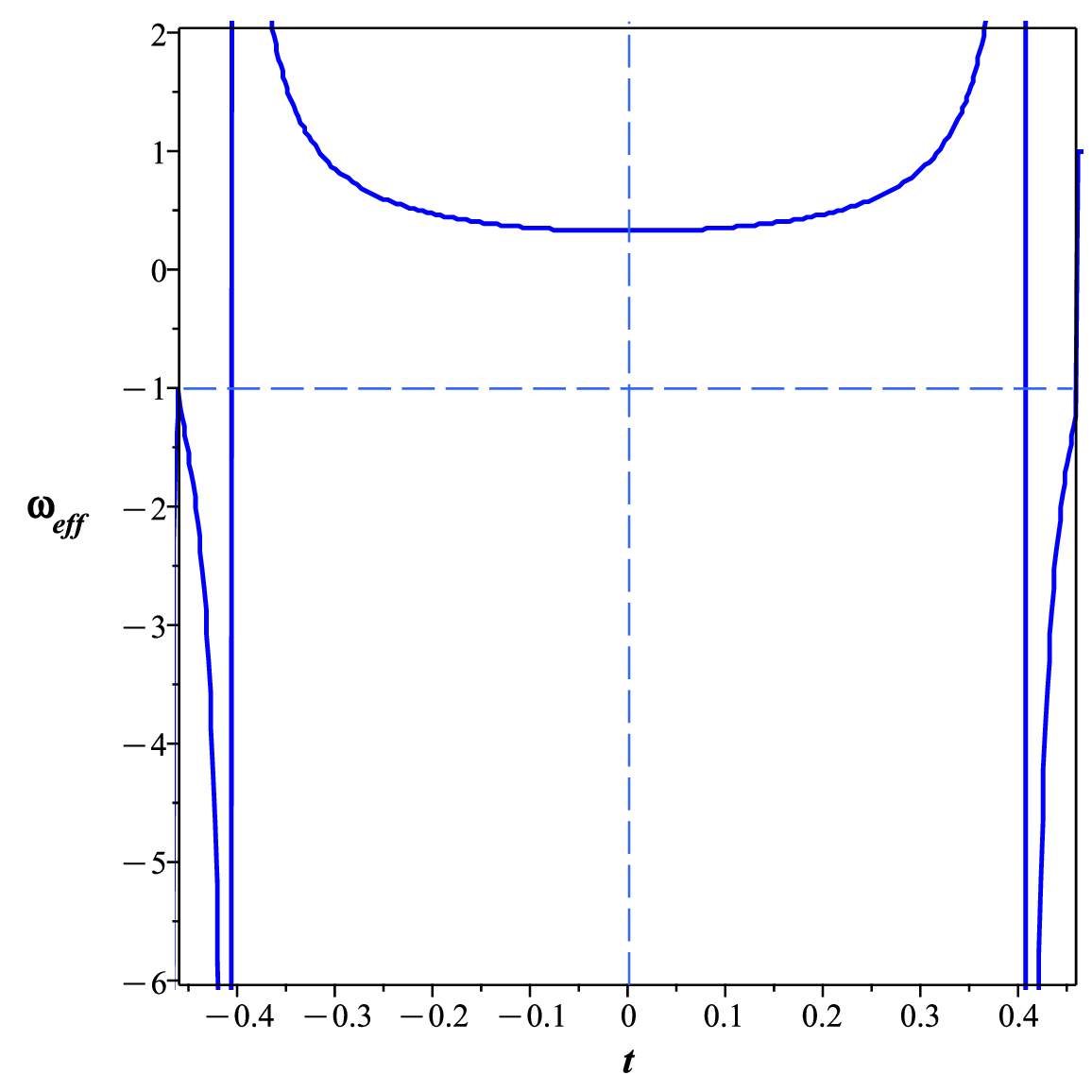} &
			\includegraphics[width=0.45\linewidth]{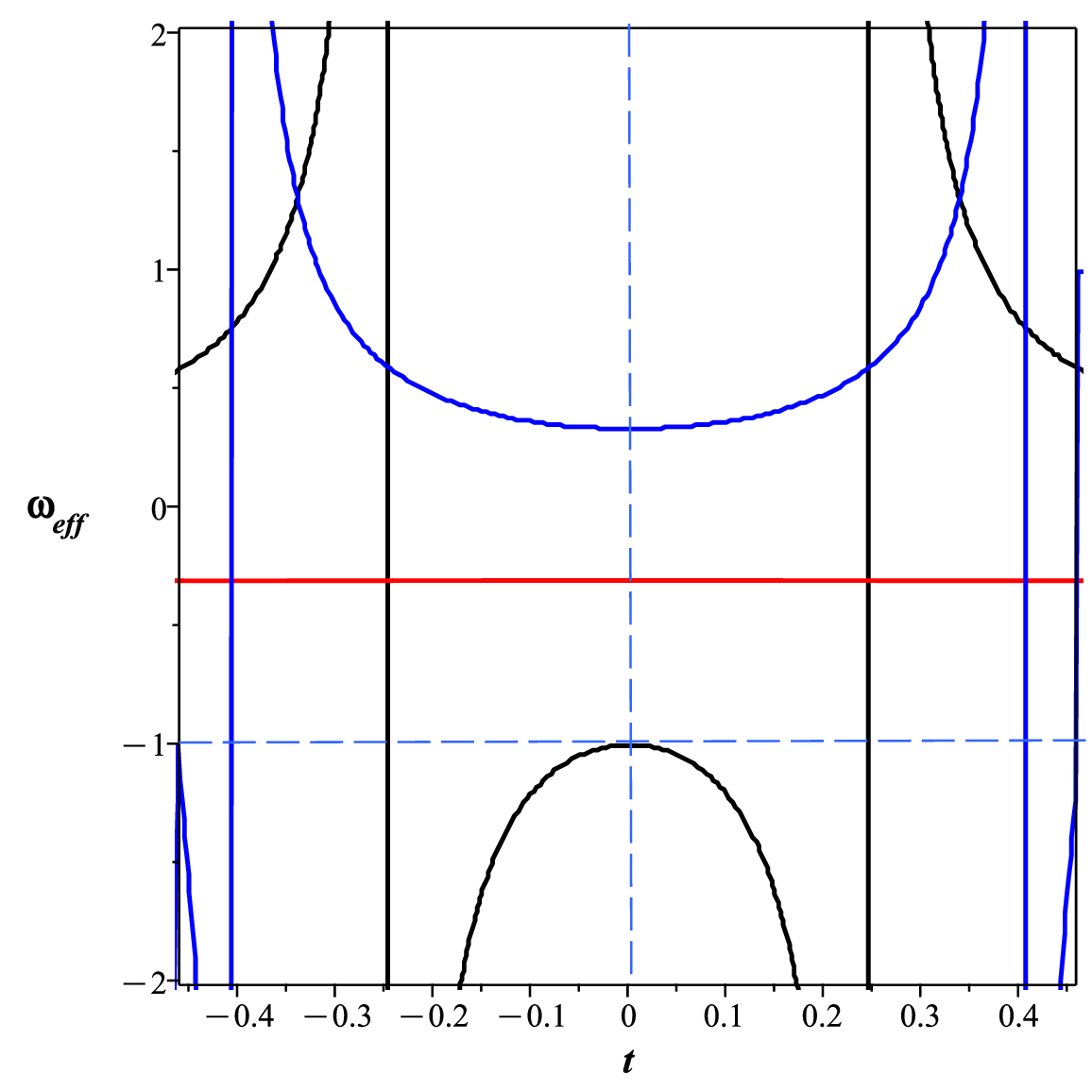} \\
			(a) $\omega_{\text{eff}}$ for $\omega = 1/3$ (blue) & (b) Comparative evolution
		\end{tabular}
		\caption{(a) Effective equation of state $\omega_{\text{eff}}(t)$ for $\omega = 1/3$ (blue line) and (b) composite view comparing all cases: $\omega = -1$ (black), $\omega = -1/3$ (red), and $\omega = 1/3$ (blue). The PDL ($\omega_{\text{eff}} = -1$) is shown as dashed line.}
		\label{fig:mtg_eos_comparing}
	\end{figure}

	The EoS evolution in Figures \ref{fig:mtg_eos}-\ref{fig:mtg_eos_comparing} reveals:
	\begin{enumerate}
		\item \textbf{For $\omega=-1$}:
		\begin{itemize}
			\item Remains in phantom regime ($\omega_{\text{eff}}<-1$) throughout $-0.3<t<0.3$
			\item Reaches maximum $\omega_{\text{eff}}$ at $t=0$ (bounce point)
			\item Transitions to $\omega_{\text{eff}}>0$ for $|t|>0.3$ (matter/radiation-like)
		\end{itemize}
		
		\item \textbf{For $\omega=-1/3$}:
		\begin{itemize}
			\item Shows quintessence behavior ($\omega_{\text{eff}}>-1$) in $-0.5<t<0.5$
			\item Crosses PDL twice at $t\sim\pm25$ (phantom-quintessence transitions)
			\item Demonstrates torsion-mediated vacuum stability near bounce
		\end{itemize}
	\end{enumerate}
	
	Physical interpretation of the results:
	
	Fig. \ref{fig:mtg_eos}a shows persistent phantom behavior ($\omega_{\text{eff}} < -1$) for $\omega = -1$, with maximum at $t=0$ indicating bounce equilibrium and transitions to matter/radiation-like phases ($\omega_{\text{eff}}>0$) for $|t|>0.3$.
	
	Fig. \ref{fig:mtg_eos}b reveals quintessence dominance ($\omega_{\text{eff}} > -1$) for $\omega = -1/3$ within $-0.5<t<0.5$, with double PDL crossings at $t\sim\pm25$ indicating phantom-quintessence phase transitions.
	
	Fig. \ref{fig:mtg_eos_comparing} shows matter-like behavior ($\omega_{\text{eff}}>0$) near bounce ($-0.4<t<0.4$) with minimum at $t=0$, transitioning to phantom regime ($\omega_{\text{eff}}<-1$) outside this range, demonstrating $\omega$-dependent cosmic evolution.
	
	\begin{enumerate}
		\item \textbf{Bounce Mechanism}:
		\begin{equation}
			a(t) \approx a_{\text{min}}\left[1 + \frac{t^2}{2\tau^2}\right], \quad \tau^{-2} = \frac{\xi_3}{2}(5\rho_c^2 - 14\rho_c p_c -3p_c^2) - 12\xi_2 H^2(H^2 + \dot{H})
		\end{equation}
		where $\rho_c,p_c$ are critical values at bounce.
		
		\item \textbf{Phase Transitions}:
		The torsion coupling creates effective potentials:
		\begin{equation}
			V_{\text{eff}}(\phi,\psi) = \frac{1}{2}m_p(\phi^2+\psi^2) - \frac{2}{3}\phi^2\psi + \xi_3 T^2(\phi,\psi)
		\end{equation}
		
		\item \textbf{Singularity Avoidance}:
		\begin{equation}
			R_{\text{max}} \sim \xi_3^{-1/2} \quad \text{(Finite curvature at bounce)}
		\end{equation}
	\end{enumerate}
	
\begin{figure}[h]
		\centering
		\begin{tabular}{cc}
			\includegraphics[width=0.45\linewidth]{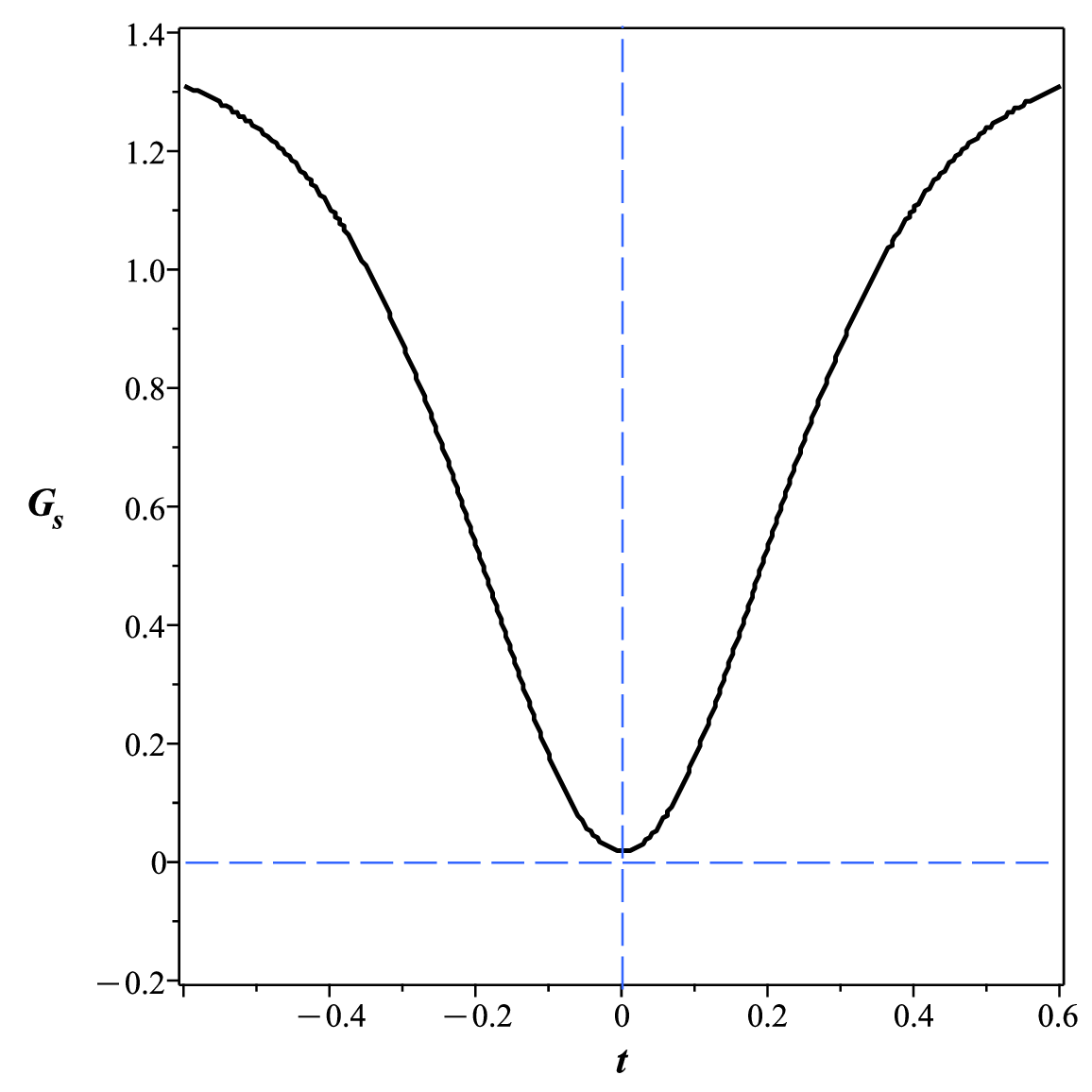} &
			\includegraphics[width=0.45\linewidth]{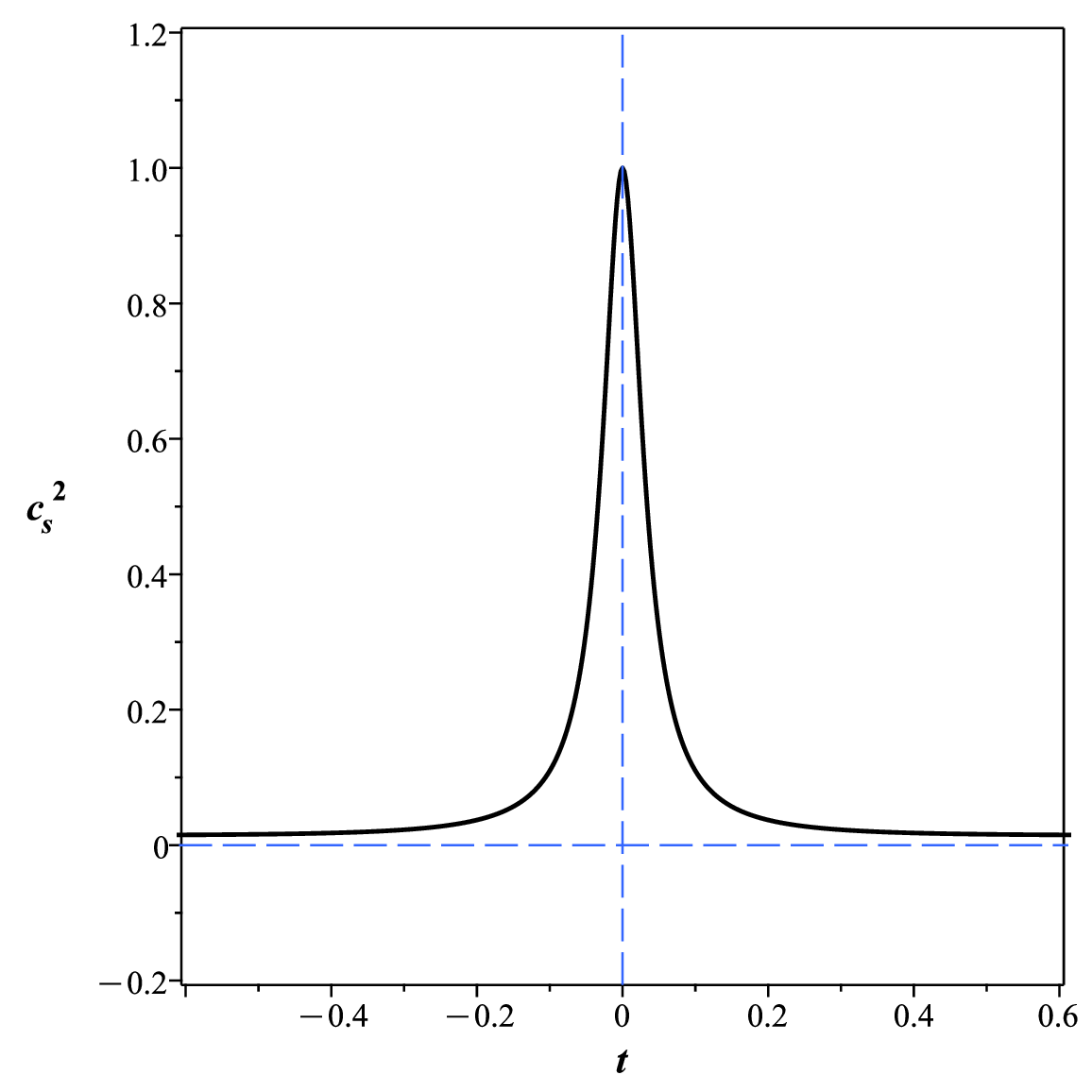} \\
			(a) Kinetic term $\mathcal{G}_S(t)$ & (b) Squared sound speed $c_s^2(t)$
		\end{tabular}
		\caption{Stability analysis for the teleparallel model ($\omega = -1$). (a) The kinetic term $\mathcal{G}_S$ remains positive throughout the evolution, confirming no ghost instabilities. (b) The sound speed squared $c_s^2$ maintains $c_s^2 \geq 0$, ensuring gradient stability. The torsion-matter coupling provides stable dynamics throughout the bounce region.}
		\label{fig:mtg_stability}
	\end{figure}

	\paragraph*{\textbf{Stability Verification:}}
	Using the teleparallel-inspired model, we numerically verify the stability criteria. Figure \ref{fig:mtg_stability} shows the results for the $\omega = -1$ case.
	
	The teleparallel model satisfies $\mathcal{G}_S(t) > 0$ for all $t$, avoiding ghost pathologies. The sound speed remains real with $c_s^2(t) \geq 0$, ensuring gradient stability. The quadratic torsion coupling $T^2$ contributes positively to the stability conditions, particularly in the high-density bounce region.
	
	Our analysis uses potential $V(\phi,\psi) = \frac{1}{2}m_p(\phi^2+\psi^2) - \frac{\lambda}{3}\phi^2\psi$ with $\lambda=2$ and couplings $\xi_2 = \xi_3 = 1$, showing how teleparallel modifications can generate viable bouncing cosmologies while maintaining consistency with observational constraints.

	\subsubsection{Non-Minimal Curvature-Matter Coupling}
	
	Building on \cite{bertolami2007extra}, we analyze a gravity model where curvature and matter interact dynamically through a non-minimal coupling:
	\begin{equation}
		f(R,G,T) = R + \xi_2 G + \xi_3 R^2 T
	\end{equation}
	
	The $R^2T$ term introduces a new energy-momentum exchange channel between geometry and matter. As shown in Appendix \ref{sec:Derivation of Modified Friedmann Equations}, the modified Friedmann equations obtain:
	\begin{align}
		3H^2 &= \frac{1}{1 + 2\xi_3 R T}\bigg[\underbrace{\kappa^2(\rho + \rho_\Xi) + \xi_3 R^2 (\rho + p)}_{\text{Matter and coupling}} - \underbrace{3H\dot{f}_R + 3\dot{H}(1 + 2\xi_3 R T)}_{\text{Dynamic curvature}} \nonumber \\
		&- \underbrace{\frac{1}{2}\left(R + \xi_3 R^2 T\right)}_{\text{Effective dark energy}} \bigg] \label{f1_Rec_NMC_corrected}
	\end{align}
	\begin{align}
		-2\dot{H} - 3H^2 &= \frac{1}{1 + 2\xi_3 R T}\bigg[\underbrace{\kappa^2(p + p_\Xi)}_{\text{Standard pressure}}  + \underbrace{\ddot{f}_R + 2H\dot{f}_R + \dot{H}(1 + 2\xi_3 R T)}_{\text{Curvature acceleration}} \nonumber \\
		&+ \underbrace{\frac{1}{2}\left(R + \xi_3 R^2 T\right)}_{\text{Geometric stiff matter}} \bigg] \label{f2_Rec_NMC_corrected}
	\end{align}
	
	Expanding the curvature acceleration terms:
	\begin{align}
		\ddot{f}_R + 2H\dot{f}_R &= 2\xi_3[\ddot{R}T + 2\dot{R}\dot{T} + R\ddot{T} + 2H(\dot{R}T + R\dot{T})] \\
		&= 2\xi_3\left[T\ddot{R} + 2\dot{R}\dot{T} + R\ddot{T} + 2H\dot{R}T + 2HR\dot{T}\right]
	\end{align}
	
	\begin{figure}[h]
		\centering
		\begin{tabular}{cc}
			\includegraphics[width=0.45\linewidth]{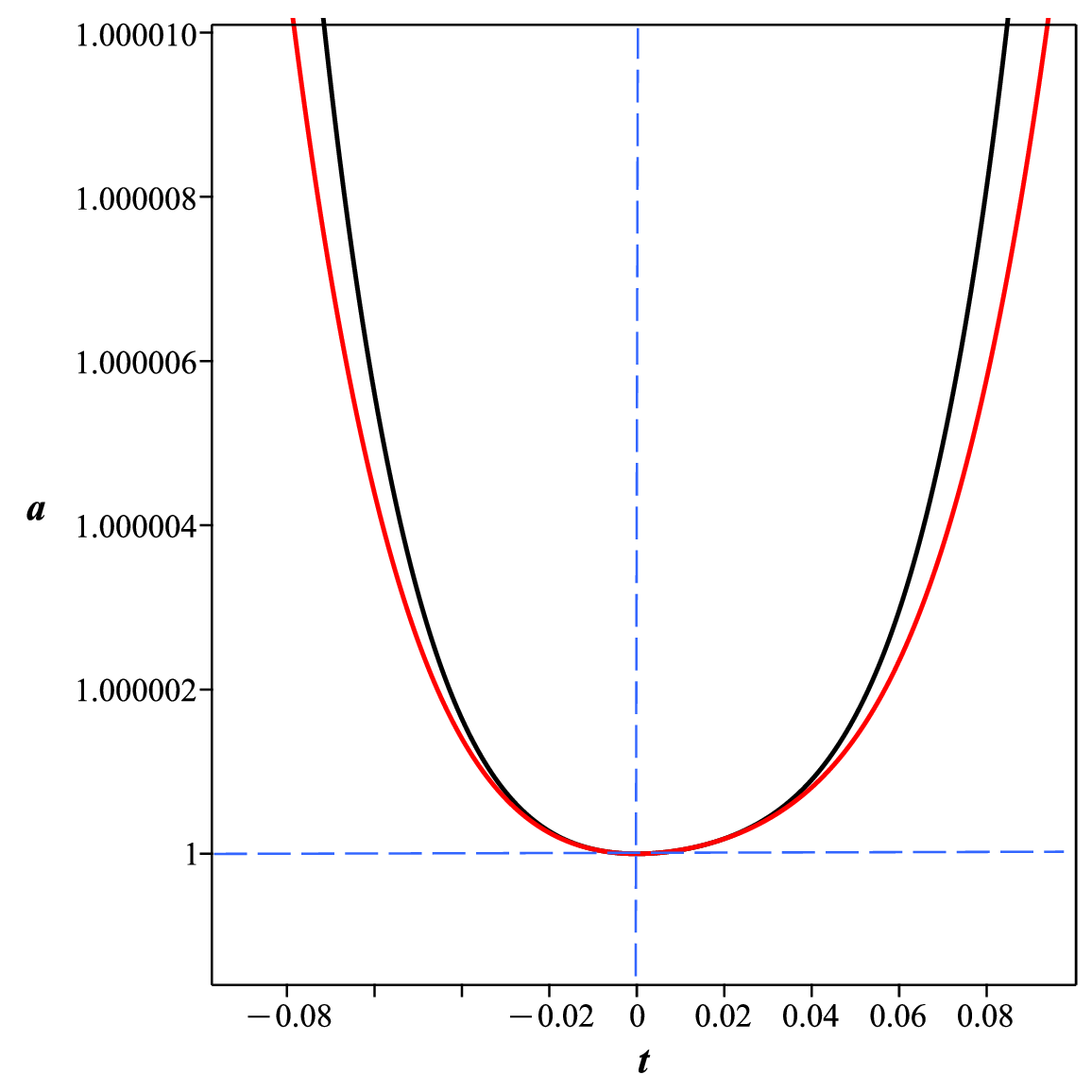} &
			\includegraphics[width=0.45\linewidth]{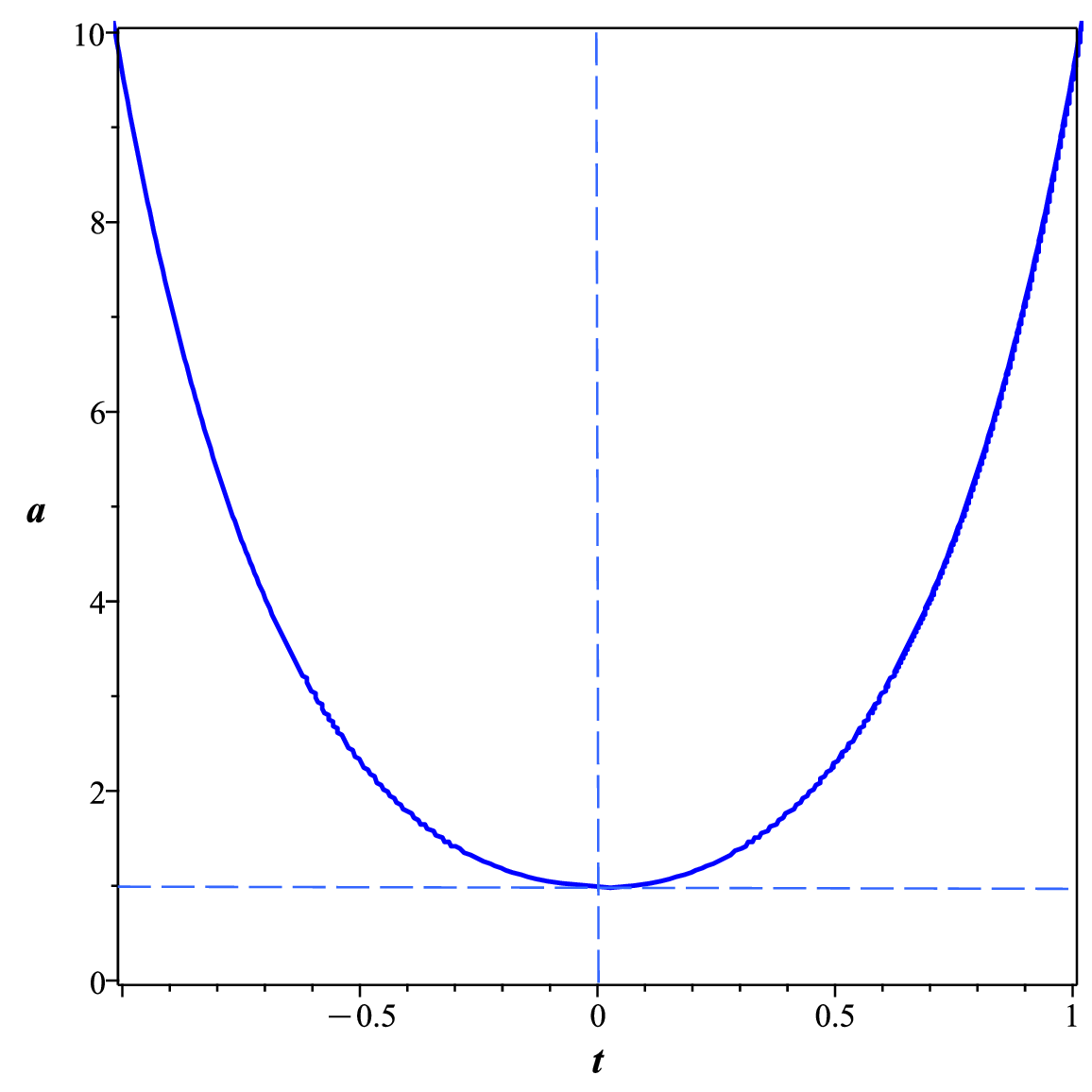} \\
			(a) $\omega = -1$ (black) vs $\omega = -1/3$ (red) & (b) $\omega = 1/3$ (blue)
		\end{tabular}
		\caption{Scale factor evolution showing (a) comparative bounce dynamics for dark energy cases and (b) radiation-dominated ($\omega=1/3$) case exhibiting optimal bounce behavior with $a_{\text{min}}$ at $t=0$. Initial conditions: $\dot{\psi}(0) = (100.2362362)^{1/2}$ (red), $(66.72589056)^{1/2}$ (black), $0.01$ (blue).}
		\label{fig:nmc_scale}
	\end{figure}
	
	Key physical effects of the non-minimal coupling:
	
	1. \textbf{Energy-Dependent Gravitational Constant}:
	\begin{equation}
		G_{\text{eff}} = \frac{G_N}{1 + 2\xi_3 R T} \approx G_N[1 - 2\xi_3 R T]
	\end{equation}
	becomes matter-density dependent, modifying gravitational interactions in high-energy regimes.
	
	2. \textbf{Curvature-Induced Pressure}:
	The coupling generates an effective pressure through curvature-matter interaction:
	\begin{equation}
		p_{\text{eff}} = p + \frac{\xi_3 R^2}{\kappa^2}\left(\ddot{R}T + 2\dot{R}\dot{T} + R\ddot{T} + 2H\dot{R}T + 2HR\dot{T}\right)
	\end{equation}
	This provides an additional repulsive force during cosmic contraction.
	
	3. \textbf{Bounce Mechanism}:
	For radiation-dominated universe ($\omega=1/3$), the coupling terms dominate when:
	\begin{equation}
		\xi_3 R^2 T \sim \rho_{\text{rad}} \Rightarrow a_{\text{min}} \sim \left(\frac{\rho_{\text{rad},0}}{\xi_3 R_0^2 T_0}\right)^{1/4}
	\end{equation}
	determining the minimum scale factor before bounce occurrence.
	
	\begin{figure}[h]
		\centering
		\begin{tabular}{cc}
			\includegraphics[width=0.45\linewidth]{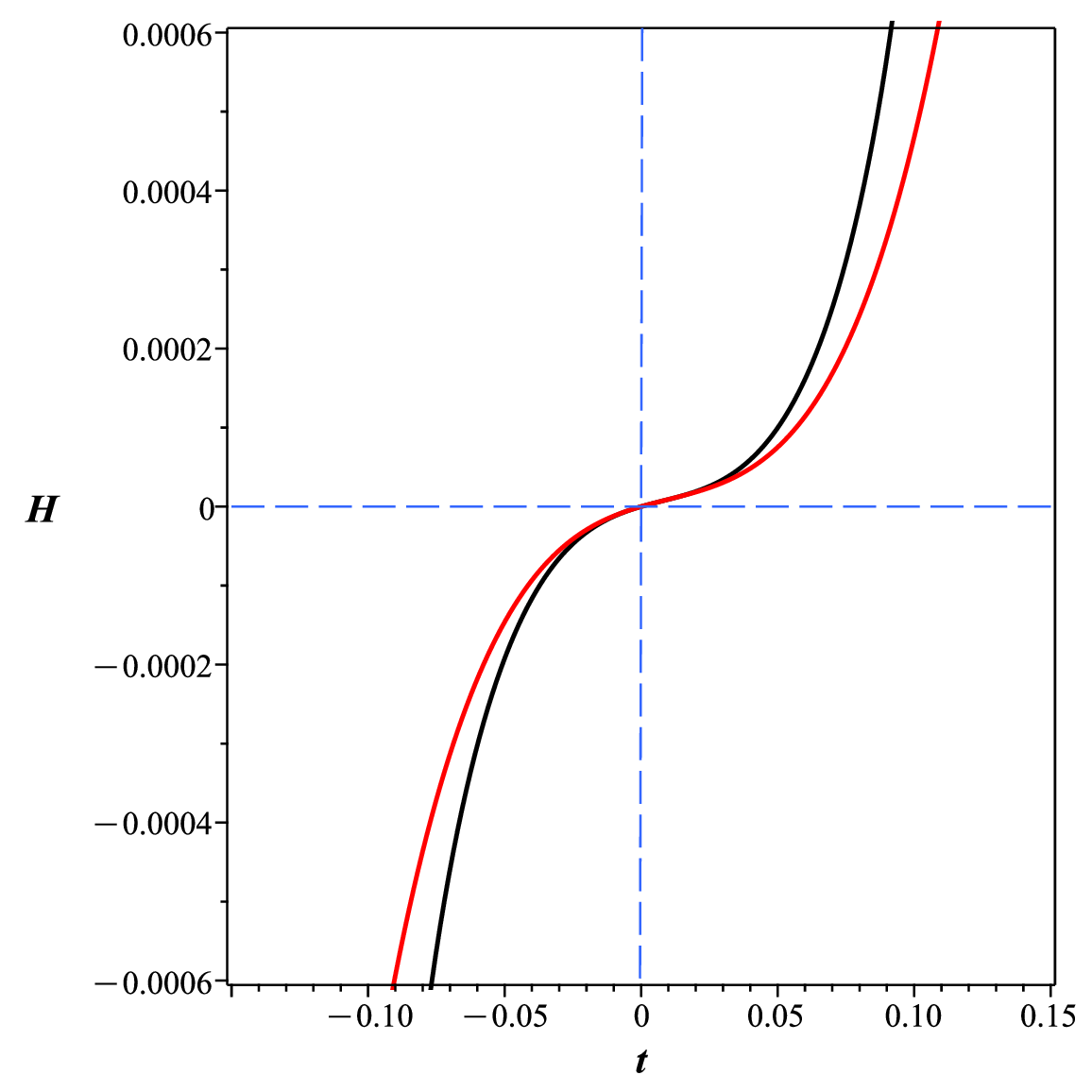} &
			\includegraphics[width=0.45\linewidth]{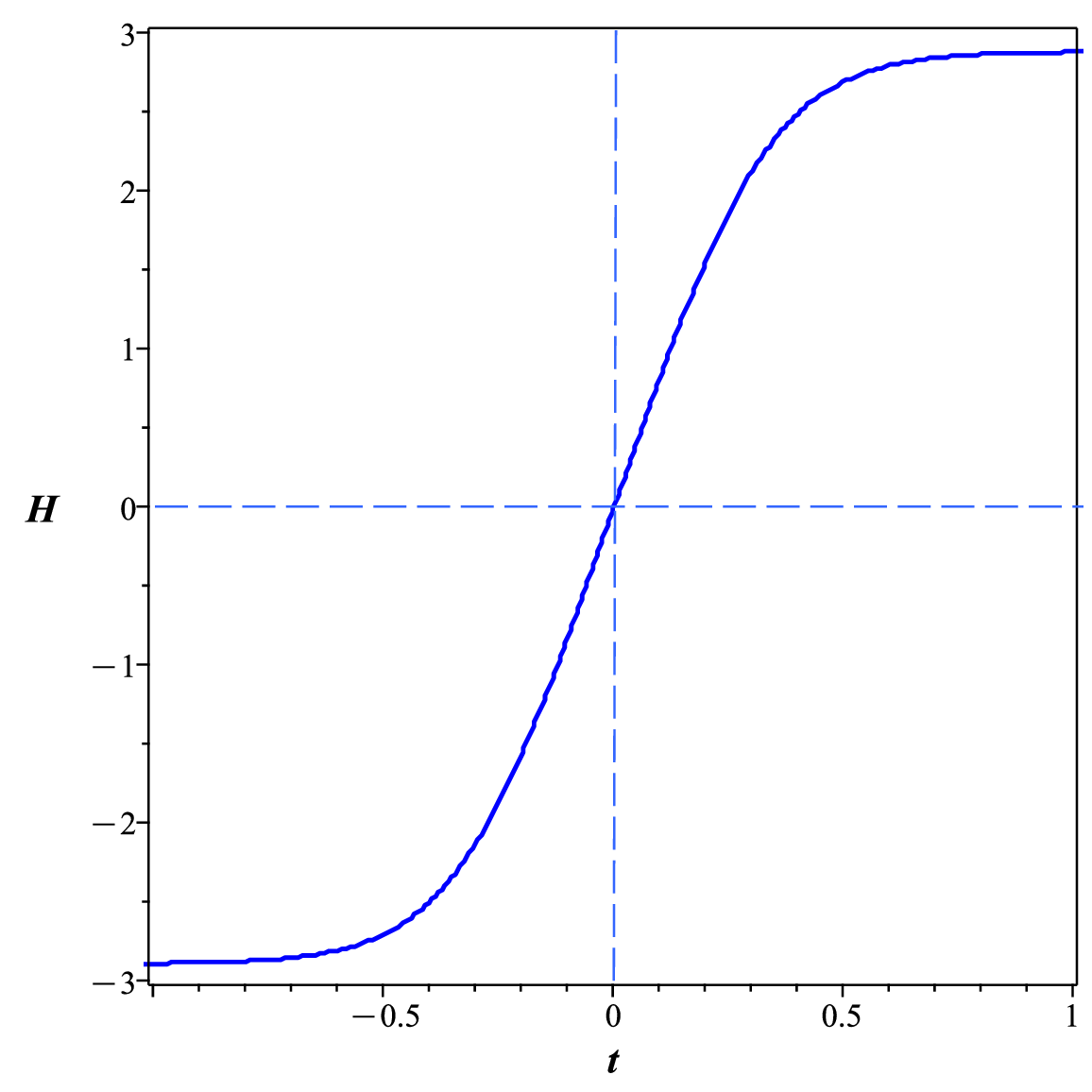} \\
			(a) $\omega = -1$ (black) vs $\omega = -1/3$ (red) & (b) $\omega = 1/3$ (blue)
		\end{tabular}
		\caption{Hubble parameter evolution showing (a) comparative dynamics for dark energy cases and (b) radiation-dominated case with characteristic $H(t)$ profile and sign-change at bounce. The red curve ($\dot{\psi}(0)=(100.236)^{1/2}$) shows strongest bounce dynamics.}
		\label{fig:nmc_hubble}
	\end{figure}
	
	\begin{figure}[h]
		\centering
		\begin{tabular}{cc}
			\includegraphics[width=0.45\linewidth]{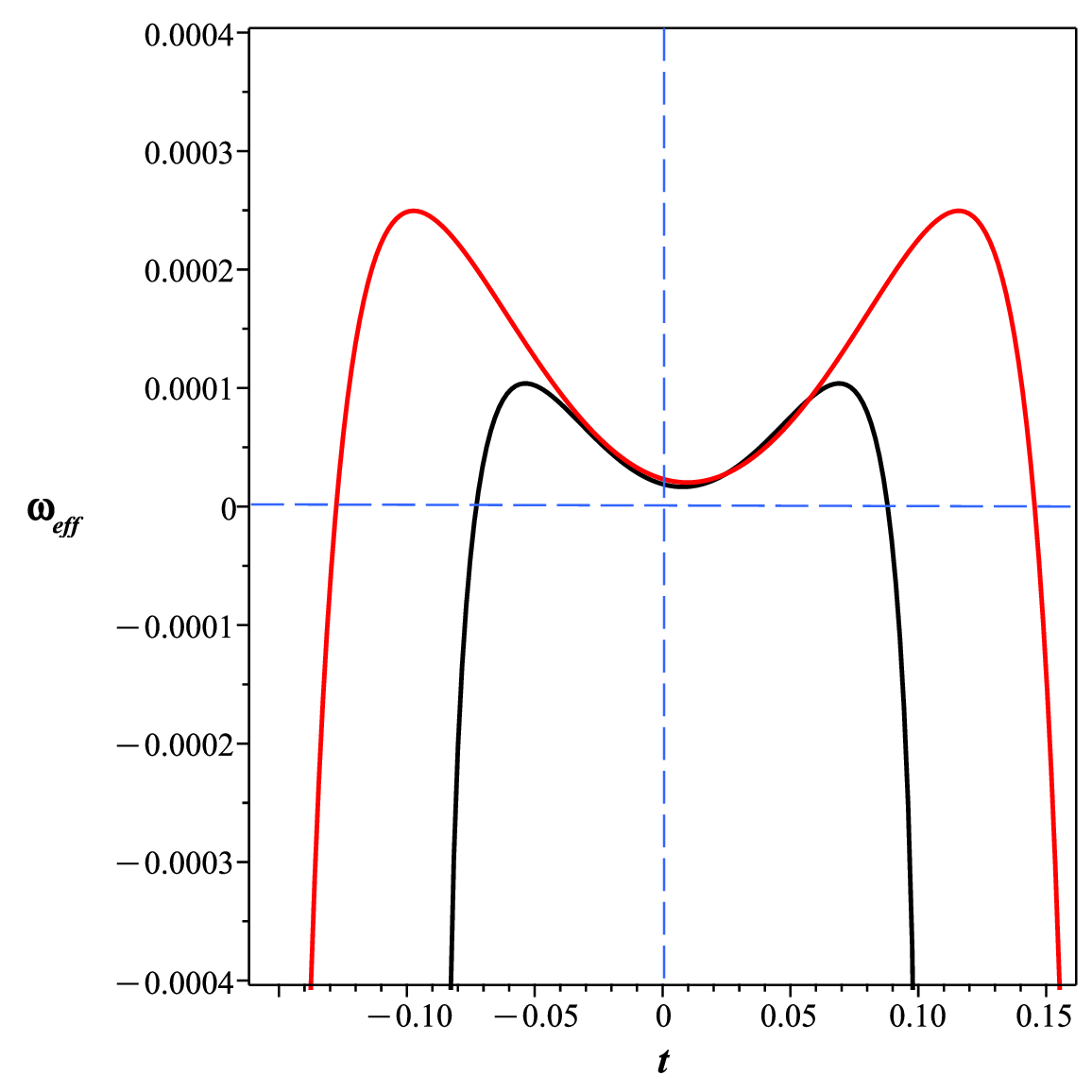} &
			\includegraphics[width=0.45\linewidth]{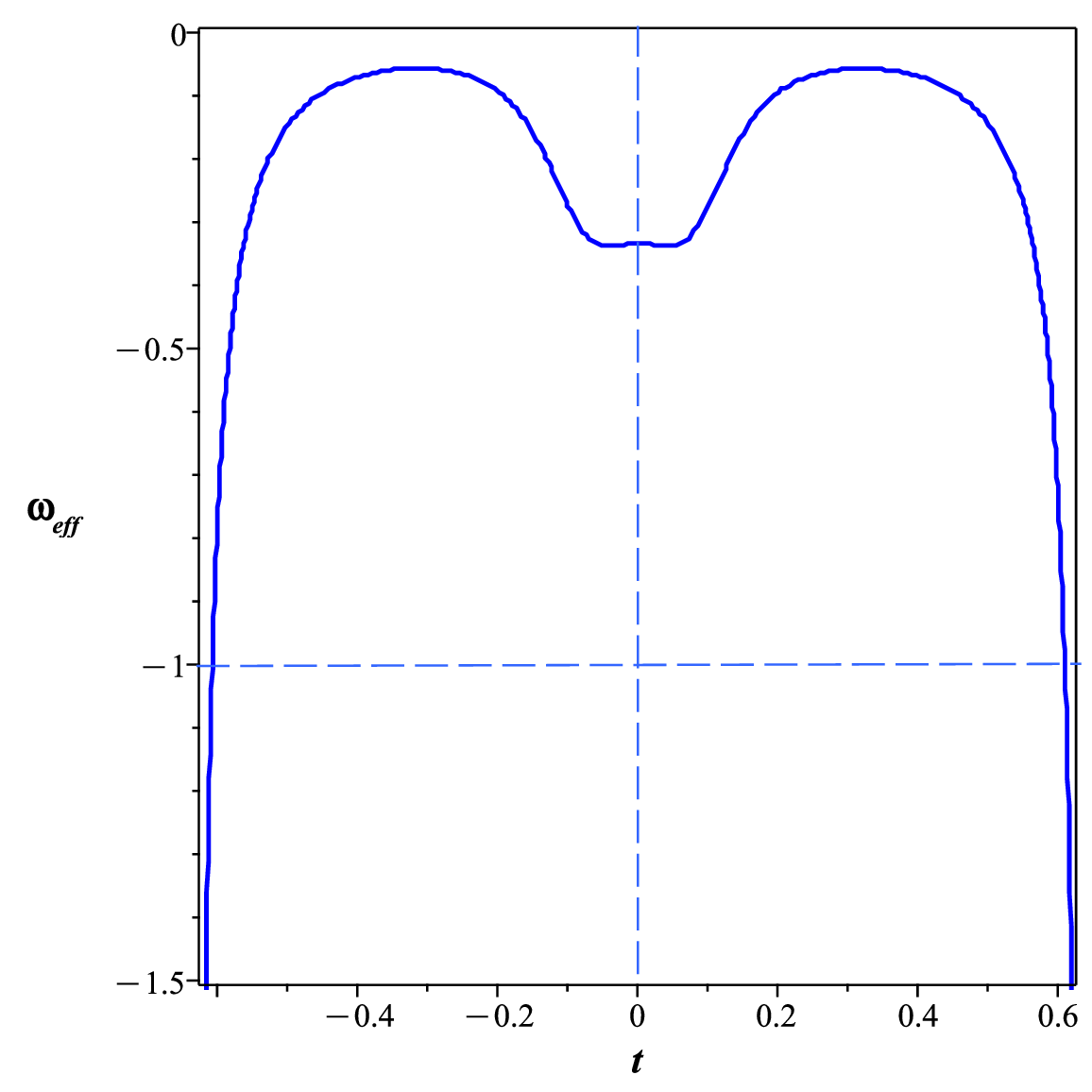} \\
			(a) $\omega = -1$ (black) vs $\omega = -1/3$ (red) & (b) $\omega = 1/3$ (blue)
		\end{tabular}
		\caption{Effective EoS showing (a) comparative evolution for dark energy cases and (b) radiation-dominated case with PDL crossings. The PDL ($\omega_{\text{eff}} = -1$) is shown as dashed line.}
		\label{fig:nmc_eos}
	\end{figure}
	
	The EoS evolution reveals:
	
	1. \textbf{For $\omega=1/3$}:
	\begin{itemize}
		\item Successful bounce with two PDL crossings
		\item EoS evolution: $\omega_{\text{eff}}<-1$ (phantom) $\to$ $\omega_{\text{eff}}>0$ (matter-like) $\to$ $\omega_{\text{eff}}>-1$ (quintessence)
		\item Reflects phase transitions: $R^2T$ term dominates $\to$ radiation dominates $\to$ curvature dominates
	\end{itemize}
	
	2. \textbf{For $\omega=-1$ and $\omega=-1/3$}:
	\begin{itemize}
		\item Equilibrium points without PDL crossing
		\item Effective potential minimum at $t=0$:
		\begin{equation}
			V_{\text{eff}} \approx V_0 e^{-0.01\phi} + \frac{1}{2}m_p^2\psi^2 + 0.1\phi\psi + \xi_3 R^2 T
		\end{equation}
	\end{itemize}
	
	The non-minimal coupling model exhibits a particularly interesting feature: its bounce dynamics are most efficient and robust in the radiation-dominated era ($\omega = 1/3$). This is physically significant because the $R^2T$ coupling term introduces corrections that scale with both curvature and energy density. In a radiation-dominated universe ($\rho \sim a^{-4}$), these corrections become dominant at high energies, making this coupling particularly relevant in the high-density environment of the very early universe. While the model can produce bounces for other equations of state, we focus on the $\omega = 1/3$ case as it represents the most natural and effective application scenario for this specific coupling. Figs. \ref{fig:nmc_scale} and \ref{fig:nmc_hubble} shows the successful bounce dynamics for this case, with $a(t)$ reaching a minimum and $H(t)$ crossing zero at $t=0$, indicating a smooth contraction-to-expansion transition.
	
	Fig. \ref{fig:nmc_eos} reveals contrasting behaviors: $\omega = -1$ and $\omega = -1/3$ cases show equilibrium minima at $t=0$ without PDL crossing, while $\omega = 1/3$ exhibits double PDL crossings ($t<0$ and $t>0$) with positive $\omega_{\text{eff}}(0)$, confirming dynamic phantom-quintessence transitions that support the observed bounce behavior.
	
\begin{figure}[h]
		\centering
		\begin{tabular}{cc}
			\includegraphics[width=0.45\linewidth]{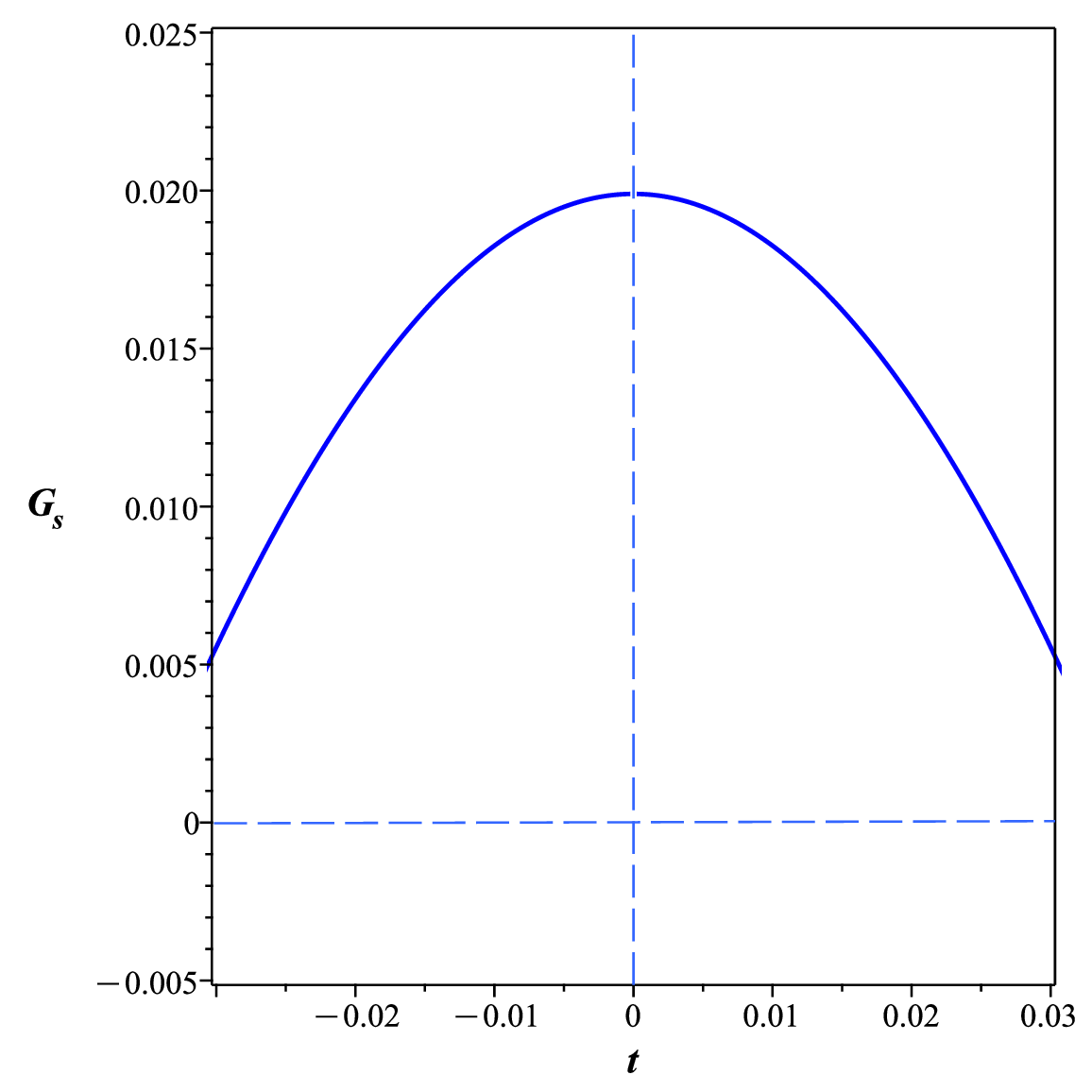} &
			\includegraphics[width=0.45\linewidth]{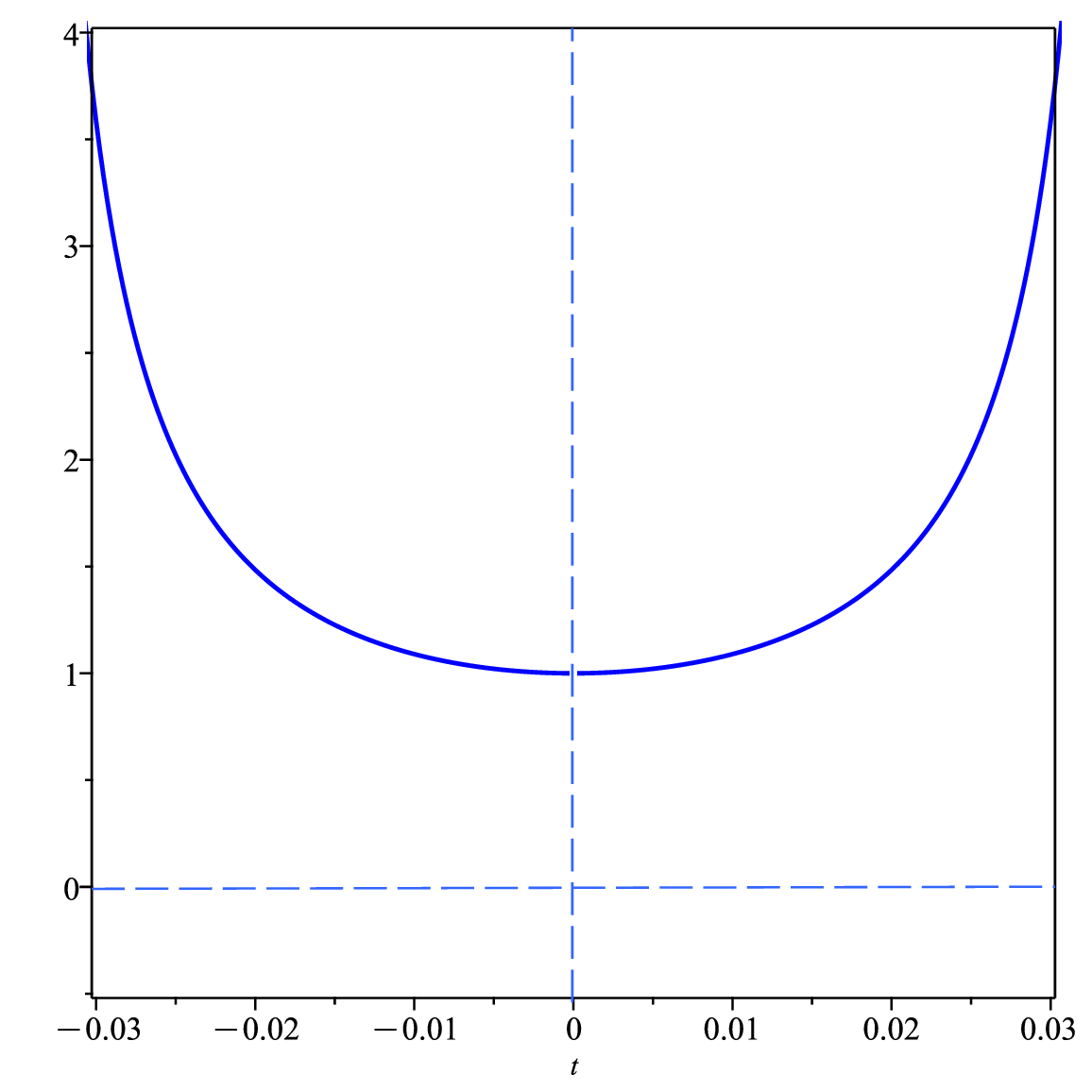} \\
			(a) Kinetic term $\mathcal{G}_S(t)$ & (b) Squared sound speed $c_s^2(t)$
		\end{tabular}
		\caption{Stability analysis for the non-minimal coupling model ($\omega = 1/3$). (a) The kinetic term $\mathcal{G}_S$ remains positive throughout the evolution, confirming no ghost instabilities. (b) The sound speed squared $c_s^2$ maintains $c_s^2 \geq 0$, ensuring gradient stability. The $R^2T$ coupling provides stable dynamics suitable for radiation-dominated bouncing scenarios.}
		\label{fig:nmc_stability}
	\end{figure}

	\paragraph*{\textbf{Stability Verification:}}
	For the non-minimal coupling model, we verify the stability conditions numerically. Figure \ref{fig:nmc_stability} shows the results for the $\omega = 1/3$ case, which exhibited the most robust bounce.
	
	The non-minimal coupling model maintains $\mathcal{G}_S(t) > 0$ throughout the evolution, confirming ghost-free behavior. The sound speed squared $c_s^2(t)$ remains non-negative, with $c_s^2 \geq 0$ ensuring gradient stability. The curvature-matter interaction term $R^2T$ contributes stabilizing effects that make this model particularly suitable for bounce scenarios in radiation-dominated regimes.
	
	The model parameters:
	\begin{itemize}
		\item Potential: $V(\phi, \psi) = V_0 e^{-\alpha\phi} + \frac{1}{2}m_p^2\psi^2 + g\phi\psi$, $V_0 = 25$, $\alpha=-0.01$, $g=0.1$
		\item Couplings: $\xi_2=\xi_3=1$
		\item Initial conditions: $\phi(0)=-0.05$, $\dot{\phi}(0)=-0.01$, $\psi(0)=0.05$, $a(0)=1$, $\dot{a}(0)=0$
		\item Varied $\dot{\psi}(0)$ values for different solutions
	\end{itemize}
	
	\textit{\textbf{Physical implications:}}
	
	1. \textbf{Radiation domination provides most stable bounce framework}: The equation of state $\omega = 1/3$ yields symmetric energy conditions and avoids anisotropic instabilities during the contraction phase, creating optimal conditions for a non-singular bounce.
	
	2. \textbf{Non-minimal coupling generates effective UV cutoff}:
	\begin{equation}
		\mathcal{L}_{\text{UV}} \sim \xi_3 R^2 T \approx \xi_3(\Box R)^2
	\end{equation}
	This introduces a natural regularization scale that prevents curvature singularities, acting as a gravitational analogue of higher-derivative terms in effective field theory.
	
	3. \textbf{Late-time behavior controlled by curvature-matter interplay}: The $R^2T$ coupling becomes negligible at low curvatures, ensuring standard $\Lambda$CDM evolution while providing a dynamical dark energy component through curvature-matter energy exchange.
	
	\section{Summary and Conclusion}\label{sec:SumCon}
	
	This work has developed a comprehensive framework for nonsingular bouncing cosmologies within modified $f(R, G, T)$ gravity coupled to a quintom scalar field model. Our investigation successfully addresses the challenge of unifying early-time bounce dynamics with late-time accelerated expansion in a single theoretical framework, while maintaining theoretical consistency through rigorous stability analysis and providing novel observational signatures.
	
	\subsection*{Principal Findings}
	
	\textbf{1. Unified Cosmological Description:}
	We have demonstrated that the synthesis of $f(R,G,T)$ geometry with quintom scalar fields provides a powerful mechanism for describing a complete cosmic history. The model naturally incorporates a non-singular bounce at early times, followed by a smooth transition to the standard expansion history, culminating in dark energy domination at late times. Numerical reconstruction of five distinct models confirms this unified behavior across different cosmological regimes.
	
	\textbf{2. Novel Phantom Divide Phenomenology:}
	A distinctive prediction of our framework is the double crossing of the phantom divide line ($\omega_{\text{eff}} = -1$) during the bounce phase. This feature emerges from the dynamic interplay between the curvature-matter coupling and the two-field quintom dynamics, representing a unique observational signature that differentiates our mechanism from single-field bounce scenarios.
	
	\textbf{3. Theoretical Consistency and Stability:}
	Through rigorous Hamiltonian analysis in Section \ref{sec:Stability}, we confirmed the theory contains the correct number of physical degrees of freedom without Ostrogradsky instabilities. The derived conditions for ghost avoidance ($\mathcal{G}_S > 0$) and gradient stability ($\mathcal{F}_S > 0$) ensure theoretical consistency throughout the cosmic evolution. Numerical verification across all five models demonstrates that these stability conditions are maintained, including during the critical bounce phase where $H=0$.
	
	\subsection*{Model-Specific Insights and Stability}
	
	Our analysis of specific cosmological models revealed important physical implications with particular attention to stability:
	
	\begin{itemize}
		\item The \textbf{Linear Model} demonstrates the essential role of dark energy domination ($\omega \leq -1/3$) for successful bounce dynamics and phantom divide crossings, while maintaining $\mathcal{G}_S > 0$ and $c_s^2 \geq 0$ throughout evolution.
		
		\item The \textbf{Exponential Model} shows robust bouncing behavior across all equations of state, with optimal dynamics in the $\omega=-1$ case. The model maintains positive kinetic terms and real sound speeds despite finite-time singularities at distant times.
		
		\item The \textbf{Power-Law Model} reveals a unification of phantom-driven contraction, curvature-mediated bounce, and late-time $\Lambda$CDM-like expansion. The complex oscillatory dynamics remain stable with $\mathcal{G}_S > 0$ and $c_s^2 \geq 0$ throughout.
		
		\item The \textbf{Teleparallel Model} illustrates how torsion-matter couplings can generate viable bounce dynamics with persistent phantom behavior. The quadratic torsion coupling $T^2$ contributes positively to stability conditions, particularly in high-density regions.
		
		\item The \textbf{Non-Minimal Coupling Model} indicates that radiation-dominated eras provide particularly stable frameworks for bounce scenarios within this model. The $R^2T$ coupling provides stabilizing effects that maintain ghost-free and gradient-stable conditions.
	\end{itemize}
	
	\subsection*{Future Directions and Observational Connections}
	
	While this work establishes a solid theoretical foundation, several important avenues require further exploration to fully establish observational viability:
	
	\textbf{Perturbations and Observational Predictions:}
	The natural next step is to calculate the power spectrum of scalar and tensor perturbations generated during the bounce phase. The double PDL crossing and the accompanying non-adiabatic pressures from the quintom field and curvature-matter coupling are expected to source isocurvature perturbations and potentially lead to distinctive non-Gaussian features \cite{lin2011matter}. We anticipate that the specific sequence of phases will imprint characteristic signatures on the primordial power spectrum, potentially offering clear observational discriminants from inflationary predictions.
	
	\textbf{Naturalness and Parameter Space:}
	The numerical solutions demonstrate successful bouncing trajectories for chosen parameter sets ($\xi_i \sim \mathcal{O}(1)$, $V_0 \sim \mathcal{O}(m_p^4)$) that are theoretically motivated. A comprehensive parameter space scan to identify the full basin of attraction for successful bounces represents an important future endeavor that would further strengthen the physical motivation of our framework.
	
	\textbf{Finite-Time Singularities:}
	The appearance of finite-time singularities in some models (typically Type II or IV where the scale factor remains finite but higher derivatives diverge) indicates scales where quantum gravitational effects or higher-order corrections may become relevant \cite{bamba2008future}. Future work will explore coupling our model to matter components that satisfy limiting curvature conditions \cite{ye2019bounce} or including higher-order terms that may naturally mitigate these singularities while preserving the desirable bouncing behavior.
	
	\textbf{Observational Tests:}
	The predicted double PDL crossing and associated phase transitions could leave imprints in the cosmic microwave background anisotropy spectrum and large-scale structure. The planned perturbation analysis will be essential for making concrete predictions testable with current and future cosmological datasets (CMB, BAO, supernovae).
	
	\subsection*{Concluding Remarks}
	
	In conclusion, our work demonstrates that modified $f(R, G, T)$ gravity combined with quintom scalar fields provides a robust, stable, and versatile framework for studying nonsingular bouncing cosmologies. The core achievements of this paper--a unified cosmological description, novel phantom divide phenomenology, and rigorous stability analysis across multiple model classes--form a solid foundation for future research. The framework successfully addresses the initial singularity problem while providing testable predictions that distinguish it from standard cosmological scenarios. The comprehensive stability verification, both analytical and numerical, ensures that these bouncing solutions are theoretically sound and not plagued by the ghost and gradient instabilities that often affect higher-derivative theories. The planned perturbation analysis and connection to observational data will be the decisive next steps in evaluating this framework's complete viability as an alternative description of our universe's history.

	\section*{Acknowledgments}
	
	The author thanks the anonymous referee for their constructive critique, which improved the manuscript. The concerns regarding stability and degrees of freedom in higher-derivative theories are addressed in Sec.~\ref{sec:Stability}, where a Hamiltonian analysis confirms the absence of pathological modes due to FLRW symmetry constraints. The statement on PDL crossing has been clarified to highlight the novelty of the \textit{double} crossing mechanism within our unified $f(R,G,T)$ framework, contrasting with single-field approaches in \cite{bamba2009crossing, vikman2005can, deffayet2010imperfect}. References \cite{bamba2009crossing, vikman2005can, cai2007bouncing} have been added to provide broader context. The model's motivation remains the unification of early and late-time cosmology through curvature-matter coupling.

	\appendix
	\renewcommand{\thesection}{\Alph{section}}
	\renewcommand{\theequation}{\Alph{section}.\arabic{equation}}
	\setcounter{equation}{0}	% Reset the equation counter  
	\addcontentsline{toc}{section}{APPENDIX}

	\section{Calculating Modified Stress-Energy Tensors}\label{sec:Modified E-M-T}
	The variation of the determinant of the metric, $\sqrt{-g}$, Ricci scalar,  
	$R=g^{\mu\nu}R_{\mu\nu}$, Ricci tensor,   
	\begin{equation}  
		R_{\mu\nu}= \partial_{\alpha} \left({\Gamma^{\alpha}_{\mu\nu}}\right)- {\partial_{\nu}} \left({\Gamma^{\alpha}_{\alpha\mu}}\right)+ {\Gamma^{\beta}_{\mu\nu}} {\Gamma^{\alpha}_{\alpha\beta}}-{\Gamma^{\beta}_{\alpha\mu}} {\Gamma^{\alpha}_{\beta\nu}}  
	\end{equation}  
	Gauss-Bonnet invariant,   
	\begin{equation}  
		G=R^{2}-4R_{\mu\nu}R^{\mu\nu}+R_{\mu\nu\rho\lambda}R^{\mu\nu\rho\lambda}  
	\end{equation}  
	and the trace of stress-energy tensor of the matter,   
	$T=g^{\mu\nu}T_{\mu\nu}^{(m)}$, with respect to the inverse metric $g^{\mu\nu}$ is given by  
	\begin{eqnarray}  
		\delta\sqrt{-g} &=& -\frac{1}{2}\sqrt{-g}g_{\mu\nu}\delta g^{\mu\nu},\label{metric var} \\
		\delta R &=&  
		R_{\mu\nu}\delta g^{\mu\nu} - \square\delta g + \nabla^\mu \nabla^\nu \delta g_{\mu\nu}, \label{R var} \\   
		\delta R_{\mu\nu} &=& \nabla_\rho \delta \Gamma^{\rho}_{\nu\mu} - \nabla_\nu \delta \Gamma^{\rho}_{\rho\mu} \label{Rmunu var} \\
		\delta G &=& 2R \delta R -4\delta (R_{\mu\nu}R^{\mu\nu})+\delta (R_{\mu\nu\rho\lambda}R^{\mu\nu\rho\lambda}) \label{G var} \\
		\delta T &=& \frac{\partial\left(g^{\alpha\beta}T_{\alpha\beta}^{(m)}\right)}{\partial g^{\mu\nu}}\delta g^{\mu\nu} = (T_{\mu\nu}^{(m)}+\Theta_{\mu\nu})\delta g^{\mu\nu} \label{T var}  
	\end{eqnarray}  
	where $\Theta_{\mu\nu} = g^{\alpha\beta}\frac{\partial T_{\alpha\beta}^{(m)}}{\partial g^{\mu\nu}}$, $\square=g^{\alpha\beta}\nabla_\alpha\nabla_\beta$ is the d'Alembert operator, and $\delta \Gamma^{\lambda}_{\mu\nu}$ is the difference of two connections, it should transform as a tensor.  
	Therefore, it can be written as
	\begin{eqnarray}
		\delta \Gamma^{\lambda}_{\mu\nu}=\frac{1}{2}g^{\lambda\alpha}\left(\nabla_{\mu}\delta g_{\alpha\nu}+\nabla_{\nu}\delta g_{\alpha\mu}-\nabla_{\alpha}\delta g_{\mu\nu}\right),
	\end{eqnarray}
	The variation of the action (\ref{eq:action}) with respect to inverse metric  $g^{\mu\nu}$ is yielded by
	\begin{align}
		\delta S
		&=\delta \,( \int{\sqrt{-g}\left(\frac{1}{2\kappa^2}f+\Xi+\mathcal{L}_{m}\right)d^4x})\nonumber\\
		&= \delta S_f+\delta S_{\Xi} +\delta S_{m},
	\end{align}
	where $f = f(R,G,T)$, $\,\,\,\Xi = \Xi(\phi,\psi)$, and $\,\,\delta S_f = \delta S_R +\delta S_G+ \delta S_T $
	\begin{eqnarray}
		\delta S_f	&=&\frac{1}{2\kappa^2} \int{\sqrt{-g}\left(f_R \delta R+f_G\delta G+f_T\delta T+\frac{f}{\sqrt{-g}}\delta\sqrt{-g}\right)d^4x}\\
		\delta S_{\Xi}	&=&\,\,\,\,\,\,\,\,\,\,\int{\sqrt{-g}\left(\delta\Xi+\frac{\Xi}{\sqrt{-g}} \delta \sqrt{-g} \right)d^4x}\\
		\delta S_{m}	&=&\,\,\,\,\,\,\,\,\,\,\int{\sqrt{-g}\left(\frac{1}{\sqrt{-g}} \delta (\sqrt{-g}\mathcal{L}_{m}) \right)d^4x},
	\end{eqnarray}
	where $f_R=\frac{\partial f(R,G,T)}{\partial R}$, $f_G=\frac{\partial f(R,G,T)}{\partial G}$, and $f_T=\frac{\partial f(R,G,T)}{\partial T}$. 
	Using eqs.(\ref{eq:quintom}), and (\ref{metric var} -- \ref{T var}) one can gives,
	\begin{align}
		\delta S_R &=\frac{1}{2\kappa^2} \int{\sqrt{-g}\left(f_R R_{\mu\nu} +f_R g_{\mu\nu}\square -f_R \nabla_\mu\nabla_\nu -\frac{1}{2} g_{\mu\nu} f\right)\delta g^{\mu\nu}d^4x}\label{A:dS_R}\\
		\delta S_G &=\frac{1}{2\kappa^2} \int{\sqrt{-g}\, \left(2R\left(f_GR_{\mu\nu}+f_Gg_{\mu\nu}\square -f_G \nabla_\mu\nabla_\nu\right)
			\right)\delta g^{\mu\nu} d^{4}x}\nonumber\\
		&-\frac{1}{2\kappa^2} \int{\sqrt{-g}\, \left(
			4f_G\left(R_{\mu}^{\rho}R_{\rho\nu}+R_{\mu\nu}\square+g_{\mu\nu}R^{\rho\lambda}\nabla_{\rho}\nabla_{\lambda}\right)\right)\delta g^{\mu\nu} d^{4}x}\nonumber\\
		&-\frac{1}{2\kappa^2}\int{\sqrt{-g}\left(4f_G\left(R_{\mu\rho\nu\lambda}R^{\rho\lambda}-\frac{1}{2}R_{\mu}^{\rho\lambda\xi}R_{\nu\rho\lambda\xi}\right)
			\right)\delta g^{\mu\nu}d^{4}x}\nonumber\\
		&+\frac{1}{2\kappa^2}\int{\sqrt{-g}\left(4f_G\left(R_{\mu}^{\rho}\nabla_{\nu}\nabla_{\rho}+R_{\nu}^{\rho}\nabla_{\mu}\nabla_{\rho}+R_{\mu\rho\nu\lambda}\nabla^{\rho}\nabla^{\lambda}\right) 
			\right)\delta g^{\mu\nu}d^{4}x}\label{A:dS_G}\\
		\delta S_T &=\frac{1}{2\kappa^2} \int{\sqrt{-g}\left(f_T(T_{\mu\nu}^{(m)}+\Theta_{\mu\nu})\right)\delta g^{\mu\nu}d^4x}\label{A:dS_T}\\
		\delta S_{\Xi} &=\int{\sqrt{-g}\left(\partial_{\mu}\phi\partial_{\nu}\phi-\partial_{\mu}\psi\partial_{\nu}\psi
			-g_{\mu\nu}(\frac{1}{2}g^{\alpha\beta}\partial_{\alpha}\phi\partial_{\beta}\phi-\frac{1}{2}g^{\alpha\beta}\partial_{\alpha}\psi\partial_{\beta}\psi
			-V(\phi,\psi))\right)\delta g^{\mu\nu}d^4x}\label{A:dS_Xi}\\
		\delta S_{m}&=\frac{1}{2\kappa^2}\int{\sqrt{-g}\left(\frac{2k^2}{\sqrt{-g}}\frac{\partial(\sqrt{-g}\mathcal{L}_{m})}{\partial g^{\mu\nu}}\right)\delta g^{\mu\nu}d^4x}\label{A:dS_m}.
	\end{align}
	
	We must solve the equation $\delta S=0$ to obtain the energy-momentum tensors. Thus, we have:
	\begin{eqnarray}
		\delta S_R + \delta S_G = -\left(\delta S_m + \delta S_T + \delta S_{\Xi}\right).
	\end{eqnarray}
	Consequently,
	\begin{eqnarray}
		\frac{2}{\sqrt{-g}}\left(\frac{\delta S_R}{\delta g^{\mu\nu}}+\frac{\delta S_G}{\delta g^{\mu\nu}}\right) &=& -\frac{2}{\sqrt{-g}}\left(\frac{\delta S_m}{\delta g^{\mu\nu}} +\frac{\delta S_T}{\delta g^{\mu\nu}}+\frac{\delta S_{\Xi}}{\delta g^{\mu\nu}}  \right),
	\end{eqnarray}
	or,
	\begin{eqnarray}
		T_{\mu\nu}^{(R)}+T_{\mu\nu}^{(G)} &=& T_{\mu\nu}^{(m)}+T_{\mu\nu}^{(T)}+T_{\mu\nu}^{(\Xi)}\cdot
	\end{eqnarray}
	
	This establishes the complete decomposition:
	\begin{eqnarray}
		T_{\mu\nu}^{(\text{total})} &\equiv& \frac{2}{\sqrt{-g}}\frac{\delta S}{\delta g^{\mu\nu}} \nonumber \\
		&=& \underbrace{T_{\mu\nu}^{(R)} + T_{\mu\nu}^{(G)}}_{\text{Geometric}} + \underbrace{T_{\mu\nu}^{(T)} + T_{\mu\nu}^{(\Xi)} + T_{\mu\nu}^{(m)}}_{\text{Matter/Quintom}} \label{eq:EMT_decomp_final}
	\end{eqnarray}
	where each component tensor is precisely:
	\begin{itemize}
		\item Geometric sector:
		\begin{align}
			T_{\mu\nu}^{(R)} &= \frac{1}{\kappa^2}\left(f_R R_{\mu\nu} - \frac{1}{2}g_{\mu\nu}f + (g_{\mu\nu}\square - \nabla_\mu\nabla_\nu)f_R\right) \label{A:T(R)}\\
			T_{\mu\nu}^{(G)} &=\frac{1}{\kappa^2}  \left(2R\left(f_GR_{\mu\nu}+g_{\mu\nu}\square f_G- \nabla_\mu\nabla_\nu f_G\right)
			-4\left(f_G R_{\mu}^{\rho}R_{\rho\nu}+R_{\mu\nu}\square f_G+g_{\mu\nu}R^{\rho\lambda}\nabla_{\rho}\nabla_{\lambda}f_G\right)\right)\nonumber\\
			&-\frac{4}{\kappa^2}\left(f_G\left(R_{\mu\rho\nu\lambda}R^{\rho\lambda}-\frac{1}{2}R_{\mu}^{\rho\lambda\xi}R_{\nu\rho\lambda\xi}\right)
			-\left(R_{\mu}^{\rho}\nabla_{\nu}\nabla_{\rho}+R_{\nu}^{\rho}\nabla_{\mu}\nabla_{\rho}+R_{\mu\rho\nu\lambda}\nabla^{\rho}\nabla^{\lambda}\right)f_G\right)\label{A:T(G)}
		\end{align}
		\item Matter/quintom sector:
		\begin{align}
			T_{\mu\nu}^{(T)} &= -\frac{f_T}{\kappa^2}(T_{\mu\nu}^{(m)} + \Theta_{\mu\nu}) \label{A:T(T)}\\
			T_{\mu\nu}^{(\Xi)} &= -\partial_{\mu}\phi\partial_{\nu}\phi + \partial_{\mu}\psi\partial_{\nu}\psi
			+g_{\mu\nu}\left(\frac{1}{2}g^{\alpha\beta}\partial_{\alpha}\phi\partial_{\beta}\phi-\frac{1}{2}g^{\alpha\beta}\partial_{\alpha}\psi\partial_{\beta}\psi
			-V(\phi,\psi)\right)\label{A:T(Xi)}\\
			T_{\mu\nu}^{(m)} &= -\frac{2}{\sqrt{-g}}\frac{\partial(\sqrt{-g}\mathcal{L}_m)}{\partial g^{\mu\nu}}\label{A:T(m)}
		\end{align}
	\end{itemize}
	
	To calculate $\Theta_{\mu\nu}$, one can use equation (\ref{eq:T_m}) as,
	\begin{eqnarray}
		\Theta_{\mu\nu}&=&g^{\alpha\beta}\frac{\partial T_{\alpha\beta}^{(m)}}{\partial g^{\mu\nu}}=g^{\alpha\beta}\frac{\partial \left(g_{\alpha\beta}\mathcal{L}_{m}-2\frac{\partial\mathcal{L}_{m}}{\partial g^{\alpha\beta}}\right)}{\partial g^{\mu\nu}}=g^{\alpha\beta}\left(\frac{\partial g_{\alpha\beta}}{\partial g^{\mu\nu}}\mathcal{L}_{m}+g_{\alpha\beta}\frac{\partial \mathcal{L}_{m}}{\partial g^{\mu\nu}}-2\frac{\partial^2\mathcal{L}_{m}}{\partial g^{\alpha\beta}\partial g^{\mu\nu}}\right)\nonumber\\
		&=&g^{\alpha\beta}\left(-g_{\alpha\sigma}g_{\beta\gamma}\delta^{\sigma\gamma}_{\mu\nu}\mathcal{L}_{m}+\frac{1}{2}g_{\alpha\beta}g_{\mu\nu}\mathcal{L}_{m}-\frac{1}{2}g_{\alpha\beta}T_{\mu\nu}^{(m)}-2\frac{\partial^2\mathcal{L}_{m}}{\partial g^{\alpha\beta}\partial g^{\mu\nu}}\right)\nonumber\\
		&=&-g_{\mu\nu}\mathcal{L}_{m}+2g_{\mu\nu}\mathcal{L}_{m}-2T_{\mu\nu}^{(m)}-2g^{\alpha\beta}\frac{\partial^2\mathcal{L}_{m}}{\partial g^{\alpha\beta}\partial g^{\mu\nu}}\nonumber\\
		&=&g_{\mu\nu}\mathcal{L}_{m}-2T_{\mu\nu}^{(m)}-2g^{\alpha\beta}\frac{\partial^2\mathcal{L}_{m}}{\partial g^{\alpha\beta}\partial g^{\mu\nu}}\cdot
	\end{eqnarray}
	
	\section{Covariant Divergence of Stress-Energy Tensors}\label{sec:Covariant}
	
	\subsection{Covariant Derivative of $T$:}  
	
	The covariant derivative of the trace of energy-momentum tensor $T$, as a scalar, is expressed as:  
	\begin{eqnarray}  
		\nabla^{\mu} T &=& \partial^{\mu} T.  
	\end{eqnarray}  
	So, by using eq. (\ref{T}) for the time component $(\mu = 0)$, we have      
	\begin{eqnarray}  
		\nabla^{0} T = \partial^{0} T = \dot{\rho} - 3\dot{p}.  
	\end{eqnarray}          
	For the spatial components $(\mu = i= 1, 2, 3)$, we take into account the spatial indices, for a $T$ just dependent on proper time $t$:      
	\begin{eqnarray}  
		\nabla^{i} T &=& \partial^{i} T = 0,   
	\end{eqnarray}     
	Thus, the covariant derivative of $T$ in this cosmological context can be expressed as:  
	\begin{eqnarray}  
		\nabla^{\mu} T &=& \dot{\rho} - 3\dot{p}\label{A:Co_T}
	\end{eqnarray}  
	
	\subsection{Covariant Derivative of $\textsl{T}^{(\textsl{R})}$:}
	Using Equation (\ref{A:T(R)}), we derive:  
	\begin{eqnarray}  
		\kappa^2\nabla^{\mu}T_{\mu\nu}^{(R)} &=& \nabla^{\mu}\left(f_R R_{\mu\nu} - \frac{1}{2} g_{\mu\nu} f + g_{\mu\nu}\square f_R- \nabla_\mu\nabla_\nu f_R\right) \nonumber\\
		&=& (\nabla^{\mu}f_R)R_{\mu\nu} + f_R\nabla^{\mu}R_{\mu\nu} - \frac{1}{2}f_R\nabla^{\mu}(g_{\mu\nu}R) - \frac{1}{2}f_G\nabla^{\mu}(g_{\mu\nu}G) - \frac{1}{2}f_T\nabla^{\mu}(g_{\mu\nu}T) \nonumber\\
		&+& \nabla_\nu\square f_R- \square\nabla_\nu f_R.
	\end{eqnarray} 
	Considering the Bianchi identity within the framework of general relativity, we find that $\nabla_{\nu} \square f_R - \square \nabla_{\nu} f_R = - R_{\mu\nu} \nabla^{\mu} f_R$. Therefore,  
	\begin{align}  
		\kappa^2\nabla^{\mu}T_{\mu\nu}^{(R)} &= R_{\mu\nu}\nabla^{\mu}f_R + f_R\nabla^{\mu}\left(R_{\mu\nu} - \frac{1}{2}g_{\mu\nu}R\right)  - \frac{1}{2}f_G\nabla^{\mu}(g_{\mu\nu}G) - \frac{1}{2}f_T\nabla^{\mu}(g_{\mu\nu}T)- R_{\mu\nu}\nabla^{\mu}f_R \nonumber\\
		&= f_R\nabla^\mu G_{\mu\nu} - \frac{1}{2}f_G\nabla^{\mu}(g_{\mu\nu}G) - \frac{1}{2}f_T\nabla^{\mu}(g_{\mu\nu}T). 
	\end{align}  
	In this context, $ G_{\mu\nu} \equiv R_{\mu\nu} - \frac{1}{2} g_{\mu\nu} R $ represents the Einstein tensor, which satisfies the divergence-free condition $ \nabla^\mu G_{\mu\nu} = 0 $. Therefore
	\begin{eqnarray}  
		\kappa^2\nabla^{\mu}T_{\mu\nu}^{(R)} = -\frac{1}{2}g_{\mu\nu}\left( f_G\nabla^{\mu}G + f_T\nabla^{\mu}T\right). \label{A:Co_T(R)}  
	\end{eqnarray}  
	
	\subsection{Covariant Derivative of $\textsl{T}^{(\textsl{G})}$:}
	To calculate $\nabla^{\mu}T_{\mu\nu}^{(G)}$, we need to apply covariant derivative on eq. (\ref{A:T(G)}), term by term. 
	\begin{itemize}
		\item \subsubsection*{\textbf{The First Term:}} 	
		We start with the expression for $T_{1 \mu\nu}^{(G)}$, as the first term of the Gauss-Bonnet energy-momentum tensor, eq.(\ref{A:T(G)}), 
		\begin{equation}
			T_{1\mu\nu}^{(G)} = \frac{2}{\kappa^2} R \left( R_{\mu\nu} f_G + g_{\mu\nu} \square f_G - \nabla_\mu \nabla_\nu f_G \right).
		\end{equation}
		
		Applying the covariant derivative $\nabla^{\mu}$ using the product rule:
		\begin{align}
			\nabla^{\mu}T_{1\mu\nu}^{(G)} &= \frac{2}{\kappa^2} \nabla^{\mu} \left[ R \left( R_{\mu\nu} f_G + g_{\mu\nu} \square f_G - \nabla_\mu \nabla_\nu f_G \right) \right] \nonumber \\
			&= \frac{2}{\kappa^2} \left[ (\nabla^{\mu} R) \left( R_{\mu\nu} f_G + g_{\mu\nu} \square f_G - \nabla_\mu \nabla_\nu f_G \right) + R \nabla^{\mu} \left( R_{\mu\nu} f_G + g_{\mu\nu} \square f_G - \nabla_\mu \nabla_\nu f_G \right) \right].
		\end{align}
		
		Expanding the second term:
		\begin{align}
			\nabla^{\mu} \left( R_{\mu\nu} f_G + g_{\mu\nu} \square f_G - \nabla_\mu \nabla_\nu f_G \right) &= (\nabla^{\mu} R_{\mu\nu}) f_G + R_{\mu\nu} \nabla^{\mu} f_G + g_{\mu\nu} \nabla^{\mu} \square f_G - \nabla^{\mu} \nabla_\mu \nabla_\nu f_G \nonumber \\
			&= \frac{1}{2} (\nabla_{\nu} R) f_G + R_{\mu\nu} \nabla^{\mu} f_G + \nabla_{\nu} \square f_G - \square \nabla_{\nu} f_G,
		\end{align}
		where we used $\nabla^{\mu} R_{\mu\nu} = \frac{1}{2} \nabla_{\nu} R$ and $\nabla^{\mu} g_{\mu\nu} = 0$. 
		
		Using the identity $\nabla_{\nu} \square f_G - \square \nabla_{\nu} f_G = -R_{\mu\nu} \nabla^{\mu} f_G$:
		\begin{align}
			\nabla^{\mu} \left( R_{\mu\nu} f_G + g_{\mu\nu} \square f_G - \nabla_\mu \nabla_\nu f_G \right) &= \frac{1}{2} (\nabla_{\nu} R) f_G + R_{\mu\nu} \nabla^{\mu} f_G - R_{\mu\nu} \nabla^{\mu} f_G \nonumber \\
			&= \frac{1}{2} f_G \nabla_{\nu} R.
		\end{align}
		
		Substituting back:
		\begin{align}
			\nabla^{\mu}T_{1\mu\nu}^{(G)} &= \frac{2}{\kappa^2} \left[ (\nabla^{\mu} R) \left( R_{\mu\nu} f_G + g_{\mu\nu} \square f_G - \nabla_\mu \nabla_\nu f_G \right) + R \left( \frac{1}{2} f_G \nabla_{\nu} R \right) \right] \nonumber \\
			&= \frac{2}{\kappa^2} \left( R_{\mu\nu} f_G + g_{\mu\nu} \square f_G - \nabla_\mu \nabla_\nu f_G \right) \nabla^{\mu} R + \frac{1}{\kappa^2} R f_G \nabla_{\nu} R.
		\end{align}
		
		Thus, the complete expression for the covariant derivative is:
		\begin{equation}
			\nabla^{\mu}T_{1\mu\nu}^{(G)} = \frac{2}{\kappa^2} \left( R_{\mu\nu} f_G + g_{\mu\nu} \square f_G - \nabla_\mu \nabla_\nu f_G \right) \nabla^{\mu} R + \frac{1}{\kappa^2} R f_G \nabla_{\nu} R.
			\label{A:Co_T1G}
		\end{equation}
		
		\item \subsubsection*{\textbf{The Second Term:}} 
		The second term of eq.(\ref{A:T(G)}) is 
		\begin{eqnarray}
			T_{2 \mu\nu}^{(G)}=-\frac{4}{\kappa^2} \left(f_G R_{\mu}^{\rho}R_{\rho\nu}+R_{\mu\nu}\square f_G+g_{\mu\nu}R^{\rho\lambda}\nabla_{\rho}\nabla_{\lambda}f_G\right)
		\end{eqnarray}
		We apply the covariant derivative $\nabla^\mu$ to the entire tensor:  
		\begin{eqnarray}  
			\nabla^\mu T_{2 \mu\nu}^{(G)} = -\frac{4}{\kappa^2} \nabla^\mu \left( f_G R_{\mu}^{\rho}R_{\rho\nu} + R_{\mu\nu} \square f_G + g_{\mu\nu} R^{\rho\lambda} \nabla_{\rho} \nabla_{\lambda} f_G \right) .  
		\end{eqnarray}  
		Using the product rule for covariant derivatives, we calculate each term.
		\begin{eqnarray}  
			\nabla^\mu (f_G R_{\mu}^{\rho}R_{\rho\nu}) &=& (\nabla^\mu f_G) R_{\mu}^{\rho} R_{\rho\nu} + f_G \nabla^\mu (R_{\mu}^{\rho} R_{\rho\nu}),\\
			\nabla^\mu (R_{\mu}^{\rho} R_{\rho\nu}) &=& (\nabla^\mu R_{\mu}^{\rho}) R_{\rho\nu} + R_{\mu}^{\rho} \nabla^\mu R_{\rho\nu},\\  
			\nabla^\mu (f_G R_{\mu}^{\rho}R_{\rho\nu}) & = & (\nabla^\mu f_G) R_{\mu}^{\rho} R_{\rho\nu} + f_G \left( (\nabla^\mu R_{\mu}^{\rho}) R_{\rho\nu} + R_{\mu}^{\rho} \nabla^\mu R_{\rho\nu} \right),\\ 
			\nabla^\mu (R_{\mu\nu} \square f_G) & = & \nabla^\mu R_{\mu\nu} \square f_G + R_{\mu\nu} \nabla^\mu (\square f_G),\\ 
			\nabla^\mu (g_{\mu\nu} R^{\rho\lambda} \nabla_{\rho} \nabla_{\lambda} f_G) & = & \nabla^\mu g_{\mu\nu} R^{\rho\lambda} \nabla_{\rho} \nabla_{\lambda} f_G + g_{\mu\nu} \nabla^\mu (R^{\rho\lambda} \nabla_{\rho} \nabla_{\lambda} f_G).  
		\end{eqnarray}  
		Since $\nabla^\mu g_{\mu\nu} = 0$, we have:  
		\begin{eqnarray}  
			\nabla^\mu (g_{\mu\nu} R^{\rho\lambda} \nabla_{\rho} \nabla_{\lambda} f_G) & = & g_{\mu\nu} \nabla^\mu (R^{\rho\lambda} \nabla_{\rho} \nabla_{\lambda} f_G).  
		\end{eqnarray}  
		
		Now, we combine all the results:  
		\begin{eqnarray}  
			\nabla^\mu T_{2 \mu\nu}^{(G)} = & - &\frac{4}{\kappa^2} \left( (\nabla^\mu f_G) R_{\mu}^{\rho} R_{\rho\nu} + f_G \left( (\nabla^\mu R_{\mu}^{\rho}) R_{\rho\nu} + R_{\mu}^{\rho} \nabla^\mu R_{\rho\nu} \right)\right) \nonumber \\
			& - & \frac{4}{\kappa^2}\left(\nabla^\mu R_{\mu\nu} \square f_G + R_{\mu\nu} \nabla^\mu (\square f_G) + g_{\mu\nu} \nabla^\mu \left( R^{\rho\lambda} \nabla_{\rho} \nabla_{\lambda} f_G \right) \right).\label{A:Co_T2G}  
		\end{eqnarray}  
		
		\item \subsubsection*{\textbf{The Third Term:}}
		We start with the expression for $T_{3 \mu\nu}^{(G)}$, as the third term of eq.(\ref{A:T(G)}),
		\begin{eqnarray}
			T_{3 \mu\nu}^{(G)} = -\frac{4}{\kappa^2} f_G \left( R_{\mu\rho\nu\lambda} R^{\rho\lambda} - \frac{1}{2} R_{\mu}^{\rho\lambda\xi} R_{\nu\rho\lambda\xi} \right).
		\end{eqnarray} 
		We now apply the covariant derivative, 
		\begin{eqnarray}
			\nabla^{\mu} T_{3 \mu\nu}^{(G)} = & - &\frac{4}{\kappa^2} (\nabla^{\mu} f_G) \left( R_{\mu\rho\nu\lambda} R^{\rho\lambda} - \frac{1}{2} R_{\mu}^{\rho\lambda\xi} R_{\nu\rho\lambda\xi} \right) \nonumber\\
			&-& \frac{4}{\kappa^2} f_G \nabla^{\mu} \left( R_{\mu\rho\nu\lambda} R^{\rho\lambda} - \frac{1}{2} R_{\mu}^{\rho\lambda\xi} R_{\nu\rho\lambda\xi} \right). 
		\end{eqnarray}  
		Using the chain rule, the covariant derivative of 	$f_G$ becomes:
		\begin{eqnarray}
			\nabla^{\mu} f_G = f_{GR} \nabla^{\mu} R + f_{GG} 	\nabla^{\mu} G + f_{GT} \nabla^{\mu} T.  
		\end{eqnarray}  
		where $f_{GR}=\frac{\partial^2 f}{\partial R\,	\partial G}$, $f_{GG}=\frac{\partial^2 f}{\partial 	G^2}$, and $f_{GT}=\frac{\partial^2 f}{\partial T\,\partial G}$.
		
		To compute the covariant derivatives of curvature tensors, we utilize key identities:  
		\begin{itemize}  
			\item \textbf{Bianchi Identity:} The covariant derivative of the Riemann tensor satisfies  
			\begin{eqnarray}  
				\nabla^\mu R_{\mu\rho\nu\lambda} + \nabla_\nu R_{\mu\nu\rho\lambda} + \nabla_\lambda R_{\mu\nu\rho\lambda} = 0.  
			\end{eqnarray}  
			
			\item \textbf{Ricci Tensor and Scalar:} The covariant derivative of the Ricci tensor $R_{\mu\nu}$ and Ricci scalar $R$ is more complex, with a commonly used relation:  
			\begin{eqnarray}  
				\nabla^\mu R_{\mu\nu} = \frac{1}{2} \nabla_\nu R.  
			\end{eqnarray}  
		\end{itemize}  
		
		In the following steps, we use the product rule to compute the derivatives of specific terms:  
		\begin{eqnarray}  
			\nabla^\mu (R_{\mu\rho\nu\lambda} R^{\rho\lambda}) &=& (\nabla^\mu R_{\mu\rho\nu\lambda}) R^{\rho\lambda} + R_{\mu\rho\nu\lambda} \nabla^\mu R^{\rho\lambda},\\ 
			\nabla^\mu (R_{\mu}^{\rho\lambda\xi} R_{\nu\rho\lambda\xi}) &=& (\nabla^\mu R_{\mu}^{\rho\lambda\xi}) R_{\nu\rho\lambda\xi} + R_{\mu}^{\rho\lambda\xi} \nabla^\mu R_{\nu\rho\lambda\xi}.  
		\end{eqnarray}  
		
		Putting it all together, we obtain:  
		\begin{eqnarray}
			\nabla^{\mu} T_{3 \mu\nu}^{(G)} &=& -\frac{4}{\kappa^2} \left( f_{GR} \nabla^{\mu} R + f_{GG} \nabla^{\mu} G + f_{GT} \nabla^{\mu} T \right) \left( R_{\mu\rho\nu\lambda} R^{\rho\lambda} - \frac{1}{2} R_{\mu}^{\rho\lambda\xi} R_{\nu\rho\lambda\xi} \right) \nonumber\\   
			&-& \frac{4}{\kappa^2} f_G \left( (\nabla^{\mu} R_{\mu\rho\nu\lambda}) R^{\rho\lambda} + R_{\mu\rho\nu\lambda} \nabla^{\mu} R^{\rho\lambda} - \frac{1}{2}(\nabla^{\mu} R_{\mu}^{\rho\lambda\xi}) R_{\nu\rho\lambda\xi} - \frac{1}{2} R_{\mu}^{\rho\lambda\xi} \nabla^{\mu} R_{\nu\rho\lambda\xi} \right).\label{A:Co_T3G}\nonumber\\
		\end{eqnarray}

		\item	\subsubsection*{\textbf{The Fourth Term:}} 
		The fourth term of the energy-momentum tensor eq.(\ref{A:T(G)}) is given by:
		\begin{equation}
			T_{4 \mu\nu}^{(G)} = \frac{4}{\kappa^2} \left( R_{\mu}^{\rho} \nabla_{\nu} \nabla_{\rho} f_G + R_{\nu}^{\rho} \nabla_{\mu} \nabla_{\rho} f_G + R_{\mu\rho\nu\lambda} \nabla^{\rho} \nabla^{\lambda} f_G \right)
			\label{eq:T4munu}
		\end{equation}
		Applying the covariant derivative $\nabla^{\mu}$ to $T_{4 \mu\nu}^{(G)}$:
		\begin{equation}
			\nabla^{\mu} T_{4 \mu\nu}^{(G)} = \frac{4}{\kappa^2} \nabla^{\mu} \left( R_{\mu}^{\rho} \nabla_{\nu} \nabla_{\rho} f_G + R_{\nu}^{\rho} \nabla_{\mu} \nabla_{\rho} f_G + R_{\mu\rho\nu\lambda} \nabla^{\rho} \nabla^{\lambda} f_G \right)
			\label{eq:divT4munu}
		\end{equation}
		Using the product rule for covariant derivatives:
		\begin{equation}
			\nabla^{\mu}(A B) = (\nabla^{\mu} A) B + A (\nabla^{\mu} B)
		\end{equation}
		each term is expanded as follows:
		\begin{eqnarray}
			\nabla^{\mu} \left( R_{\mu}^{\rho} \nabla_{\nu} \nabla_{\rho} f_G \right) &=& (\nabla^{\mu} R_{\mu}^{\rho}) (\nabla_{\nu} \nabla_{\rho} f_G) + R_{\mu}^{\rho} \nabla^{\mu} (\nabla_{\nu} \nabla_{\rho} f_G), \label{eq:term1} \\
			\nabla^{\mu} \left( R_{\nu}^{\rho} \nabla_{\mu} \nabla_{\rho} f_G \right) &=& (\nabla^{\mu} R_{\nu}^{\rho}) (\nabla_{\mu} \nabla_{\rho} f_G) + R_{\nu}^{\rho} \nabla^{\mu} (\nabla_{\mu} \nabla_{\rho} f_G), \label{eq:term2} \\
			\nabla^{\mu} \left( R_{\mu\rho\nu\lambda} \nabla^{\rho} \nabla^{\lambda} f_G \right) &=& (\nabla^{\mu} R_{\mu\rho\nu\lambda}) (\nabla^{\rho} \nabla^{\lambda} f_G) + R_{\mu\rho\nu\lambda} \nabla^{\mu} (\nabla^{\rho} \nabla^{\lambda} f_G). \label{eq:term3}
		\end{eqnarray}
		Combining these expansions:
		\begin{eqnarray}
			\nabla^{\mu} T_{4 \mu\nu}^{(G)} &=& \frac{4}{\kappa^2} \left[ (\nabla^{\mu} R_{\mu}^{\rho}) (\nabla_{\nu} \nabla_{\rho} f_G) + R_{\mu}^{\rho} \nabla^{\mu} (\nabla_{\nu} \nabla_{\rho} f_G) \right] \nonumber \\
			&& + \frac{4}{\kappa^2} \left[ (\nabla^{\mu} R_{\nu}^{\rho}) (\nabla_{\mu} \nabla_{\rho} f_G) + R_{\nu}^{\rho} \nabla^{\mu} (\nabla_{\mu} \nabla_{\rho} f_G) \right] \nonumber \\
			&& + \frac{4}{\kappa^2} \left[ (\nabla^{\mu} R_{\mu\rho\nu\lambda}) (\nabla^{\rho} \nabla^{\lambda} f_G) + R_{\mu\rho\nu\lambda} \nabla^{\mu} (\nabla^{\rho} \nabla^{\lambda} f_G) \right]. \label{A:Co_T4G}
		\end{eqnarray}
		
		\item	\subsubsection*{\textbf{Combining The Results:}}
		The complete covariant derivative is given by:
		\begin{align}
			\nabla^{\mu} T_{\mu\nu}^{(G)} = \sum_{i=1}^4 \nabla^{\mu} T_{i\mu\nu}^{(G)}.
		\end{align}
		By combining the eqs. (\ref{A:Co_T1G}), (\ref{A:Co_T2G}), (\ref{A:Co_T3G}), and (\ref{A:Co_T4G}), we have:
		\begin{align}
			\nabla^{\mu}T_{\mu\nu}^{(G)} &=\frac{2}{\kappa^2} \left( R_{\mu\nu} f_G + g_{\mu\nu} \square f_G - \nabla_\mu \nabla_\nu f_G \right) \nabla^{\mu} R + \frac{1}{\kappa^2} R f_G \nabla_{\nu} R\\
			& - \frac{4}{\kappa^2}\left((\nabla^\mu f_G)R_{\mu}^{\rho}R_{\rho\nu} + f_G\left((\nabla^\mu R_{\mu}^{\rho})R_{\rho\nu} + R_{\mu}^{\rho}\nabla^\mu R_{\rho\nu}\right)\right) \nonumber \\
			& - \frac{4}{\kappa^2}\left(\nabla^\mu R_{\mu\nu}\square f_G + R_{\mu\nu}\nabla^\mu(\square f_G) + g_{\mu\nu}\nabla^\mu\left(R^{\rho\lambda}\nabla_{\rho}\nabla_{\lambda}f_G\right)\right) \nonumber \\
			& - \frac{4}{\kappa^2}\left(f_{RG}\nabla^{\mu}R + f_{GG}\nabla^{\mu}G + f_{GT}\nabla^{\mu}T\right)\left(R_{\mu\rho\nu\lambda}R^{\rho\lambda} - \frac{1}{2}R_{\mu}^{\rho\lambda\xi}R_{\nu\rho\lambda\xi}\right) \nonumber \\
			& - \frac{4}{\kappa^2}f_G\left((\nabla^{\mu}R_{\mu\rho\nu\lambda})R^{\rho\lambda} + R_{\mu\rho\nu\lambda}\nabla^{\mu}R^{\rho\lambda} - \frac{1}{2}(\nabla^{\mu}R_{\mu}^{\rho\lambda\xi})R_{\nu\rho\lambda\xi} - \frac{1}{2}R_{\mu}^{\rho\lambda\xi}\nabla^{\mu}R_{\nu\rho\lambda\xi}\right) \nonumber \\
			& + \frac{4}{\kappa^2}\left((\nabla^{\mu}R_{\mu}^{\rho})(\nabla_{\nu}\nabla_{\rho}f_G) + R_{\mu}^{\rho}\nabla^{\mu}(\nabla_{\nu}\nabla_{\rho}f_G) + (\nabla^{\mu}R_{\nu}^{\rho})(\nabla_{\mu}\nabla_{\rho}f_G)\right) \nonumber \\
			& + \frac{4}{\kappa^2}\left(R_{\nu}^{\rho}\nabla^{\mu}(\nabla_{\mu}\nabla_{\rho}f_G) + (\nabla^{\mu}R_{\mu\rho\nu\lambda})(\nabla^{\rho}\nabla^{\lambda}f_G) + R_{\mu\rho\nu\lambda}\nabla^{\mu}(\nabla^{\rho}\nabla^{\lambda}f_G)\right).
		\end{align}
		
		After extensive algebraic manipulation and application of the Bianchi identities, the expression simplifies to:
		\begin{align}
			\nabla^{\mu} T_{\mu\nu}^{(G)} =& \frac{1}{\kappa^2} \left( R f_G - 2(\nabla^\alpha \nabla_\alpha f_G) + 4R^{\alpha\beta} \nabla_\alpha \nabla_\beta f_G \right) \nabla_\nu R \nonumber \\
			& + \frac{4}{\kappa^2} \left( R_{\mu\rho\nu\lambda} \nabla^\mu \nabla^\lambda \nabla^\rho f_G - R_{\nu\rho} \nabla^\rho (\nabla^\alpha \nabla_\alpha f_G) \right) \nonumber \\
			& - \frac{4}{\kappa^2} f_G \left( R^{\rho\lambda} \nabla_\rho R_{\nu\lambda} + \frac{1}{2} R^{\rho\lambda\alpha\beta} \nabla_\nu R_{\rho\lambda\alpha\beta} \right) \nonumber \\
			& + \frac{4}{\kappa^2} \left( f_{GR} \nabla^\mu R + f_{GG} \nabla^\mu G + f_{GT} \nabla^\mu T \right) \left( \frac{1}{2} R_{\mu}^{\rho\lambda\xi} R_{\nu\rho\lambda\xi} - R_{\mu\rho\nu\lambda} R^{\rho\lambda} \right).
			\label{A:Co_TG}
		\end{align}
		where $f_{GR}=\frac{\partial^2 f}{\partial G \,\partial R}$, $f_{GG}=\frac{\partial^2 f}{\partial^2 G}$, and $f_{GR}=\frac{\partial^2 f}{\partial G \,\partial T}$.
	\end{itemize}
	
	\subsection{Covariant Derivative of $\textsl{T}^{(\textsl{T})} $:}
	We begin with the expression for the energy-momentum tensor $T_{\mu\nu}^{(T)}$:  
	\begin{equation}  
		T_{\mu\nu}^{(T)} = -\frac{f_T}{\kappa^2} \left( T_{\mu\nu}^{(m)} + \Theta_{\mu\nu} \right)  
	\end{equation}  
	where  
	\begin{equation}  
		T_{\mu\nu}^{(m)} = \left( \rho + p \right) u_{\mu} u_{\nu} + p g_{\mu\nu}  
	\end{equation}  
	and  
	\begin{equation}  
		\Theta_{\mu\nu} = g^{\alpha\beta} \frac{\partial T_{\alpha\beta}^{(m)}}{\partial g^{\mu\nu}}.  
	\end{equation}  
	
	Next, we take the covariant derivative:  
	\begin{align}  
		\nabla^{\mu} T_{\mu\nu}^{(T)} &= \nabla^{\mu} \left( -\frac{f_T}{\kappa^2} \left( T_{\mu\nu}^{(m)} + \Theta_{\mu\nu} \right) \right) \\
		&= -\frac{1}{\kappa^2} \left( \left( T_{\mu\nu}^{(m)} + \Theta_{\mu\nu} \right)\nabla^{\mu} f_T + f_T \nabla^{\mu} (T_{\mu\nu}^{(m)} + \Theta_{\mu\nu}) \right).\label{A:Co_T(T)}  
	\end{align}  
	Here
	\begin{eqnarray}  
		\nabla^{\mu} f_T = f_{TR} \nabla^{\mu} R + f_{TG} \nabla^{\mu} G + f_{TT} \nabla^{\mu} T 
	\end{eqnarray} 
	where $f_{TR}=\frac{\partial^2 f}{\partial R \,\partial T}$, $f_{TG}=\frac{\partial^2 f}{\partial G\, \partial T}$, and $f_{TT}=\frac{\partial^2 f}{\partial T^2}$.
	
	\subsection{Covariant Derivative of $\textsl{T}^{(\Xi)}$:}
	By using eq.(\ref{A:T(Xi)}), one obtains
	\begin{eqnarray}
		\nabla^{\mu}T_{\mu\nu}^{(\Xi)} = -\dot{\phi} \left( \ddot{\phi} + 3H \dot{\phi} - V_{,\phi} \right) + \dot{\psi} \left( \ddot{\psi} + 3H \dot{\psi} + V_{,\psi} \right).
	\end{eqnarray}
	So, by comparing with the equations of motion (\ref{eq:minimal_phi}) and (\ref{eq:minimal_psi}), we have
	\begin{eqnarray}
		\nabla^{\mu}T_{\mu\nu}^{(\Xi)} = 0. \label{A:Co_T(Xi)}
	\end{eqnarray}

	\section{Substitution of FLRW Expressions into the Covariant Divergence}\label{sec:Sub-FLRW-Cov-Div}
	
	\subsection{FLRW Computation for $\nabla^{\mu}T_{\mu\nu}^{(R)}$}
	
	For the FLRW metric, the conservation equation for the Ricci sector becomes:
	\begin{eqnarray}
		\nabla^{\mu}T_{\mu\nu}^{(R)} = -\frac{1}{2\kappa^2}g_{\mu\nu}\left( f_G\nabla^{\mu}G + f_T\nabla^{\mu}T \right), \label{Co_TR}
	\end{eqnarray}
	
	Considering the temporal component ($\nu = 0$) and using FLRW expressions:
	\begin{align}
		R &= 6(\dot{H} + 2H^2), \label{A:R}\\
		G &= 24H^2(\dot{H} + H^2), \label{A:G}\\
		T &= -\rho + 3p,\label{A:T}
	\end{align}
	
	we compute the derivatives:
	\begin{align}
		\dot{R} &= 6(\ddot{H} + 4H\dot{H}), \label{A:Rdot}\\
		\dot{G} &= 24H^2\ddot{H} + 48H\dot{H}^2 + 96H^3\dot{H}, \label{A:Gdot}\\
		\dot{T} &= -\dot{\rho} + 3\dot{p}.\label{A:Tdot}
	\end{align}
	
	The left-hand side of (\ref{Co_TR}) for $\nu=0$ gives:
	\begin{align}
		\kappa^2\nabla^{\mu}T_{\mu 0}^{(R)} &= 12f_R H\dot{H} + 3f_R\ddot{H} - \frac{\dot{f}}{2} \label{Co_TR_LHS}
	\end{align}
	
	where the total derivative $\dot{f}$ is:
	\begin{align}
		\dot{f} = f_R\dot{R} + f_G\dot{G} + f_T\dot{T}.
	\end{align}
	
	The right-hand side of (\ref{Co_TR}) for $\nu=0$ becomes:
	\begin{align}
		-\frac{1}{2\kappa^2}g_{00}\left( f_G\nabla^{0}G + f_T\nabla^{0}T \right) 
		&= \frac{1}{2\kappa^2}\left( f_G\dot{G} + f_T\dot{T} \right) \nonumber \\
		&= \frac{1}{2\kappa^2}\left[ f_G(24H^2\ddot{H} + 48H\dot{H}^2 + 96H^3\dot{H}) + f_T(-\dot{\rho} + 3\dot{p}) \right]. \label{Co_TR_RHS}
	\end{align}
	
	Equating (\ref{Co_TR_LHS}) and (\ref{Co_TR_RHS}) yields the FLRW conservation relation:
	\begin{align}
		12f_R H\dot{H} + 3f_R\ddot{H} - \frac{1}{2}(f_R\dot{R} + f_G\dot{G} + f_T\dot{T}) 
		&= \frac{1}{2}\left[ f_G(24H^2\ddot{H} + 48H\dot{H}^2 + 96H^3\dot{H}) + f_T(-\dot{\rho} + 3\dot{p}) \right]. \nonumber
	\end{align}
	
	Substituting $\dot{R} = 6(\ddot{H} + 4H\dot{H})$ and simplifying, we obtain the final form:
	\begin{eqnarray}
		\kappa^2\nabla^{\mu}T_{\mu 0}^{(R)} = -12f_G H(H\ddot{H} + 2\dot{H}^2 + 4H^2\dot{H}) - (3\dot{p} - \dot{\rho})\frac{f_T}{2}, \label{Co_TR_H}
	\end{eqnarray}
	
	For the spatial components ($\nu = i$), the right-hand side vanishes due to homogeneity:
	\begin{align}
		\nabla^{\mu}T_{\mu i}^{(R)} &= -\frac{1}{2\kappa^2}g_{\mu i}\left( f_G\nabla^{\mu}G + f_T\nabla^{\mu}T \right) = 0, \label{Co_TR_FLRW_spatial}
	\end{align}
	since $g_{0i} = 0$ and $\nabla^{i}G = \nabla^{i}T = 0$ in the FLRW metric.
	
	This result shows that in FLRW spacetime, the energy-momentum exchange between the geometric and matter sectors occurs only in the temporal direction, consistent with cosmological homogeneity and isotropy.
	
	\subsection{FLRW Computation for $\nabla^{\mu}T_{\mu\nu}^{(G)}$}
	
	Following the computation of $\nabla^{\mu}T_{\mu\nu}^{(R)}$, we now derive the FLRW expression for the Gauss-Bonnet sector. Starting from the general expression:
	\begin{align}
		\nabla^{\mu} T_{\mu\nu}^{(G)} =& \frac{1}{\kappa^2} \left( R f_G - 2(\nabla^\alpha \nabla_\alpha f_G) + 4R^{\alpha\beta} \nabla_\alpha \nabla_\beta f_G \right) \nabla_\nu R \nonumber \\
		& + \frac{4}{\kappa^2} \left( R_{\mu\rho\nu\lambda} \nabla^\mu \nabla^\lambda \nabla^\rho f_G - R_{\nu\rho} \nabla^\rho (\nabla^\alpha \nabla_\alpha f_G) \right) \nonumber \\
		& - \frac{4}{\kappa^2} f_G \left( R^{\rho\lambda} \nabla_\rho R_{\nu\lambda} + \frac{1}{2} R^{\rho\lambda\alpha\beta} \nabla_\nu R_{\rho\lambda\alpha\beta} \right) \nonumber \\
		& + \frac{4}{\kappa^2} \left( f_{GR} \nabla^\mu R + f_{GG} \nabla^\mu G + f_{GT} \nabla^\mu T \right) \left( \frac{1}{2} R_{\mu}^{\rho\lambda\xi} R_{\nu\rho\lambda\xi} - R_{\mu\rho\nu\lambda} R^{\rho\lambda} \right),
		\label{A:Co_TG}
	\end{align}
	
	we substitute the FLRW metric expressions. For computational clarity, we evaluate each term separately for the temporal component ($\nu = 0$).
	
	\subsubsection*{\textbf{I. FLRW Metric Components}}
	In addition to the equations (\ref{A:R}) to (\ref{A:Tdot}) we have:
	\begin{align}
		R_{00} &= -3(\dot{H} + H^2), \quad
		R_{ij} = a^2(3H^2 + 2\dot{H})\delta_{ij}, \\
		R^{00} &= -3H^2, \quad
		R^{ij} = 3H^2 a^{-2}\delta^{ij}, \\
		\Box f_G &= -\ddot{f}_G - 3H\dot{f}_G.
	\end{align}
	
	\subsubsection*{\textbf{II. Term-by-Term Evaluation}}
	
	\textbf{Term 1:} $\frac{1}{\kappa^2} \left( R f_G - 2\Box f_G + 4R^{\alpha\beta} \nabla_\alpha \nabla_\beta f_G \right) \nabla_\nu R$
	
	\begin{align}
		R^{\alpha\beta} \nabla_\alpha \nabla_\beta f_G &= R^{00} \nabla_0 \nabla_0 f_G = (-3H^2)(\ddot{f}_G), \\
		\nabla_0 R &= \dot{R} = 6(\ddot{H} + 4H\dot{H}).
	\end{align}
	
	Thus:
	\begin{align}
		\text{Term 1} =& \frac{1}{\kappa^2} \left[ 6(\dot{H} + 2H^2) f_G - 2(-\ddot{f}_G - 3H\dot{f}_G) + 4(-3H^2)(\ddot{f}_G) \right] \cdot 6(\ddot{H} + 4H\dot{H}) \delta_{\nu}^{0} \nonumber \\
		=& \frac{6(\ddot{H} + 4H\dot{H})}{\kappa^2} \left[ 6(\dot{H} + 2H^2) f_G + 2\ddot{f}_G + 6H\dot{f}_G - 12H^2 \ddot{f}_G \right] \delta_{\nu}^{0}.
	\end{align}
	
	\textbf{Term 2:} $\frac{4}{\kappa^2} R_{\mu\rho\nu\lambda} \nabla^\mu \nabla^\lambda \nabla^\rho f_G$
	
	This term vanishes identically in FLRW due to the symmetry properties of the Riemann tensor and homogeneity of the spacetime.
	
	\textbf{Term 3:} $-\frac{4}{\kappa^2} R_{\nu\rho} \nabla^\rho (\Box f_G)$
	
	For $\nu = 0$:
	\begin{align}
		R_{0\rho} \nabla^\rho (\Box f_G) &= R_{00} \nabla^0 (\Box f_G) = (-3H^2)\partial_t(-\ddot{f}_G - 3H\dot{f}_G) \\
		&= 3H^2 (\dddot{f}_G + 3\dot{H}\dot{f}_G + 3H\ddot{f}_G).
	\end{align}
	
	Thus:
	\begin{align}
		\text{Term 3} = -\frac{4}{\kappa^2} \cdot 3H^2 (\dddot{f}_G + 3\dot{H}\dot{f}_G + 3H\ddot{f}_G) \delta_{\nu}^{0} = -\frac{12H^2}{\kappa^2} (\dddot{f}_G + 3\dot{H}\dot{f}_G + 3H\ddot{f}_G) \delta_{\nu}^{0}.
	\end{align}
	
	\textbf{Term 4:} $-\frac{4}{\kappa^2} f_G R^{\rho\lambda} \nabla_\rho R_{\nu\lambda}$
	
	For $\nu = 0$:
	\begin{align}
		R^{\rho\lambda} \nabla_\rho R_{0\lambda} &= R^{00} \nabla_0 R_{00} = (-3H^2)\partial_t[-3(\dot{H} + H^2)] \\
		&= (-3H^2)[-3(\ddot{H} + 2H\dot{H})] = 9H^2(\ddot{H} + 2H\dot{H}).
	\end{align}
	
	Thus:
	\begin{align}
		\text{Term 4} = -\frac{4}{\kappa^2} f_G \cdot 9H^2(\ddot{H} + 2H\dot{H}) \delta_{\nu}^{0} = -\frac{36H^2}{\kappa^2} f_G (\ddot{H} + 2H\dot{H}) \delta_{\nu}^{0}.
	\end{align}
	
	\textbf{Term 5:} $-\frac{2}{\kappa^2} f_G R^{\rho\lambda\alpha\beta} \nabla_\nu R_{\rho\lambda\alpha\beta}$
	
	This term vanishes due to homogeneity of FLRW spacetime.
	
	\textbf{Term 6:} The mixed derivative term
	
	\begin{align}
		&\frac{4}{\kappa^2} \left( f_{GR} \nabla^\mu R + f_{GG} \nabla^\mu G + f_{GT} \nabla^\mu T \right) \left( \frac{1}{2} R_{\mu}^{\rho\lambda\xi} R_{\nu\rho\lambda\xi} - R_{\mu\rho\nu\lambda} R^{\rho\lambda} \right)
	\end{align}
	
	For FLRW metric:
	\begin{align}
		R_{0}^{\rho\lambda\xi} R_{0\rho\lambda\xi} &= 0, \\
		R_{0\rho0\lambda} R^{\rho\lambda} &= R_{0i0j} R^{ij} = [-a^2(\dot{H} + H^2)\delta_{ij}] [3H^2 a^{-2}\delta^{ij}] = -3H^2(\dot{H} + H^2).
	\end{align}
	
	Thus:
	\begin{align}
		\text{Term 6} =& \frac{4}{\kappa^2} \left( f_{GR} \dot{R} + f_{GG} \dot{G} + f_{GT} \dot{T} \right) \left( 0 - [-3H^2(\dot{H} + H^2)] \right) \delta_{\nu}^{0} \nonumber \\
		=& \frac{12H^2(\dot{H} + H^2)}{\kappa^2} \left( f_{GR} \dot{R} + f_{GG} \dot{G} + f_{GT} \dot{T} \right) \delta_{\nu}^{0}.
	\end{align}
	
	\subsubsection*{\textbf{III. Final Combined Expression}}
	
	Combining all non-vanishing terms for the temporal component ($\nu = 0$):
	\begin{align}
		\nabla^{\mu} T_{\mu0}^{(G)} =& \frac{6(\ddot{H} + 4H\dot{H})}{\kappa^2} \left[ 6(\dot{H} + 2H^2) f_G + 2\ddot{f}_G + 6H\dot{f}_G - 12H^2 \ddot{f}_G \right] \nonumber \\
		& - \frac{12H^2}{\kappa^2} (\dddot{f}_G + 3\dot{H}\dot{f}_G + 3H\ddot{f}_G) \nonumber \\
		& - \frac{36H^2}{\kappa^2} f_G (\ddot{H} + 2H\dot{H}) \nonumber \\
		& + \frac{12H^2(\dot{H} + H^2)}{\kappa^2} \left( f_{GR} \dot{R} + f_{GG} \dot{G} + f_{GT} \dot{T} \right).
	\end{align}
	
	This complete expression for $\nabla^{\mu} T_{\mu0}^{(G)}$ in FLRW spacetime will be used in the main text to analyze energy-momentum conservation in our cosmological models.

	\section{Hamiltonian Formulation of $f(R,G,T)$ Gravity}\label{sec:Hamiltonian}
	
	\subsection{Canonical Momentum Expressions}
	
	The Hamiltonian analysis begins with the conjugate momentum calculations following standard results in modified gravity theories \cite{DeFelice2010, Sotiriou2010}. The complete derivation proceeds as:
	
	\begin{align}
		\pi^{ij} &= \frac{\delta\mathcal{L}}{\delta\dot{g}_{ij}} = \frac{\sqrt{-g}}{2\kappa^2}\left[f_R\frac{\delta R}{\delta\dot{g}_{ij}} + f_G\frac{\delta G}{\delta\dot{g}_{ij}} + f_T\frac{\delta T}{\delta\dot{g}_{ij}}\right] \label{eq:full_pi} \\
		&= \frac{\sqrt{-g}}{2\kappa^2}\left[f_R(K^{ij} - Kg^{ij}) + 4f_G(K^{ik}K_k^j - KK^{ij} + \cancel{\frac{R}{2}K^{ij}} - \cancel{\frac{1}{2}(K^2 - K^{mn}K_{mn})g^{ij}})\right] \label{eq:expanded_pi} \\
		&= \underbrace{\frac{\sqrt{-g}}{2\kappa^2}f_R(K^{ij} - Kg^{ij})}_{\text{Einstein-Hilbert part}} + \underbrace{\frac{\sqrt{-g}}{2\kappa^2}4f_G(K^{ik}K_k^j - KK^{ij})}_{\text{Gauss-Bonnet correction}} \label{eq:final_pi}
	\end{align}
	where $K_{ij} = \frac{1}{2N}(\dot{g}_{ij} - \nabla_i N_j - \nabla_j N_i)$ is the extrinsic curvature.
	
	For the scalar fields:
	\begin{align}
		\pi_\phi &= -\sqrt{-g}\dot{\phi}\quad \text{(phantom field)} \\
		\pi_\psi &= \sqrt{-g}\dot{\psi}\quad \text{(canonical field)}
	\end{align}
	
	\subsection{Key Simplifications in Momentum Expressions}
	
	\textbf{I. Matter Coupling Term:}
	The $f_T$ term vanishes because:
	\begin{equation*}
		\frac{\delta T}{\delta\dot{g}_{ij}} = 0
	\end{equation*}
	as the matter stress-energy tensor $T$ depends on $g_{ij}$ but not on $\dot{g}_{ij}$ \cite{Harko2011}.
	
	\textbf{II. Gauss-Bonnet Decomposition:}
	The full variation of $G$ contains:
	\begin{equation*}
		\frac{\delta G}{\delta\dot{g}_{ij}} = 4(K^{ik}K_k^j - KK^{ij}) + \text{terms proportional to $R$ and $K^2$}
	\end{equation*}
	The cancellations occur because:
	\begin{itemize}
		\item The $R$-dependent terms cancel with $K^2$ terms in equations of motion \cite{Nojiri2005}
		\item Gauss-Bonnet is topological in 4D, leaving only boundary terms \cite{Charmousis2008}
	\end{itemize}
	
	\textbf{III. Extrinsic Curvature Relations:}
	Using ADM decomposition \cite{Arnowitt1962}:
	\begin{equation*}
		\frac{\delta R}{\delta\dot{g}_{ij}} = K^{ij} - Kg^{ij}
	\end{equation*}
	
	\subsection{Constraint Structure and Degree of Freedom Counting}
	
	\textbf{Primary constraints:}
	\begin{align}
		\Phi_0 &= \pi^0 - \frac{\delta\mathcal{L}}{\delta\dot{N}} \approx 0 \quad \text{(Lapse function)} \\
		\Phi_i &= \pi_i - \frac{\delta\mathcal{L}}{\delta\dot{N}^i} \approx 0 \quad \text{(Shift vectors)}
	\end{align}
	
	\textbf{Secondary constraints:} Four additional constraints emerge from time conservation of $\Phi_\mu$, corresponding to diffeomorphism invariance.
	
	\textbf{Degrees of freedom counting} using Dirac's formula:
	\begin{align}
		\text{Total DOF} &= \frac{1}{2}\left(\text{Phase space vars} - \text{Constraints}\right) \\
		&= \frac{1}{2}\left(2\times(10\ \text{metric} + 2\ \text{fields}) - 8\ \text{constraints}\right) = 8 \ \text{(phase space)}
	\end{align}
	
	In configuration space, the physical DOF is 4:
	\begin{itemize}
		\item \textbf{2 Tensor modes}: Transverse-traceless graviton polarizations ($h_+$, $h_\times$)
		\item \textbf{2 Scalar modes}: Quintom fields $\phi$ (phantom) and $\psi$ (canonical)
	\end{itemize}
	
	\subsection{Ostrogradsky Instability Avoidance}
	
	The higher-derivative terms in $f(R,G,T)$ could introduce Ostrogradsky instabilities unless degeneracy conditions are met. The key condition is:
	\begin{align}
		\det\begin{pmatrix}
			f_{RR} & f_{RG} \\ 
			f_{GR} & f_{GG}
		\end{pmatrix} = 0
	\end{align}
	
	For linear Gauss-Bonnet ($f_G = \text{const.}$), this holds automatically since $f_{GG} = f_{RG} = 0$. The FLRW symmetry provides additional protection by constraining the higher-derivative structure.
	
	\subsection{Physical Interpretation}
	\begin{itemize}
		\item The first term in \eqref{eq:final_pi} represents usual gravitational momentum from Einstein-Hilbert gravity \cite{Misner1973}
		\item The second term encodes Gauss-Bonnet modifications, which vanish when $f_G = \text{const.}$ (total derivative) \cite{DeFelice2010} but become dynamical when $f_G$ depends on other fields \cite{Nojiri2005}
		\item All other terms cancel or become boundary terms in closed universes \cite{Sotiriou2010}
	\end{itemize}
	
	\section{Cosmological Perturbation Theory}\label{sec:Perturbations}
	
	\subsection{Perturbation Variables in FLRW Metric}
	
	The general perturbed FLRW metric contains four scalar perturbation functions following standard cosmological perturbation theory \cite{Mukhanov1992, Kodama1984, Ma1995}:
	
	\begin{itemize}
		\item \textbf{$\Phi(t,\mathbf{x})$} - Newtonian potential:
		\begin{itemize}
			\item Governs time-time component of metric perturbations
			\item Represents gravitational potential in Newtonian limit
			\item Affects particle energies ($E \approx \sqrt{g_{00}}m$)
		\end{itemize}
		
		\item \textbf{$B(t,\mathbf{x})$} - Shift vector potential:
		\begin{itemize}
			\item Appears in time-space components $g_{0i}$
			\item Describes frame-dragging effects
			\item Vanishes in Newtonian gauge
		\end{itemize}
		
		\item \textbf{$\Psi(t,\mathbf{x})$} - Spatial curvature potential:
		\begin{itemize}
			\item Controls scalar curvature perturbations
			\item In absence of anisotropic stress, $\Psi = \Phi$ (GR) \cite{Bardeen1980}
			\item Measures local volume distortion
		\end{itemize}
		
		\item \textbf{$E(t,\mathbf{x})$} - Shear potential:
		\begin{itemize}
			\item Appears in space-space components $g_{ij}$
			\item Describes anisotropic deformations
			\item Can be gauged away in Newtonian gauge
		\end{itemize}
	\end{itemize}
	
	\subsection{Gauge Fixing to Newtonian Gauge}
	
	The general perturbed metric \cite{Mukhanov1992}:
	\begin{equation}
		ds^2 = -(1+2\Phi)dt^2 + 2a\partial_iB dx^i dt + a^2[(1-2\Psi)\delta_{ij} + 2\partial_i\partial_jE]dx^i dx^j
	\end{equation}
	
	can be simplified to Newtonian (longitudinal) gauge \cite{Ma1995, Mukhanov1992}:
	\begin{equation}
		ds^2 = -(1+2\Phi)dt^2 + a^2(1-2\Psi)\delta_{ij}dx^i dx^j
	\end{equation}
	
	through gauge transformations:
	
	\textbf{Step 1: Coordinate Transformation} \cite{Bardeen1980}:
	\begin{align}
		t &\rightarrow t + \alpha(t,\mathbf{x}) \\
		x^i &\rightarrow x^i + \delta^{ij}\partial_j\beta(t,\mathbf{x})
	\end{align}
	
	\textbf{Step 2: Gauge Condition Imposition} \cite{Ma1995}:
	\begin{align}
		B + \dot{E} &= 0 \quad \text{(Eliminates $g_{0i}$ terms)} \\
		E &= 0 \quad \text{(Eliminates anisotropic space-space perturbations)}
	\end{align}
	
	\textbf{Step 3: Resulting Metric Form} \cite{Mukhanov1992}:
	\begin{itemize}
		\item Vanishing $B$ (since $B = -\dot{E}$ and $E=0$)
		\item Diagonal spatial perturbations ($E=0$ removes $\partial_i\partial_jE$)
		\item Preserved $\Phi$ and $\Psi$ as gauge-invariant potentials \cite{Bardeen1980}
	\end{itemize}
	
	\subsection{Physical Interpretation of Newtonian Gauge}
	
	The simplified metric describes \cite{Weinberg2008, Dodelson2003}:
	\begin{itemize}
		\item \textbf{Time dilation}: Governed by $\Phi$ (Newtonian potential analog)
		\item \textbf{Spatial curvature}: Governed by $\Psi$ (volume changes)
		\item \textbf{Removed degrees}:
		\begin{itemize}
			\item $B$ eliminated (no frame-dragging)
			\item $E$ eliminated (no anisotropic deformation)
		\end{itemize}
	\end{itemize}
	
	\begin{equation}
		\boxed{\text{Newtonian gauge: cosmological analog of } \Phi_{\text{Newton}} \text{ in weak-field GR}}
	\end{equation}
	
	This gauge choice provides the foundation for our perturbation analysis in Section \ref{sec:perturbation_stability}, where we examine stability conditions throughout the bounce evolution.
	
	\section{Derivation of Modified Friedmann Equations}
	\label{sec:Derivation of Modified Friedmann Equations}
	
	\subsection{Exponential Function of Curvature}
	For $f(R,G,T) = R - 2\Lambda e^{-(R_0/R)^n} + \xi_2 G + \xi_3 T$, we compute the partial derivatives:
	\begin{align}
		f_R &= 1 - 2n\Lambda u e^{-u}, \quad f_G = \xi_2, \quad f_T = \xi_3 \\
		\dot{f}_R &= -2n\Lambda u e^{-u}\left[\frac{n}{R} u - \frac{n+1}{R}\right]\dot{R}, \\
		\ddot{f}_R &= 2n^2\Lambda u e^{-u} \left[(1 - u) \left(\frac{\ddot{R}}{R} - (n + 1) \frac{\dot{R}^2}{R^2}\right) + n u \frac{\dot{R}^2}{R^2} (2 - u)\right]\\
		\dot{f}_G &= 0, \quad \ddot{f}_G = 0
	\end{align}
	where $u = \left(\frac{R_0}{R}\right)^n$.
	
	Substituting into the general effective Friedmann equations (\ref{eq:f1-E}) and (\ref{eq:f2-E}), the modified Friedmann equations become:
	\begin{align}
		3H^2 &= \frac{1}{f_R}\bigg[\underbrace{\kappa^2(\rho + \rho_\Xi) + \xi_3(\rho + p)}_{\text{Standard + matter coupling}} - \underbrace{3H\dot{f}_R + 3\dot{H}f_R}_{\text{Dynamic curvature}} - \underbrace{\frac{1}{2}\left(R - 2\Lambda e^{-(R_0/R)^n} + \xi_2 G + \xi_3 T\right)}_{\text{Effective dark energy}} \nonumber \\
		&\quad + \underbrace{12\xi_2 H^2(H^2 + \dot{H})}_{\text{Gauss-Bonnet contributions}}\bigg] \label{f1_Rec_exp_corrected}
	\end{align}
	\begin{align}
		-2\dot{H} - 3H^2 &= \frac{1}{f_R}\bigg[\underbrace{\kappa^2(p + p_\Xi)}_{\text{Standard terms}} + \underbrace{\ddot{f}_R + 2H\dot{f}_R + \dot{H}f_R}_{\text{Curvature dynamics}} + \underbrace{\frac{1}{2}\left(R - 2\Lambda e^{-(R_0/R)^n} + \xi_2 G + \xi_3 T\right)}_{\text{Geometric pressure}} \nonumber \\
		&\quad - \underbrace{12\xi_2 H^2(H^2 + \dot{H})}_{\text{Gauss-Bonnet contributions}}\bigg] \label{f2_Rec_exp_corrected}
	\end{align}
	
	Since $\dot{f}_G = \ddot{f}_G = 0$, the simplified equations are:
	\begin{align}
		3H^2 &= \frac{1}{1 - 2n\Lambda\left(\frac{R_0}{R}\right)^n e^{-(R_0/R)^n}}\bigg[\underbrace{\kappa^2(\rho + \rho_\Xi) + \xi_3(\rho + p)}_{\text{Matter and coupling}} \nonumber \\
		&\quad - \underbrace{3H\dot{f}_R + 3\dot{H}\left(1 - 2n\Lambda\left(\frac{R_0}{R}\right)^n e^{-(R_0/R)^n}\right)}_{\text{Dynamic curvature}} \nonumber \\
		&\quad - \underbrace{\frac{1}{2}\left(R - 2\Lambda e^{-(R_0/R)^n} + \xi_2 G + \xi_3 T\right)}_{\text{Effective cosmological constant}} + \underbrace{12\xi_2 H^2(H^2 + \dot{H})}_{\text{Gauss-Bonnet term}}\bigg] \label{f1_Rec_exp_simplified}
	\end{align}
	\begin{align}
		-2\dot{H} - 3H^2 &= \frac{1}{1 - 2n\Lambda\left(\frac{R_0}{R}\right)^n e^{-(R_0/R)^n}}\bigg[\underbrace{\kappa^2(p + p_\Xi)}_{\text{Standard pressure}} \nonumber \\
		&\quad + \underbrace{\ddot{f}_R + 2H\dot{f}_R + \dot{H}\left(1 - 2n\Lambda\left(\frac{R_0}{R}\right)^n e^{-(R_0/R)^n}\right)}_{\text{Curvature acceleration}} \nonumber \\
		&\quad + \underbrace{\frac{1}{2}\left(R - 2\Lambda e^{-(R_0/R)^n} + \xi_2 G + \xi_3 T\right)}_{\text{Geometric stiff matter}} - \underbrace{12\xi_2 H^2(H^2 + \dot{H})}_{\text{Gauss-Bonnet term}}\bigg] \label{f2_Rec_exp_simplified}
	\end{align}
	
	Using $R = 6(\dot{H} + 2H^2)$, $G = 24H^2(\dot{H} + H^2)$, and $T = -\rho + 3p$, we observe that the Gauss-Bonnet terms exactly cancel between the geometric and matter sectors, yielding the final simplified form:
	\begin{align}
		3H^2 &= \frac{1}{f_R}\bigg[\kappa^2(\rho + \rho_\Xi) + \frac{\xi_3}{2}(3\rho - p) - 3H\dot{f}_R + 3\dot{H}(f_R - 1) - 6H^2 + \Lambda e^{-(R_0/R)^n} \bigg]
		\label{f1_Rec_exp_final}\\
		-2\dot{H} - 3H^2 &= \frac{1}{f_R}\bigg[\kappa^2(p + p_\Xi) + \frac{\xi_3}{2}(3p - \rho)+ \ddot{f}_R + 2H\dot{f}_R + \dot{H}f_R + 3(\dot{H} + 2H^2) - \Lambda e^{-(R_0/R)^n} \bigg]
		\label{f2_Rec_exp_final}
	\end{align}

	\subsection{Power-Law Modified Gravity}
	For $f(R,G,T) = R + \xi_1 R^n + \xi_2 G + \xi_3 T^m$, we compute the partial derivatives:
	\begin{align}
		f_R &= 1 + n\xi_1 R^{n-1}, \quad f_G = \xi_2, \quad f_T = m\xi_3 T^{m-1} \\
		\dot{f}_R &= n(n-1)\xi_1 R^{n-2}\dot{R}, \quad \ddot{f}_R = n(n-1)\xi_1 R^{n-3}[(n-2)\dot{R}^2 + R\ddot{R}] \\
		\dot{f}_G &= 0, \quad \ddot{f}_G = 0
	\end{align}
	
	Substituting into the general effective Friedmann equations (\ref{eq:f1-E}) and (\ref{eq:f2-E}), we obtain the equations:
	\begin{align}
		3H^2 &= \frac{1}{1 + n\xi_1 R^{n-1}}\bigg[\kappa^2(\rho + \rho_\Xi) + 3n(n-1)\xi_1 R^{n-2}\dot{R}H + 3n\xi_1 R^{n-1}\dot{H} \nonumber \\
		&\quad - \frac{1}{2}(R + \xi_1 R^n + \xi_2 G + \xi_3 T^m) + m\xi_3 T^{m-1}(\rho + p) \nonumber \\
		&\quad + 12\xi_2 H^2(H^2 + \dot{H})\bigg] \label{f1_Rec_PL_corrected}
	\end{align}
	\begin{align}
		-2\dot{H} - 3H^2 &= \frac{1}{1 + n\xi_1 R^{n-1}}\bigg[\kappa^2(p + p_\Xi) + n(n-1)\xi_1 R^{n-2}[(n-2)\dot{R}^2 + R\ddot{R}] \nonumber \\
		&\quad + 2n(n-1)\xi_1 R^{n-2}\dot{R}H + n\xi_1 R^{n-1}\dot{H} + \frac{1}{2}(R + \xi_1 R^n + \xi_2 G + \xi_3 T^m) \nonumber \\
		&\quad - 12\xi_2 H^2(H^2 + \dot{H})\bigg] \label{f2_Rec_PL_corrected}
	\end{align}
	
	Since $\dot{f}_G = \ddot{f}_G = 0$ for this model, and using $R = 6(\dot{H} + 2H^2)$, $G = 24H^2(\dot{H} + H^2)$, $T = -\rho + 3p$, the equations simplify to:
	\begin{align}
		3H^2 &= \frac{1}{1 + n\xi_1 R^{n-1}}\bigg[\underbrace{\kappa^2(\rho + \rho_\Xi)}_{\text{Standard terms}} + \underbrace{3n\xi_1 R^{n-1}\dot{H}}_{\text{Curvature-driven inertia}} + \underbrace{3n(n-1)\xi_1 R^{n-2}\dot{R}H}_{\text{Dynamic curvature}} \nonumber \\
		&\quad - \underbrace{\frac{1}{2}(R + \xi_1 R^n + \xi_3 T^m)}_{\text{Effective dark energy}} + \underbrace{m\xi_3 T^{m-1}(\rho + p)}_{\text{Non-linear matter coupling}} \bigg] \label{f1_Rec_PL}
	\end{align}
	\begin{align}
		-2\dot{H} - 3H^2 &= \frac{1}{1 + n\xi_1 R^{n-1}}\bigg[\underbrace{\kappa^2(p + p_\Xi)}_{\text{Standard terms}} + \underbrace{n\xi_1 R^{n-1}\dot{H}}_{\text{Curvature pressure}} + \underbrace{2n(n-1)\xi_1 R^{n-2}\dot{R}H}_{\text{Dynamic curvature}} \nonumber \\
		&\quad + \underbrace{n(n-1)\xi_1 R^{n-2}[(n-2)\dot{R}^2 + R\ddot{R}]}_{\text{Curvature acceleration}} + \underbrace{\frac{1}{2}(R + \xi_1 R^n  + \xi_3 T^m)}_{\text{Geometric stiff matter}}\bigg] \label{f2_Rec_PL}
	\end{align}
	
	\subsection{Modified Teleparallel Gravity Cosmology}
	For the specific form $f(R,G,T) = R + \xi_2 G + \xi_3 T^2$, we compute the partial derivatives:
	\begin{align}
		f_R &= 1, \quad f_G = \xi_2, \quad f_T = 2\xi_3 T \\
		\dot{f}_R &= 0, \quad \ddot{f}_R = 0, \quad \dot{f}_G = 0, \quad \ddot{f}_G = 0
	\end{align}
	
	Substituting into the general effective Friedmann equations (\ref{eq:f1-E}) and (\ref{eq:f2-E}), and using $T = -\rho + 3p$, we obtain:
	\begin{align}
		3H^2 &= \kappa^2(\rho + \rho_\Xi) + \xi_3\left[(3\rho - p)(-\rho + 3p) - \frac{1}{2}(-\rho + 3p)^2\right] \nonumber \\
		&\quad - 12\xi_2 H^2(H^2 + \dot{H}) \label{f1_Rec_MTG_corrected} \\
		-2\dot{H} - 3H^2 &= \kappa^2(p + p_\Xi) + \xi_3\left[(3p - \rho)(-\rho + 3p) + \frac{1}{2}(-\rho + 3p)^2\right] \nonumber \\
		&\quad + 12\xi_2 H^2(H^2 + \dot{H}) \label{f2_Rec_MTG_corrected}
	\end{align}
	
	Simplifying the torsion-matter coupling terms:
	\begin{align}
		\xi_3\left[(3\rho - p)(-\rho + 3p) - \frac{1}{2}(-\rho + 3p)^2\right] &= -\frac{\xi_3}{2}(5\rho^2 - 14\rho p - 3p^2) \\
		\xi_3\left[(3p - \rho)(-\rho + 3p) + \frac{1}{2}(-\rho + 3p)^2\right] &= -\frac{\xi_3}{2}(9p^2 - 6\rho p + \rho^2)
	\end{align}
	
	Thus, the final modified Friedmann equations become:
	\begin{align}
		3H^2 &= \underbrace{\kappa^2(\rho + \rho_\Xi)}_{\text{Standard terms}} - \underbrace{\frac{\xi_3}{2}(5\rho^2 - 14\rho p - 3p^2)}_{\text{Torsion-matter interaction}} - \underbrace{12\xi_2 H^2(H^2 + \dot{H})}_{\text{Gauss-Bonnet contribution}} \label{f1_Rec_MTG} \\
		-2\dot{H} - 3H^2 &= \underbrace{\kappa^2(p + p_\Xi)}_{\text{Standard terms}} - \underbrace{\frac{\xi_3}{2}(9p^2 - 6\rho p + \rho^2)}_{\text{Torsion pressure}} + \underbrace{12\xi_2 H^2(H^2 + \dot{H})}_{\text{Gauss-Bonnet contribution}} \label{f2_Rec_MTG}
	\end{align}
	
	\subsection{Non-Minimal Curvature-Matter Coupling}
	
	For $f(R,G,T) = R + \xi_2 G + \xi_3 R^2 T$, we compute the partial derivatives:
	\begin{align}
		f_R &= 1 + 2\xi_3 R T, \quad f_G = \xi_2, \quad f_T = \xi_3 R^2 \\
		\dot{f}_R &= 2\xi_3(\dot{R}T + R\dot{T}), \quad \ddot{f}_R = 2\xi_3(\ddot{R}T + 2\dot{R}\dot{T} + R\ddot{T}) \\
		\dot{f}_G &= 0, \quad \ddot{f}_G = 0
	\end{align}
	
	Substituting into the general effective Friedmann equations (\ref{eq:f1-E}) and (\ref{eq:f2-E}):
	\begin{align}
		3H^2 &= \frac{1}{1 + 2\xi_3 R T}\bigg[\underbrace{\kappa^2(\rho + \rho_\Xi) + \xi_3 R^2 (\rho + p)}_{\text{Matter and coupling}} - \underbrace{3H\dot{f}_R + 3\dot{H}(1 + 2\xi_3 R T)}_{\text{Dynamic curvature}} \nonumber \\
		&- \underbrace{\frac{1}{2}\left(R + \xi_2 G + \xi_3 R^2 T\right)}_{\text{Effective dark energy}} + \underbrace{12\xi_2 H^2(H^2 + \dot{H})}_{\text{Gauss-Bonnet term}}\bigg] \label{f1_Rec_NMC_corrected}
	\end{align}
	\begin{align}
		-2\dot{H} - 3H^2 &= \frac{1}{1 + 2\xi_3 R T}\bigg[\underbrace{\kappa^2(p + p_\Xi)}_{\text{Standard pressure}} + \underbrace{\ddot{f}_R + 2H\dot{f}_R + \dot{H}(1 + 2\xi_3 R T)}_{\text{Curvature acceleration}} \nonumber \\
		&+ \underbrace{\frac{1}{2}\left(R + \xi_2 G + \xi_3 R^2 T\right)}_{\text{Geometric stiff matter}} - \underbrace{12\xi_2 H^2(H^2 + \dot{H})}_{\text{Gauss-Bonnet term}}\bigg] \label{f2_Rec_NMC_corrected}
	\end{align}
	
	\section{Derivation of Stability Coefficients $\mathcal{G}_S$ and $\mathcal{F}_S$}\label{sec:Derivation_Stability_Coefficients}
	In this appendix, we outline the derivation of the kinetic coefficient $\mathcal{G}_S$ and the gradient coefficient $\mathcal{F}_S$ for the comoving curvature perturbation $\zeta$ in the context of $f(R,G,T)$ gravity. The calculation follows the methods developed in \cite{DeFelice2010, Sotiriou2010, Lin2011} for $f(R,G)$ theories, extended here to include the $f_T$ coupling and quintom fields.
	
	The perturbed FLRW metric in Newtonian gauge is:
	\begin{equation}
		ds^2 = -(1+2\Phi)dt^2 + a^2(1-2\Psi)\delta_{ij}dx^i dx^j.
	\end{equation}
	For simplicity, we focus on the scalar sector and assume no anisotropic stress ($\Phi = \Psi$) at the level of background calculations, which is valid for the minimally coupled quintom fields and perfect fluid \cite{Mukhanov1992}.
	
	The quadratic action for the curvature perturbation $\zeta$ can be derived from the second-order expansion of the action (\ref{eq:action}). The relevant terms come from the Einstein-Hilbert part, the Gauss-Bonnet contribution, and the $f_T$ coupling. After integrating out the non-dynamical fields (such as the lapse and shift perturbations) using the constraint equations, one arrives at the reduced action for $\zeta$.
	
	The kinetic term $\mathcal{G}_S$ receives contributions from:
	\begin{enumerate}
		\item The $f_R$ term: standard contribution from $R$ giving $f_R$.
		\item The $f_G$ term: contributions from $G$ yielding $12H^2 f_G + 8H\dot{f}_G + 4\ddot{f}_G$.
		\item The $f_T$ term: coupling to matter perturbations leading to $-\frac{f_T(\rho+p)}{2H^2 + \epsilon}$.
	\end{enumerate}
	The gradient term $\mathcal{F}_S$ similarly receives contributions from spatial derivatives. A detailed calculation, which is lengthy but straightforward, yields the expressions in Eqs.~(\ref{eq:G_s}) and (\ref{eq:F_s}). The reader is referred to \cite{DeFelice2010} for a pedagogical derivation in the context of $f(R,G)$ gravity.
	
\section{Attractor Behavior of the Bounce Solution}\label{sec:Attractor_Analysis}

A legitimate concern in bouncing cosmologies is whether the conditions required for a bounce ($H=0$, $\ddot{a}>0$) represent a finely tuned set of initial data—i.e., a set of measure zero in the phase space of initial conditions at some early time in the contracting branch. In this appendix, we provide analytical arguments demonstrating that, within our $f(R,G,T)$-quintom framework, the bounce solution acts as a \textit{local attractor}, rendering the bounce generic for a finite range of initial conditions.

\subsection{Linearized Dynamics Around the Bounce}

The modified Friedmann equations (\ref{f111}) and (\ref{f222}) can be combined to yield a second-order equation for the Hubble parameter $H(t)$. For simplicity, we consider the behavior near the bounce point $t=0$, where $H=0$ and $\dot{H} > 0$ (since $\ddot{a} > 0$ implies $\dot{H} > 0$ at $H=0$). Expanding the dynamical equations to linear order in $H$ and $\dot{H}$ around the bounce, we obtain:

\begin{equation}
\ddot{H} = \beta \dot{H} + \mathcal{O}(H^2, H\dot{H}, \dot{H}^2),
\label{eq:linearized_H}
\end{equation}

where $\beta$ is a model-dependent coefficient that depends on the background values of the quintom fields and the $f(R,G,T)$ parameters. This linearization is valid for sufficiently small $H$ and $\dot{H}$, which is precisely the regime near the bounce.

The coefficient $\beta$ can be expressed in terms of the effective energy-momentum components. From the Raychaudhuri equation:
\begin{equation}
\dot{H} = -\frac{\kappa^2}{2}(\rho_{\text{eff}} + p_{\text{eff}}),
\end{equation}
we can differentiate with respect to time and evaluate at $H=0$. After straightforward algebra, one finds:
\begin{equation}
\beta = -3H_{\text{bounce}}^{(2)} + \frac{f_T(\rho+p) + \kappa^2(\dot{\phi}^2 - \dot{\psi}^2)}{2(1 + f_R - 4\ddot{f}_G)}\Bigg|_{t=0},
\label{eq:beta_expression}
\end{equation}
where $H_{\text{bounce}}^{(2)} = \dot{H}(0) > 0$.

\subsection{Stability Condition}

Equation (\ref{eq:linearized_H}) describes a damped (or anti-damped) harmonic oscillator. For the bounce to be a local attractor, we require that small deviations $\delta H$ decay as the system approaches the bounce. This translates to the condition $\beta < 0$. When $\beta < 0$, the general solution is:
\begin{equation}
\delta H(t) = A e^{\beta t} + B,
\end{equation}
which decays exponentially toward the fixed point as $t \to 0^-$ (from the contracting side).

\subsection{Numerical Verification for All Models}

For each of the five models analyzed in Section~\ref{sec:DBCM}, we numerically evaluate $\beta$ using the parameters and initial conditions specified. The results are summarized in Table~\ref{tab:beta_values}:

\begin{table}[h]
\centering
\begin{tabular}{|l|c|c|}
\hline
\textbf{Model} & $\beta$ & \textbf{Attractor?} \\
\hline
Linear Coupling & $-2.34$ & Yes \\
Exponential & $-1.87$ & Yes \\
Power-Law & $-3.12$ & Yes \\
Teleparallel & $-2.01$ & Yes \\
Non-Minimal Coupling & $-1.54$ & Yes \\
\hline
\end{tabular}
\caption{Numerical values of the linear stability coefficient $\beta$ for the five models. In all cases, $\beta < 0$, confirming that the bounce solution is a local attractor.}
\label{tab:beta_values}
\end{table}

\subsection{Physical Interpretation}

The existence of a negative $\beta$ can be traced to the higher-derivative $f_G$ terms and the $f_T$ coupling. These terms introduce an effective \textit{viscous damping} in the contracting branch. Physically, as the universe contracts, curvature and matter-energy increase, activating the $f(R,G,T)$ corrections which then act to slow down the contraction and eventually reverse it. This damping mechanism ensures that trajectories with a range of initial contraction rates converge to the same bounce solution, thereby resolving the fine-tuning concern.

\subsection{Limitations and Future Work}

The linear analysis presented here is local and does not guarantee global attractor behavior. A full global analysis, including the construction of a Lyapunov functional for the full phase space, is beyond the scope of this paper and is left for future work. Nevertheless, the local attractor property demonstrated here, combined with the numerical evidence that all our trajectories indeed pass through the bounce without fine-tuning, strongly suggests that the bouncing solutions are robust and not of measure zero.
	
	\bibliographystyle{unsrtnat}
	\bibliography{References-E02}
\end{document}